\documentclass[%
reprint,
superscriptaddress,
showpacs,preprintnumbers,
 amsmath,amssymb,
 aps,
 pra,
]{revtex4-1}

\usepackage{graphicx}
\usepackage{dcolumn}
\usepackage{bm}

\usepackage{amsmath,amsfonts,amssymb,amsthm}
\usepackage{txfonts}
\usepackage{enumerate}
\usepackage{color}
\usepackage{cases}
\usepackage{comment}
\usepackage{braket}

\makeatletter
\renewcommand{\theequation}{\arabic{section}.\arabic{equation}}
\@addtoreset{equation}{section}
\makeatother

\begin{document}


\title{Integrable model for density-modulated quantum condensates: \\ solitons passing through a soliton lattice}
\author{Daisuke A. Takahashi}
\email{daisuke.takahashi.ss@riken.jp}
\affiliation{Department of Basic Science, The University of Tokyo, Tokyo 153-8902, Japan}
\affiliation{Research and Education Center for Natural Sciences, Keio University, Hiyoshi 4-1-1, Yokohama, Kanagawa 223-8521, Japan}
\affiliation{Department of Theoretical Physics, Research School of Physics and Engineering, Australian National University, Canberra ACT 0200, Australia}
\affiliation{RIKEN Center for Emergent Matter Science (CEMS), Wako, Saitama 351-0198, Japan}
\altaffiliation[Present address: 1. RIKEN Center for Emergent Matter Science (CEMS), Wako, Saitama 351-0198, Japan. 2. Research and Education Center for Natural Sciences, Keio University, Hiyoshi 4-1-1, Yokohama, Kanagawa 223-8521, Japan.]{}

\date{\today}

\begin{abstract}
An integrable model possessing inhomogeneous ground states is proposed as an effective model of non-uniform quantum condensates such as supersolids and Fulde--Ferrell--Larkin--Ovchinnikov superfluids. 
The model is a higher-order analog of the nonlinear Schr\"odinger equation. 
We derive an $n$-soliton solution via the inverse scattering theory with elliptic-functional background, 
and reveal various kinds of soliton dynamics such as dark soliton billiards, dislocations, gray solitons, and envelope solitons. 
We also provide the exact bosonic and fermionic quasiparticle eigenstates and show their tunneling phenomena. 
The solutions are expressed by a determinant of theta functions.
\end{abstract}

\maketitle

\section{Introduction}
	Spatially inhomogeneous quantum condensates have been attracting a lot of attention for a long time. For bosonic condensates, the supersolid phase, which was originally discussed four decades ago \cite{AndreevLifshitz1969,Chester1970,Leggett1970}, has received a renewed interest since the torsional oscillator experiments of ${}^4$He \cite{EKimMHWChanN,EKimMHWChanS}. While the most recent work \cite{DYKimMHWChan} has concluded the absence of supersolidity, the candidate of supersolid is also proposed in Rydberg matters \cite{HenkelNathPohl,CintiJainBoninsegniMicheliZollerPupillo}.  For fermionic condensates, the realization and observation of Fulde--Ferrell (FF) \cite{FuldeFerrell} and Larkin--Ovchinnikov (LO) \cite{LarkinOvchinnikov} states have been a long-standing topic. 
	Within a framework of self-consistent Bogoliubov--de Gennes (BdG) formalism, the LO state is shown to be a ground state in the presence of a magnetic field or a population imbalance \cite{Machida:1984zz,MizushimaMachidaIchioka}. 
	There are also various experimental candidates, for example, CeCoIn${}_5$ in condensed matters \cite{Radovan2003,KakuyanagiPRL2005}. In ultracold atoms, the spin-imbalanced superfluid $ {}^6 $Li has been investigated as a candidate \cite{ZwierleinScience2006,PartridgeScience2006,LiaoNature2010}. While other phases have been reported \cite{PhysRevLett.91.247002,PhysRevA.69.063602,PhysRevLett.97.070402,1742-5468-2007-11-P11012,ShinNature2008}, the high controllability of system parameters and rich atomic species in ultracold atomic systems still provide good opportunities to investigate these nonuniform phases. 
	The problem equivalent to the BdG systems also appears in high-energy physics. Modulated phases in the Nambu--Jona-Lasinio or the Gross--Neveu (GN) model \cite{Nambu:1961tp,Gross:1974jv} are studied \cite{Thies:2003kk,Schnetz:2004vr,Basar:2008im,Basar:2008ki,Basar:2009fg,Correa:2009xa} as an effective model of quantum chromodynamics \cite{Hatsuda:1994pi}. \\ 
	\indent To study the quantum condensates, in addition to the density-functional approach or the Thomas Fermi approximation \cite{PhysRevA.78.043616}, the nonlinear Schr\"odinger (NLS) equation and its generalizations are often used, and referred to as the Gross--Pitaevskii or the Ginzburg--Landau (GL) equation for bosonic or fermionic systems. Though Gorkov's original derivation justifies the GL description only near $ T=T_c $, the recent studies show that the gap function obeys the NLS equation with higher-order corrections even near $ T=0 $ \cite{PieriStrinati,Basar:2008im,Basar:2008ki,Basar:2009fg,Correa:2009xa}. \\
	\indent Many theoretical studies have established a common and model-independent understanding for the mechanism of spontaneous modulation in the ground states and the low-energy excitations around them. Compared to stationary states, however, the nature of nonlinear excitations such as solitons or vortices passing through these modulated condensates has not been fully investigated yet, because of the difficulty to treat time-dependent phenomena. 
	Solitons are also important to understand transport phenomena past an obstacle in non-stationary regimes \cite{Hakim1997,PhysRevLett.99.160405}. 
	To investigate these issues, an integrable model will play a prominent role, 
	since we can access the various kinds of dynamics exactly. 
	We also mention that the chiral soliton-lattice structure in a chiral helimagnet has been directly observed by Lorenz microscopy \cite{PhysRevLett.108.107202}, and the sine-Gordon soliton running through this lattice has been investigated \cite{PhysRevB.79.134436}. The collision between the soliton and the surface in mixed phases \cite{ZwierleinScience2006,PartridgeScience2006,PhysRevLett.91.247002,PhysRevA.69.063602,PhysRevLett.97.070402,1742-5468-2007-11-P11012} will also be important. The ``supersolitons'' in two-component Bose condensates are proposed in Ref.~\cite{PhysRevLett.101.144101}. Thus, understanding the soliton motion with pattern-formed background is becoming more important today.\\ 
	\indent In this paper, we propose an integrable model of non-uniform quantum condensates using the higher-order differential equations in the NLS hierarchy. Solving it by the inverse scattering theory (IST) with soliton-lattice background, we obtain an $n$-soliton solution written by elliptic theta functions. 
	The obtained soliton solutions are classified based on the shape of the background lattices and the eigenvalues of solitons, and we propose the following: If the background lattice is almost an array of well-separated dark solitons, (in other words, if the elliptic parameter of the modulated condensate is nearly $ m\simeq 1 $), the system exhibits three kinds of solitons: \textit{the dark soliton billiards, the static dislocations, and the gray solitons.} If the background lattice has rather trigonometric shape (if $  m\simeq 0 $), we observe \textit{the envelope solitons.} The behavior of the envelope soliton is similar to those observed in supersolid theoretical models. 
	Furthermore, we also provide exact eigenstates for bosonic and fermionic Bogoliubov quasiparticles. The bosonic ones are essential in investigation of Nambu-Goldstone (NG) modes and linear stability. \\ 
	\indent Note that the solitons given here are different from gap solitons (See, e.g., \cite{MuruganandamAdhikari} and references therein.). The system forms a pattern not by a periodic external force but by itself, and hence the modulated background and the solitons influence each other.\\ 
	\indent The organization of this paper is as follows. Section~\ref{sec:mainresult} summarizes the main result of this paper. The idea of finding a model, the determination of density-modulated ground states, the eigenstates for bosonic and fermionic Bogoliubov quasiparticles, and the expressions of $n$-soliton solutions, their classifications and the animation examples, are included in this section. In Secs.~\ref{sec:bdgwithakns1bg}-\ref{sec:timeevo}, mathematical details of formulations are presented. Section~\ref{sec:bdgwithakns1bg} provides fermionic eigenstates of the BdG operator for the elliptic-functional background. In Sec.~\ref{sec:ist}, we formulate the IST with the soliton-lattice background. In Sec.~\ref{sec:timeevo2}, we describe a general criterion in order for the higher-order NLS equations to have the solution of the lower-order ones. In Sec.~\ref{sec:timeevo}, we determine the time evolution of general higher-order NLS equations with elliptic background. 
	In Sec.~\ref{sec:summary}, we give a summary and perspective. Appendices provide details of calculations and conventions and formulas of elliptic functions.

\section{Summary of Main Result}\label{sec:mainresult}
	\indent The energy functional of the model proposed in this paper is
	\begin{align}
		H&=c_3I_3+c_5I_5, \label{eq:Hami}
	\end{align}
	where $ c_3, c_5 $ are real and  $ I_3=\int\mathrm{d}x\,\bigl(|\psi_x|^2+|\psi|^4\bigr) $ and $ I_5=\int\!\mathrm{d}x\,\Bigl\{|\psi_{xx}|^2+6|\psi|^2|\psi_x|^2+[(|\psi|^2)_x]^2+2|\psi|^6\Bigr\} $ are the third and fifth conserved quantities in the NLS hierarchy \cite{ZakharovShabat2,FaddeevTakhtajan}. 
	We are interested in the soliton motion with the finite-density background, so we consider $ H-\mu N $, where $ \mu $ is a chemical potential and $ N=I_1=\int\mathrm{d}x|\psi|^2 $ is the particle number. The resulting partial differential equation $ \mathrm{i}\partial_t\psi=\delta (H-\mu N)/\delta \psi^* $ is given by
	\begin{align}
		\mathrm{i}\psi_t=& -\mu \psi+c_3\bigl(-\psi_{xx}+2|\psi|^2\psi\bigr)+ \nonumber \\
		&\qquad c_5\bigl[\psi_{xxxx}-2(|\psi|^2)_{xx}\psi-3\psi^*(\psi^2)_{xx}+6|\psi|^4\psi\bigr], \label{eq:I3I5q}
	\end{align}
	where the subscripts $ t $ and $ x $ denote the differentiation. 

\subsection{Idea of model construction}\label{subsec:ideainmainresult}
	Let us see how to find an integrable model of density-modulated condensates. We first demonstrate that the model with a non-local interaction such as soft-core bosons \cite{PomeauRica,JosserandPomeauRica,SepulvedaJosserandRica,KunimiNagaiKato,KunimiConf,PhysRevB.86.060510}, which are used as a model of supersolid, can be approximated by a higher-order differential equation. Consider, for example, the Gaussian-type two-body interaction $ V(x)=\frac{V_0(x)}{2a\sqrt{\pi}}\mathrm{e}^{-x^2/(4a^2)} $, where $ a>0 $ is an interaction length and  $ V_0(x) $ is a slowly-varying even function. Using the expansion $ \frac{1}{2a\sqrt{\pi}}\mathrm{e}^{-x^2/(4a^2)}=\delta(x)+a^2\delta''(x)+\frac{a^4}{2}\delta''''(x)+\dotsb $, 
	the NLS equation for the soft-core model $ \mathrm{i}\partial_t\psi(x,t)=-\partial_x^2\psi(x,t)+\int\mathrm{d}yV(x-y)|\psi(y,t)|^2\psi(x,t) $ can be approximated as 
	\begin{align}
		\mathrm{i}\psi_t=&-\psi_{xx}+\left[ \tilde{V}_0|\psi|^2+\tilde{V}_2(|\psi|^2)_{xx}+\tilde{V}_4(|\psi|^2)_{xxxx} \right]\psi \label{eq:appsoftcore}
	\end{align}
	up to $ O(a^4) $, where $ \tilde{V}_0=V_0(0)+a^2V_0''(0)+\frac{a^4}{2}V_0''''(0),\, \tilde{V}_2=a^2(V_0(0)+3a^2V_0''(0)),\, \text{and } \tilde{V}_4=\frac{a^4}{2}V_0(0) $. 
	Even though Eq. (\ref{eq:appsoftcore}) is too rough an approximation for the original non-local model, it exhibits a roton minimum in the Bogoliubov spectrum and has an inhomogeneous ground state in certain parameter regions, as similar to Ref.~\cite{SepulvedaJosserandRica}. 
	It is reasonable that the higher-order derivative can induce a spatial order, because the energy of the system should have a minimum at a non-zero momentum, and the simplest such example is given by $ E(p) \sim -p^2+p^4 $. 
	In fact, many pattern-forming models 
	have higher-order derivatives, such as the convective instability \cite{SwiftHohenberg}, the magnetic fluids \cite{RichterBarashenkov}, and the generalized GL theory \cite{BuzdinKachkachi}. \\
	\indent While Eq. (\ref{eq:appsoftcore}) is not integrable, we can construct an integrable model including higher-order derivatives by using the higher-order conserved quantities in the NLS hierarchy. 
	Since the even-number $ I_n $'s break a parity symmetry \cite{ZakharovShabat2}, the minimal model including higher-order derivatives is given by $ H=c_3I_3+c_5I_5 $, that is, the model (\ref{eq:Hami}).\\
	\indent The system is unstable if $ c_5<0 $ since the dispersion of the linearized operator becomes $ \epsilon\sim -k^4 $. We can also confirm that the ground state becomes a trivial uniform state if both $ c_3 $ and  $ c_5 $ are positive. Thus, the non-trivial physics arises when $ c_3<0 $ and $ c_5>0 $. So, we mainly consider this case. 

\subsection{Density-modulated ground state}\label{subsec:maings}
\begin{figure}[tb]
		\begin{center}
		\includegraphics[scale=1.0]{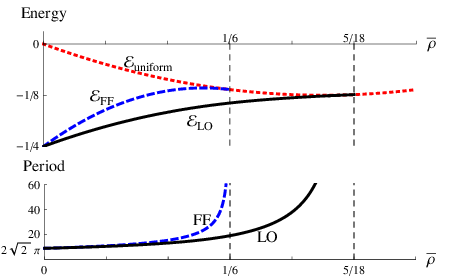}
		\caption{\label{fig:energycompare} (Color online) The energies per particle and the periods for FF and LO states. We set $ (c_3,c_5)=(-1,1) $. For reference, we also show the energy for the uniform state $ \psi=\sqrt{\bar{\rho}} $. 
		}
		\end{center}
	\end{figure}

	Let us begin the analysis of the model (\ref{eq:Hami}) in detail. We first determine the static ground state. Although the general stationary solution to Eq.~(\ref{eq:I3I5q}) is the quasi-periodic Riemann theta function with genus $ g=3 $ \cite{Krichever1977,BelokolosBobenkoEnolskiiItsMatveev,TanakaDate,GesztesyHolden}, 
	here we assume that higher-genus solutions are energetically unfavored, 
	and only consider the two candidates, i.e., the FF and LO states:
	\begin{align}
		\psi_{\text{FF}}(x)&=\sqrt{\bar{\rho}}\mathrm{e}^{\mathrm{i}px} \\
		\psi_{\text{LO}}(x)&=\mathrm{i}\sqrt{m}\alpha\operatorname{sn}(\alpha x|m),\quad \alpha=\sqrt{\bar{\rho}/Q(m)}, \label{eq:LO}  
	\end{align}
	where $ Q(m):=\sqrt{1-\frac{E(m)}{K(m)}} $ with $ K(m) $ and $ E(m) $ being the complete elliptic integral of the first and second kind, respectively. 
	$ \bar{\rho} $ is an average of particle number density, and $ p $ and $ m $ are variational parameters chosen to minimize the energy. 
	These states solve Eq.~(\ref{eq:I3I5q}) and chemical potentials are determined as $ \mu_{\text{FF}}=c_3(p^2+2\bar{\rho})+c_5(p^4+12\bar{\rho}p^2+6\bar{\rho}^2) $ for  $ \psi_{\text{FF}} $ and $ \mu_{\text{LO}}=c_3(m+1)\alpha^2+c_5(m^2+4m+1)\alpha^4 $ for $ \psi_{\text{LO}} $. 
	Let $ \mathcal{E}_{\text{FF}}(\bar{\rho}) $ and $ \mathcal{E}_{\text{LO}}(\bar{\rho}) $ be the energies per particle for FF and LO states, in which the variational parameters $ p $ and $ m $ are chosen to minimize the energy for fixed $ \bar{\rho} $. 
	See Appendix~\ref{sec:minimizefflo} for their evaluation.
	Figure~\ref{fig:energycompare} shows the plot of $ \mathcal{E}_{\text{FF}}(\bar{\rho}) $ and $ \mathcal{E}_{\text{LO}}(\bar{\rho}) $ and corresponding periods. 
	From Fig.~\ref{fig:energycompare}, we can conclude that the density-modulated LO state becomes the lowest-energy state if the particle density is small $ (\bar{\rho}<\frac{-5c_3}{18c_5}) $. 
	Note also that if the density becomes smaller, the period becomes shorter. This behavior is similar to the gap function of the BdG/GN models \cite{Machida:1984zz,Thies:2003kk}. 
	As shown in Subsec.~\ref{subsec:bosonicbogoliubov}, this LO state is linearly stable. 

\subsection{AKNS form}\label{subsec:mainakns}
	Next, we want to present the quasiparticle wavefunctions and soliton dynamics  in the presence of the LO background. To do this, we introduce a few theoretical tools from soliton theory, that is, the Ablowitz--Kaup--Newell--Segur (AKNS) representation and the uniformization variable of the genus-1 Riemann surface. \\
	\indent Equation (\ref{eq:I3I5q}) enjoys the AKNS representation \cite{AKNS1974,FaddeevTakhtajan}: 
	\begin{align}
		\partial_xf=U(x,t,\lambda)f,\quad \partial_tf=V(x,t,\lambda) f, \label{eq:AKNS}
	\end{align}
	where  $ \lambda $ is a spectral parameter and $ f $ is a two-component vector, called the Jost function. The matrices $ U $ and $ V $ for Eq. (\ref{eq:I3I5q}) are given by \cite{FaddeevTakhtajan,Takahashi:2012aw}
	\begin{align}
		U=\begin{pmatrix} -\mathrm{i}\lambda & q \\ r & \mathrm{i}\lambda \end{pmatrix}, \quad V=-\mu V^{(1)}+c_3V^{(3)}+c_5V^{(5)},
	\end{align}
	where $q=r^*=-\mathrm{i}\psi $, $ V^{(n)}=\sum_{j=0}^{n-1}(-2\lambda)^{n-j-1}M^{(j)} $, and $ M^{(j)} $'s are the formal Laurent expansion solution of $ M_x=[U,M] $ (see Sec.~\ref{sec:timeevo2}). The explicit forms of $ M^{(j)} $ for $ j\le 4 $ are given in Ref.~\cite{Takahashi:2012aw} with $ \psi=\mathrm{i}q $ and $ \psi^*=-\mathrm{i}r $. 
	The compatibility condition $ U_t-V_x+[U,V]=0 $ yields Eq.~(\ref{eq:I3I5q}). 
	\\ 
	\indent It is known that quasi-periodic solutions in integrable equations have an associated higher-genus Riemann surface, which plays an essential role in the algebro-geometric formulation \cite{Krichever1977,BelokolosBobenkoEnolskiiItsMatveev,TanakaDate,GesztesyHolden}. Defining $ V_3:=\left.V\right|_{c_3=1,\ c_5=0} $, the genus-1 Riemann surface for the sn function (\ref{eq:LO}) is given by 
	\begin{align}
		\omega^2=\left.\det V_3\right|_{\psi=\psi^{}_{\text{LO}}}=4\lambda^4-2\alpha^2(1+m)\lambda^2+\!\tfrac{1}{4}\alpha^4(1-m)^2. \label{eq:genus1RS}
	\end{align}
	This Riemann surface is parametrized by the following uniformization variable $ z $ \cite{BelokolosBobenkoEnolskiiItsMatveev}: 
	\begin{align}
		\lambda(z)&=-\frac{\alpha}{2}\operatorname{dn}(\mathrm{i}z)\operatorname{dn}(\mathrm{i}z'), \label{eq:uniformization} \\
		\omega(z)&=\alpha\lambda'(z)=\frac{\alpha^2}{2}\left[ \operatorname{dn}^2(\mathrm{i}z')-\operatorname{dn}^2(\mathrm{i}z) \right].
	\end{align}
	Here and hereafter, the elliptic parameter $ m $ is omitted, and we write $ z'=K'-z $. The convention of elliptic functions is summarized in Appendix~\ref{app:ellipconv}. The time evolution of Jost functions are described by $ \omega_3^2:=\left.\det V\right|_{\psi=\psi^{}_{\text{LO}}}= [4c_5\lambda^2+c_5\alpha^2(m+1)+c_3]^2\omega^2 $, which is parametrized in the same way:
	\begin{align}
		\omega_3(z)=[4c_5\lambda(z)^2+c_5\alpha^2(m+1)+c_3]\omega(z). \label{eq:mainomega5}
	\end{align}
	Using these tools, we obtain the eigenfunctions for Bogoliubov quasiparticles and soliton solutions shown below. The usage of these tools is demonstrated in Secs. \ref{sec:bdgwithakns1bg}-\ref{sec:timeevo}.

\subsection{Fermionic BdG quasiparticle eigenstates}
	\begin{figure}[tb]
		\begin{center}
		\includegraphics{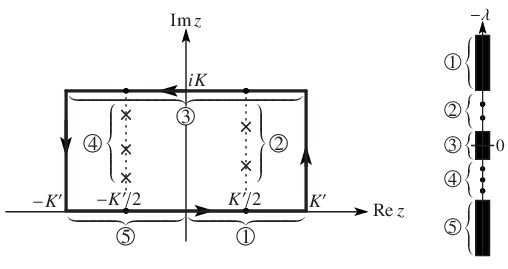}
		\caption{\label{fig:univar}Relation between the uniformization variable $ z $ and the spectral parameter $ -\lambda $. 
		Regions with the same circled numbers correspond to each other. The scattering states exist on the lines $ \operatorname{Im}z=0 $ and $ \operatorname{Im}z=K $. The cross marks on the lines $ \operatorname{Re}z=\pm K'/2 $ represent discrete eigenvalues for bound states. 
		The rectangular contour shown by the bold line is used to derive the Gelfand-Levitan-Marchenko (GLM) equation in the IST.
		}
		\end{center}
	\end{figure}
	The first equation of Eq.~(\ref{eq:AKNS}) is equivalent to the \textit{fermionic} BdG equation for the quasiparticle with eigenenergy $ -\lambda $:
	\begin{align}
		\begin{pmatrix} -\mathrm{i}\partial_x & \psi
		 \\ \psi^* & \mathrm{i}\partial_x \end{pmatrix}f=-\lambda f. \label{eq:ZSBdG}
	\end{align}
	 Then, the two linearly independent solutions for Eq.~(\ref{eq:ZSBdG}) with $ \lambda=\lambda(z) $ and  $ \psi=\psi_{\text{LO}} $ are given by \cite{BrazovskiiGordyuninKirova,Horovitz:1981,HaraNagai,Basar:2008ki}
	\begin{align}
		&f_0(t,x,z)=\nonumber \\
		&\frac{\mathrm{i}\alpha\vartheta_2\vartheta_4\mathrm{e}^{\mathrm{i}[k(z)-(\pi\alpha)/(4K)]x}\mathrm{e}^{\mathrm{i}\omega_3(z)t}}{\vartheta_3\vartheta_4(\frac{\alpha x}{2K})}\begin{pmatrix} \vartheta_1(\frac{\alpha x-\mathrm{i}z}{2K})/\vartheta_4(\frac{\mathrm{i}z}{2K}) \\ \vartheta_4(\frac{\alpha x-\mathrm{i}z}{2K})/\vartheta_1(\frac{\mathrm{i}z}{2K})\end{pmatrix} \label{eq:Jost0}
	\end{align}
	and  $ f_0(t,x,z') $, where $ \vartheta_a(u)=\vartheta_a(u,q) $ is the theta function with $ q=\mathrm{e}^{-\pi K'/K} $,  $ \vartheta_a=\vartheta_a(0) $, and 
	\begin{align}
	 	k(z):=-\frac{\mathrm{i}\alpha}{2}\left[Z(\mathrm{i}z)-Z(\mathrm{i}z')\right].
	\end{align}
	is a crystal momentum, with $ Z(u) $ being the Jacobi zeta function (see Appendix~\ref{app:ellipconv} for their definition). 
	The fermionic spectrum is given by the condition $ \omega^2>0 $ in Eq.~(\ref{eq:genus1RS}) (Ref. \cite{Takahashi:2012aw}), i.e.,  $ |\lambda|>\frac{\alpha(1+\sqrt{m})}{2} $  and $ |\lambda|<\frac{\alpha(1-\sqrt{m})}{2} $, corresponding to  $ \operatorname{Im}z=nK, n\in\mathbb{Z} $ in $ z $ plane. 
	The bound states appear in the energy gap, which corresponds to $ \operatorname{Re}z=\pm\frac{K'}{2} $. 
	See Fig.~\ref{fig:univar}. 

\subsection{Bosonic quasiparticles, NG modes, and linear stability}\label{subsec:bosonicbogoliubov}
	\begin{figure}[tb]
		\begin{center}
		\includegraphics[scale=.9]{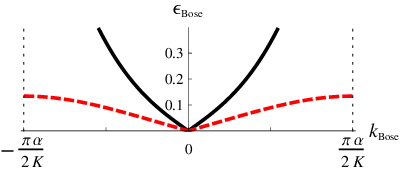} \\
		\caption{\label{fig:bsbg} (Color online) Bosonic Bogoliubov spectrum for the sn state [Eq.~(\ref{eq:LO})]. Here $ (k_{\text{Bose}},\epsilon_{\text{Bose}})=(2k(z),-2\omega_3(z)) $. We set $ (c_3,c_5)=(-1,1),\, m=0.3 $, and $ \alpha=0.638 $.  The red dashed (black solid) line represents the lattice-vibration (Bogoliubov) phonons and corresponds to {\normalsize\textcircled{\small 3}} ({\normalsize\textcircled{\small 1}} and {\normalsize\textcircled{\small 5}}) in Fig.~\ref{fig:univar}.}
		\end{center}
	\end{figure}
	\begin{figure}[tb]
		\begin{center}
		\includegraphics[scale=.9]{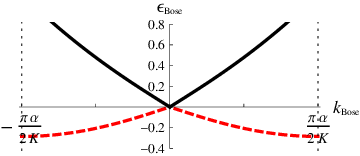} \\
		\caption{\label{fig:bsbgc3} (Color online) The same plot as Fig.~\ref{fig:bsbg}, but we set $ (c_3,c_5)=(1,0) $, which corresponds to the ordinary NLS system.  $ (k_{\text{Bose}},\epsilon_{\text{Bose}})=(2k(z),-2\omega(z)) $.}
		\end{center}
	\end{figure}
	Next, we derive the \textit{bosonic} Bogoliubov spectrum by regarding  $ \psi $ as a \textit{bosonic} condensate. 
	The bosonic Bogoliubov equation is obtained by linearization of Eq.~(\ref{eq:I3I5q}) (see e.g., Ref.~\cite{DalfovoGiorginiPitaevskiiStringari}); writing the linearized fields $ (\delta \psi,\delta \psi^*)=(u,v) $, we obtain 
	\begin{align}
		\mathrm{i}u_t&=-\mu u+c_3[-u_{xx}+2(2|\psi|^2u+\psi^2v)]+c_5\bigl[u_{xxxx} \nonumber \\ 
		&\quad-2(\psi^*u+\psi v)_{xx}\psi-2(|\psi|^2)_{xx}u-3(\psi^2)_{xx}v \nonumber \\
		&\quad-6\psi^*(\psi u)_{xx}+6|\psi|^2(3|\psi|^2u+2\psi^2v)\bigr], \label{eq:bosonicBdGu} \\ 
		\mathrm{i}v_t&=\mu v-c_3[-v_{xx}+2(2|\psi|^2v+\psi^{*2}u)]-c_5\bigl[v_{xxxx} \nonumber \\ 
		&\quad-2(\psi v+\psi^*u)_{xx}\psi^*-2(|\psi|^2)_{xx}v-3(\psi^{*2})_{xx}u \nonumber \\
		&\quad-6\psi(\psi^*v)_{xx}+6|\psi|^2(3|\psi|^2v+2\psi^{*2}u)\bigr]. \label{eq:bosonicBdGv} 
	\end{align}
	The stationary Bogoliubov equation with the eigenenergy $ \epsilon $ is obtained by substitution $ (u(x,t),v(x,t))=(u(x),v(x))\mathrm{e}^{-\mathrm{i}\epsilon t} $, and its spectrum determines the linear stability of a given stationary state. 
	We can solve the above equation by the squared eigenfunctions \cite{Kaup1976,ChenChenHuang}.
	Let $ f=(u_{\text{Fermi}},v_{\text{Fermi}})^T $ be a solution of Eq. (\ref{eq:AKNS}). Then, 
	\begin{align}
		\begin{pmatrix}u_{\text{Bose}} \\ v_{\text{Bose}} \end{pmatrix} = \begin{pmatrix}u_{\text{Fermi}}^2 \\ v_{\text{Fermi}}^2 \end{pmatrix} 
	\end{align}
	 solves the bosonic Bogoliubov equation (\ref{eq:bosonicBdGu}) and (\ref{eq:bosonicBdGv}). 
	 Therefore, we can draw the dispersion relation of linearized waves by plotting $(2k(z),-2\omega_3(z))$; see Fig.~\ref{fig:bsbg}. 
	Since the condensate breaks two continuous symmetries, i.e., the U(1)-gauge and the translational symmetries, we observe two NG modes, the Bogoliubov phonon and the lattice-vibration phonon. 
	We can confirm that the two zero modes $ (u_{\text{Bose}},v_{\text{Bose}})=(\mathrm{i}\psi,-\mathrm{i}\psi^*) $ and $ (\psi_x,\psi_x^*) $ originating from U(1) and translational symmetry breaking are orthogonal with respect to $\sigma$-inner products \cite{Takahashi2015101}, and thus they independently form type-I NG modes with linear dispersion. This is consistent with the counting theory of NG modes based on the Bogoliubov theory \cite{Takahashi2015101,PhysRevD.91.025018,PhysRevB.91.184501}. If the counting theory is formulated based on the Lie algebra \cite{Watanabe:2011ec,Watanabe:2012hr,Hidaka:2012ym}, the commutativity is to be checked in the sense of the centrally-extended algebra \cite{PhysRevLett.112.191804,PhysRevLett.113.120403}. \\
	\indent Figure~\ref{fig:bsbg} also proves that there is no negative or complex eigenvalue. Thus, the LO state is stable. 
	On the other hand, if we plot the same relation for the $ (c_3,c_5)=(1,0) $ system, i.e., for the ordinary NLS system, we find that the lattice-vibration mode has the negative dispersion, as shown in Fig.~\ref{fig:bsbgc3}. The presence of negative energy dispersion suggests that the LO state is at least metastable at zero temperature, but it may become unstable if the system is thermally excited, for example, if the finite-temperature effect is included.\\
	\indent Note that the bosonic Bogoliubov equation always exhibits positive- and negative-energy eigenstates in pairs. However, when we plot the dispersion relation, we must use the physical solutions satisfying $ \int\mathrm{d}x(|u_{\text{Bose}}|^2-|v_{\text{Bose}}|^2)\ge0 $, since only these solutions are used in the definition of the bosonic Bogoliubov transformation. The solutions with $ \int\mathrm{d}x(|u_{\text{Bose}}|^2-|v_{\text{Bose}}|^2)<0 $ are regarded as unphysical.
%
%
\subsection{$n$-soliton solution}\label{sec:nsolitonrealcase}
	\begin{figure}[t]
		\begin{center}
		\includegraphics[width=0.98\linewidth]{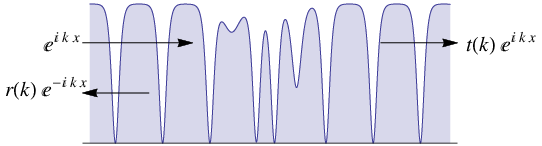} 
		\caption{\label{fig:istschematic} (Color online) The IST with elliptic background. We impose the boundary condition such that $ \psi(x) $ asymptotically tends to the soliton lattice, i.e.,  $ \psi(x\to-\infty)=\psi_{\text{LO}}(x) $ and $ \psi(x\to+\infty)=\psi_{\text{LO}}(x-x_0)\mathrm{e}^{2\mathrm{i}\varphi_0} $, where $ x_0,\varphi_0 $ represents the shift induced by solitons and radiations. The inverse problem of the ZS operator, i.e., determination of the potential $ \psi(x) $ from the scattering data, is solved by the GLM equation (Sec.~\ref{sec:ist}).} 
		\end{center}
	\end{figure}

	\begin{figure}[t]
		\begin{center}
		\includegraphics[scale=.75]{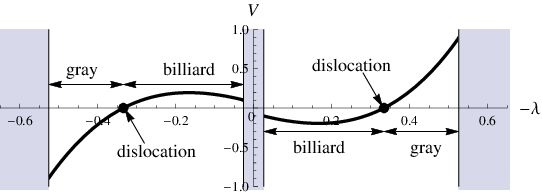}
		\caption{\label{fig:onesolvelo}(Color online) Relation between the velocity of the soliton $ V $ and the spectral parameter $ -\lambda $. We use $ (c_3,c_5)=(-1,1), \ m=0.82, $ and $  \alpha=0.525 $.  For the eigenvalue $ -\lambda(z_j) $, the soliton velocity is given by $ V_j=-\frac{\operatorname{Im}\omega_3(z_j)}{\operatorname{Im}k(z_j)} $.  The shaded areas represent the continuous spectra. When $ m\simeq 1 $, the width of the central band becomes very narrow. The soliton with zero velocity is a static dislocation. The other two represent the dark soliton billiard (Figure~\ref{fig:onesoliton01}(a), animation1-1.gif) and the gray soliton that has a small dip (animation1-2.gif). 
		}
		\end{center}
	\end{figure}
	\begin{figure}[t]
		\begin{center}
		\includegraphics[scale=.75]{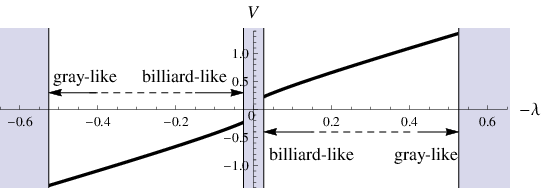}
		\caption{\label{fig:onesolvelo2}(Color online) The same figure as Fig.~\ref{fig:onesolvelo}, but we consider $ (c_3,c_5)=(1,0) $, i.e., the ordinary NLS system. In this case, the velocity $ V_j=-\frac{\operatorname{Im}\omega(z_j)}{\operatorname{Im}k(z_j)} $ becomes a monotonic function, and there is no zero-velocity soliton (i.e., no static dislocation). The shape of the soliton continuously changes from the dark soliton billiard to the gray soliton. 
		}
		\end{center}
	\end{figure}
	\begin{figure}[t]
		\begin{center}
		\includegraphics[scale=1]{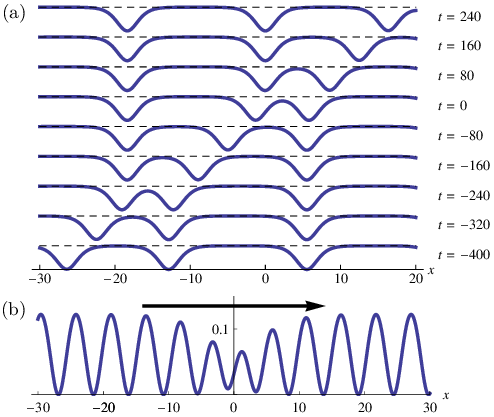}
		\caption{\label{fig:onesoliton01}(Color online) Examples of the one-soliton solution [Eq.~(\ref{eq:mainnsol}) with $ n=1 $]. The plot shows  the amplitude  $ |\psi(t,x)|^2 $. We set $ (c_3,c_5)=(-1,1) $. (a)~Dark soliton billiard. Parameters are $ m=0.999,\, \alpha=0.527,\, z_1= -0.5K'+0.3\mathrm{i}K, $ and $ C_1=1.16 $.  (b)~Snapshot of the envelope soliton when the background is almost trigonometric. The arrow shows the direction of the soliton propagation. Parameters are  $ m=0.3,\, \alpha=0.638,\, z_1= -0.5K'+0.55\mathrm{i}K, $ and $ C_1=4.39 $. See also animation files in Ref.~\cite{SeeAnimation}, where animation1-1.gif corresponds to (a) and 1-5.gif to (b).}
		\end{center}
	\end{figure}

	We now show the soliton dynamics in the presence of density-modulated background. 
	We can formulate the IST for the elliptic-function background (see Fig. \ref{fig:istschematic}). The GLM equation can be derived in the same way as the uniform background, and the reflectionless potentials can be constructed as a special solution (Sec.~\ref{sec:ist}). For these potentials, the time-evolution problem under the AKNS system [Eq.~(\ref{eq:AKNS})] can be solved for the higher-order NLS equations (Sec.~\ref{sec:timeevo}). In fact, Sec.~\ref{sec:timeevo} provides a more general solution --- We solve not only the AKNS$_3$ but also  the general AKNS$_n$ equation.  We mention that the KdV equation with elliptic background has been solved in Ref.~\cite{KuznetsovMikhailov}. \\ 
	\indent Here we extract the main result from Secs.~\ref{sec:ist}-\ref{sec:timeevo}. Let us assume that the potential has  $ n $ discrete eigenvalues $ \lambda(z_j), j=1,\dots,n $, with $ z_j=s_j\frac{K'}{2}+\mathrm{i}\eta_j,\ s_j=\pm1 $, and $  0<\eta_j<K $ (see Fig.~\ref{fig:univar}). We also write $ z_j'=K'-z_j $.  Then, the determinant expression of the $n$-soliton solution is given by
	\begin{align}
		\psi(t,x)=\psi_{\text{LO}}(x)\frac{\det[I_n+\mathcal{E\tilde{A}E}]}{\det[I_n+\mathcal{EME}]}, \label{eq:mainnsol}
	\end{align}
	where the $ n\times n $ matrices $ \mathcal{E}(t,x),\ \mathcal{M}(x), $ and $ \mathcal{\tilde{A}}(x) $ are defined as follows. $ \mathcal{E}(t,x)=\operatorname{diag}[e_1(t,x),\dots,e_n(t,x)] $ with
	\begin{align}
		e_j(t,x)=C_j \mathrm{e}^{-\mathrm{i}\omega_3(z_j)t-\mathrm{i}k(z_j)x},\quad C_j>0.
	\end{align}
	The $(i,j)$-components of $ \mathcal{M} $ and $ \mathcal{\tilde{A}} $ are defined by
	\begin{align}
		\mathcal{M}_{ij}(x)&=-2\alpha\frac{ \vartheta_2\vartheta_4\vartheta_4(\frac{\alpha x+\mathrm{i}(z_i-z_j')}{2K})}{\vartheta_3\vartheta_4(\frac{\alpha x}{2K})\vartheta_1(\frac{\mathrm{i}(z_i-z_j')}{2K})}, \\
		\mathcal{\tilde{A}}_{ij}(x)&=-2\alpha \frac{\vartheta_2\vartheta_4\vartheta_1(\frac{\alpha x+\mathrm{i}(z_i-z_j')}{2K})}{\vartheta_3\vartheta_1(\frac{\alpha x}{2K})\vartheta_1(\frac{\mathrm{i}(z_i-z_j')}{2K})}\frac{\vartheta_4(\frac{\mathrm{i}z_i}{2K})\vartheta_1(\frac{\mathrm{i}z_j'}{2K})}{\vartheta_1(\frac{\mathrm{i}z_i}{2K})\vartheta_4(\frac{\mathrm{i}z_j'}{2K})}.
	\end{align}
	The velocity of the $j$-th soliton is given by $ V_j=-\frac{\operatorname{Im}\omega_3(z_j)}{\operatorname{Im}k(z_j)} $. The value of $ C_j $ determines the initial position of this soliton.  $ \psi(t,x) $ has the asymptotic form
	\begin{align}
		\psi(t,x)\to \begin{cases} \psi_{\text{LO}}(x) &  (x\to-\infty) \\ \psi_{\text{LO}}(x-x_0)\mathrm{e}^{2\mathrm{i}\varphi_0} & (x\to+\infty) \end{cases}
	\end{align}
	with
	\begin{align}
		x_0&=\frac{2\sum_j\operatorname{Im}z_j}{\alpha}, \\
		\mathrm{e}^{2\mathrm{i}\varphi_0}&=\prod_j\frac{\mathrm{e}^{\frac{2 \pi\mathrm{i}s_j\eta_j}{2K}}\vartheta_1(\frac{\mathrm{i}z_j^*}{2K})^2}{\vartheta_1(\frac{\mathrm{i}z_j}{2K})^2},
	\end{align}
	which represent the lattice translation and the phase shift induced by the interaction between the moving solitons and the soliton-lattice background. 
	
	Writing  $ f_0=(u_0,v_0)^T $ in Eq.~(\ref{eq:Jost0}), the fermionic eigenstates are given by
	\begin{align}
		f(t,x,z')=\frac{1}{\det[I_n+\mathcal{EME}]} \begin{pmatrix} u_0(t,x,z')\det[I_n+\mathcal{E\tilde{U}E}] \\ v_0(t,x,z')\det[I_n+\mathcal{E\tilde{V}E}] \end{pmatrix}
	\end{align}
	with
	\begin{align}
		\mathcal{\tilde{U}}_{ij}(x)&=-2\alpha \frac{\vartheta_2\vartheta_4\vartheta_1(\frac{\alpha x-\mathrm{i}(z'-z_i+z_j')}{2K})}{\vartheta_3\vartheta_1(\frac{\alpha x-\mathrm{i}z'}{2K})\vartheta_1(\frac{\mathrm{i}(z_i-z_j')}{2K})} \frac{\vartheta_4(\frac{\mathrm{i}z_i}{2K})\vartheta_1(\frac{\mathrm{i}(z'-z_j')}{2K})}{\vartheta_1(\frac{\mathrm{i}(z'-z_i)}{2K})\vartheta_4(\frac{\mathrm{i}z_j'}{2K})},  \\
		\mathcal{\tilde{V}}_{ij}(x)&=-2\alpha \frac{\vartheta_2\vartheta_4\vartheta_1(\frac{\alpha x+\mathrm{i}(z+z_i-z_j')}{2K})}{\vartheta_3\vartheta_1(\frac{\alpha x-\mathrm{i}z'}{2K})\vartheta_1(\frac{\mathrm{i}(z_i-z_j')}{2K})}\frac{\vartheta_1(\frac{\mathrm{i}z_i}{2K})\vartheta_4(\frac{\mathrm{i}(z+z_j')}{2K})}{\vartheta_4(\frac{\mathrm{i}(z+z_i)}{2K})\vartheta_1(\frac{\mathrm{i}z_j'}{2K})}.
	\end{align}
	$ \operatorname{Im}z=0 $ and $ K $ correspond to scattering states, and $ C_jf_-(t,x,z_j'),\ j=1,\dots,n $ are the normalized bound states. The square of them gives bosonic Bogoliubov quasiparticle eigenstates.

	Let us see the one-soliton solution in detail. The solution shows a variety of behaviors dependent on the choice of parameters. 
	When $ m\simeq 1 $, we can broadly classify it into three categories by its velocity: dark soliton billiards, stationary dislocations, and gray solitons; see Fig.~\ref{fig:onesolvelo}. 
	In this case, the soliton propagation can be understood as a successive collision between the moving soliton and the array of static dark solitons. 
	Figure~{\ref{fig:onesoliton01}}(a) shows an  example of the dark soliton billiard. 
	The LO background experiences a position shift $ \Delta x=\frac{2\operatorname{Im}z_1}{\alpha} $ after the passing of the soliton. 
	Such behavior is different from the soliton train, which is the sliding of the whole soliton lattice. 
	The gray soliton has a more shallow shape and its lattice-shifting effect is weaker than that of the dark soliton billiard.
	The zero-velocity soliton can be interpreted as a static dislocation. Their animation examples are animation1-1, 1-2, 1-3, and 1-4.gif in Ref.~\cite{SeeAnimation}. A static dislocation can appear only for the higher-order NLS system, because the soliton velocity becomes monotonic function  for the ordinary NLS system $(c_3,c_5)=(1,0)$, as shown in Fig~\ref{fig:onesolvelo2}. \\  
	\indent When the background lattice is almost trigonometric $ (m\simeq 0) $,  the distinction between billiards and gray solitons becomes obscure, and any soliton is observed as an envelope soliton (Figure~{\ref{fig:onesoliton01}}(b) and animation1-5 and 1-6.gif). This behavior is similar to the solitons observed in the soft-core bosons \cite{KunimiConf}.  
	
\begin{figure}[tb]
		\begin{center}
		\includegraphics[scale=.9]{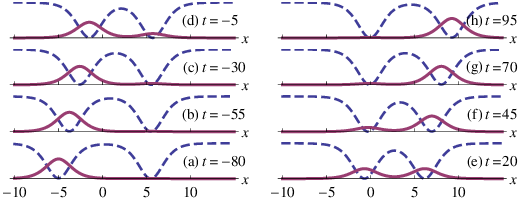}
		\caption{\label{fig:bound} (Color online) Transport and tunneling of the quasiparticle bound state in the dark soliton billiard. Figures should be seen in alphabetical order. The parameters are the same as Fig.~{\ref{fig:onesoliton01}}(a). The solid line represents the amplitude of the bound state $ |C_1f(t,x,z_1')|^2 $, and the dashed line is the soliton. See also animation2-x.gif.}
		\end{center}
	\end{figure}

	Plotting the accompanying quasiparticle bound state is also interesting. For the dark soliton billiard, the transport of quasiparticle wave packet during the collision of solitons is not a simple translation but rather a ``tunneling'' from one soliton to another; see Fig.~\ref{fig:bound} and animation2-1.gif. 
	The other animation examples are also available in \cite{SeeAnimation}. \\
	\indent We note that the ordinary  NLS equation also has the same soliton solutions, which can be obtained by setting $ (c_3,c_5)=(1,0) $. In this case, $ \omega_3(z)=\omega(z) $ [Eq.~(\ref{eq:mainomega5})].
	As discussed in Subsec.~\ref{subsec:bosonicbogoliubov}, the linear stability analysis of the density-modulated state shows the negative spectrum (Fig.~\ref{fig:bsbgc3}), since the ground state of the ordinary NLS system is a uniform state.
	However, if we can prepare a low-temperature environment and can suppress thermal instability, the metastable soliton dynamics with modulated background will be observed even in this system. This will be realized in the Bose condensates of typical ultracold atomic experiments by phase imprint \cite{PhysRevLett.83.5198}. When the soliton lattice consists of sufficiently separated dark solitons, its life time due to the effects of finite temperature and radial confinements can be approximated by that of a single dark soliton, and estimated by the methods in Refs.~\cite{PhysRevA.60.R2665,PhysRevA.60.3220,PhysRevLett.89.110401}.
%


\subsection{Current-carrying (twisted-kink crystal) background}\label{subsec:nsolitoncomplexcase}
The soliton dynamics can be generalized for the case where the background condensate is the FFLO state or the twisted-kink crystal. Here we give a brief summary. \\ 
\indent Equation (\ref{eq:I3I5q}) has the stationary solution
\begin{align}
	\psi_{\text{FFLO}}(x)=\mathrm{e}^{\mathrm{i}[\tilde{p}-\mathrm{i}\alpha Z(2\mathrm{i}z_0)]x}\frac{\mathrm{i}\alpha\vartheta_2\vartheta_4\vartheta_1(\frac{\alpha x-2\mathrm{i}z_0}{2K})}{\vartheta_3\vartheta_4(\frac{2\mathrm{i}z_0}{2K})\vartheta_4(\frac{\alpha x}{2K})}.
\end{align}
Here, $z_0$ and $ \tilde{p} $ are real parameters satisfying $ -\frac{K'}{2}<z_0<\frac{K'}{2} $ and $ \tilde{p}^3+\left( \tfrac{c_3}{2c_5}+S_1 \right)\tilde{p}+S_3=0 $ with writing $ s=-\mathrm{i}\alpha\sqrt{m}\operatorname{sn}(2\mathrm{i}z_0),\ c=\alpha \sqrt{m}\operatorname{cn}(2\mathrm{i}z_0),\ d=\alpha \operatorname{dn}(2\mathrm{i}z_0) $, $ S_1=s^2+c^2+d^2 $, $ S_2=s^2c^2+c^2d^2+d^2s^2 $, and $ S_3=scd $. The chemical potential is given by $ \mu=\mu_{\text{FFLO}}=c_3(S_1-7\tilde{p}^2)+c_5(S_1^2+2S_2-10S_1\tilde{p}^2-15\tilde{p}^4) $. 
Both the density and the phase are modulated in this state:
\begin{align}
	|\psi_{\text{FFLO}}|^2&=m\alpha^2[\operatorname{sn}^2(\alpha x)-\operatorname{sn}^2(2\mathrm{i}z_0)], \\
	\arg\psi_{\text{FFLO}}&=\frac{\operatorname{cn}(2\mathrm{i}z_0)\operatorname{dn}(2\mathrm{i}z_0)}{-\mathrm{i}\operatorname{sn}(2\mathrm{i}z_0)}\Pi(\operatorname{sn}(2\mathrm{i}z_0)^{-2};\operatorname{am}(\alpha x)|m)\nonumber \\
	&\quad+\tilde{p}x+\text{const}. \label{eq:mainfflophase}
\end{align}
If $ z_0=\tilde{p}=0 $, $ \psi_{\text{FFLO}} $ reduces to the real lattice $ \psi_{\text{LO}} $. This solution can be found by following the general argument on the stationary solutions in the higher-order and lower-order NLS equations in  Sec.~\ref{sec:timeevo2}.  \\ 
\indent Although such current-carrying states are not the ground state, the linear stability analysis for the bosonic Bogoliubov quasiparticle $(u_{\text{Bose}},v_{\text{Bose}})$ suggests that these states are metastable, if $ z_0 $ is not too large. Therefore, the soliton dynamics with these backgrounds will be stably observable. \\
\indent With this FFLO background, the $ n $-soliton solution is given as follows. The uniformization variable of the Riemann surface is given by $ (\lambda,\omega)=(\tilde{\lambda}(z),\omega(z)) $  with $ \tilde{\lambda}(z)=\lambda(z)-\frac{\tilde{p}}{2} $ and Eqs. (\ref{eq:applambdazn1}) and (\ref{eq:unifoomegan1}). The crystal momentum $ k(z) $ of BdG eigenstates is given by Eq.~(\ref{eq:appcrystammomenn1}). The time evolution in the AKNS$_3$ equation is described by $ \omega_3(z)=[c_3+c_5(4\tilde{\lambda}(z)^2-4\tilde{p}\tilde{\lambda}(z)+3\tilde{p}^2+S_1)]\omega(z) $. The $n$-soliton solution is 
	\begin{align}
		\psi(t,x)=\psi_{\text{FFLO}}(x)\frac{\det[I_n+\mathcal{E\tilde{A}E}]}{\det[I_n+\mathcal{EME}]},
	\end{align}
	where the definitions of $ \mathcal{E} $ and $ \mathcal{M} $ are the same as in the previous subsection, but we must use new $ \omega_3(z) $ and $ k(z) $ mentioned above. The matrix $ \tilde{\mathcal{A}} $ is modified to be
	\begin{align}
		\tilde{\mathcal{A}}_{ij}=-2\alpha\frac{\vartheta_2\vartheta_4\vartheta_1(\frac{\alpha x-\mathrm{i}(2z_0-z_i+z_j')}{2K})}{\vartheta_3\vartheta_1(\frac{\alpha x-2\mathrm{i}z_0}{2K})\vartheta_1(\frac{\mathrm{i}(z_i-z_j')}{2K})}\frac{\vartheta_4(\frac{\mathrm{i}(z_0+z_i)}{2K})\vartheta_1(\frac{\mathrm{i}(z_0-z_j')}{2K})}{\vartheta_1(\frac{\mathrm{i}(z_0-z_i)}{2K})\vartheta_4(\frac{\mathrm{i}(z_0+z_j')}{2K})}.
	\end{align}
	The asymptotics of this solution is given in Subsec.~\ref{subsec:asymptotics}, where we  write $ \psi_0=\psi_{\text{FFLO}} $. The fermionic eigenstates in the absence and presence of solitons are given by Eqs.~(\ref{eq:fermionicsoln1}) and (\ref{eq:fermiondet21}), respectively. 

%
%

\subsection{Parameters for gif animation files}
	Here we show the parameters used in gif animation files in the Supplemental Material \cite{SeeAnimation}. The animation1-x.gif (x$=1,\dots,6$) provide soliton dynamics. The animation2-x.gif (x$=1,\dots,4$) draw the accompanying bound states.
	\begin{itemize}
		\item animation1-1.gif: Dark soliton billiard.\\ The parameters are the same as Fig.~\ref{fig:onesoliton01}(a). The spectral parameter and the soliton velocity are evaluated as $ \lambda(z_1)=0.0576 $ and $ V_1=0.0711 $. 
		\item animation1-2.gif: Gray soliton.\\  $ z_1=-0.5K'+0.05\mathrm{i}K,\ C_1=0.893 $, and the other parameters are the same as Fig.~\ref{fig:onesoliton01}(a).  The spectral parameter and the soliton velocity are evaluated as $ \lambda(z_1)=0.471 $ and $ V_1=-0.469 $. 
		\item animation1-3.gif: Static dislocation.\\  $ z_1=-0.5K'+0.1066\mathrm{i}K,\ C_1=0.0336 $, and the other parameters are the same as Fig.~\ref{fig:onesoliton01}(a).  The spectral parameter and the soliton velocity are evaluated as $ \lambda(z_1)=0.333 $ and $ V_1=0 $. 
		\item animation1-4.gif: Example of 3-soliton solution.\\  $ z_1=-0.5K'+0.3\mathrm{i}K,\ z_2=-0.5K'+0.1066\mathrm{i}K,\ z_3=-0.5K'+0.05\mathrm{i}K,\ C_1=0.467,\ C_2=0.0336, $ $ C_3=176 $, and  the other parameters are the same as Fig.~\ref{fig:onesoliton01}(a).
		\item animation1-5.gif: Envelope soliton.\\ The parameters are the same as Fig.~\ref{fig:onesoliton01}(b). The spectral parameter and the soliton velocity are evaluated as $ \lambda(z_1)=0.243 $ and $ V_1=0.239 $. 
		\item animation1-6.gif: Another envelope soliton.\\  $ z_1=-0.5K'+0.34\mathrm{i}K,\ C_1=4.13 $, and the other parameters are the same as Fig.~\ref{fig:onesoliton01}(b). The spectral parameter and the soliton velocity are evaluated as $ \lambda(z_1)=0.357 $ and $ V_1=-0.0477 $. 
	\end{itemize}
	The parameters of animation2-1.gif, 2-2.gif, 2-3.gif, and 2-4.gif, showing the dynamics of the bound states, are the same as 1-1.gif, 1-2.gif, 1-5.gif, and 1-6.gif. 

\section{Fermionic eigenstates for AKNS$_{\text{1}}$ background}\label{sec:bdgwithakns1bg}
	Sections~\ref{sec:bdgwithakns1bg}-\ref{sec:timeevo} are devoted to the detail of the formulation and calculation. \\ 
	\indent In order to formulate the IST with soliton-lattice background in Sec.~\ref{sec:ist}, 
	we first summarize the eigenstates of the BdG equation
	\begin{align}
		\begin{pmatrix} -\mathrm{i}\partial_x & \psi_0 \\ \psi_0^* & \mathrm{i}\partial_x \end{pmatrix}\begin{pmatrix}u \\ v \end{pmatrix}= \epsilon \begin{pmatrix} u \\ v \end{pmatrix}, \label{eq:zsbdgn1}
	\end{align}
	when $ \psi_0 $ satisfies the AKNS$_1$ equation
	\begin{align}
		d_1\psi_0+d_2(-\mathrm{i}\partial_x\psi_{0})+d_3(-\partial_x^2\psi_{0}+2|\psi|^2\psi)=0, \label{eq:akns1n1}
	\end{align}
	where the coefficients $d_i$'s are real. 
	The solutions expressed by the Weierstrass functions are given in Ref.~\cite{Basar:2008ki}. We give an expression using the Jacobi theta functions. Derivation based on Ref. \cite{Takahashi:2012aw} is given in Appendix~\ref{app:bdgforakns1}. The convention of elliptic functions is summarized in Appendix~\ref{app:ellipconv}. 
\subsection{Solutions}
	\indent The general bounded solution of Eq.~(\ref{eq:akns1n1}) is
	\begin{align}
		\psi_0(x)&=\mathrm{e}^{\mathrm{i}px}\frac{\mathrm{i}\alpha\vartheta_2\vartheta_4\vartheta_1(\frac{\alpha x-2\mathrm{i}z_0}{2K})}{\vartheta_3\vartheta_4(\frac{2\mathrm{i}z_0}{2K})\vartheta_4(\frac{\alpha x}{2K})}, \label{eq:akns1psi0n1} \\
		p&=-\mathrm{i}\alpha Z(2\mathrm{i}z_0)+\tilde{p},
	\end{align}
	where $ \tilde{p}\in \mathbb{R},\ \alpha>0, m \in [0,1] $, and $ -\frac{K'}{2}<z_0<\frac{K'}{2} $. 
	Here and hereafter, the omitted elliptic parameter and nome are always $ m $ and $ q=\mathrm{e}^{-\pi K'/K} $.  These parameters are related to $ d_i $'s as
	\begin{align}
		\frac{d_2}{d_3}=-2\tilde{p},\quad \frac{d_1}{d_3}=\tilde{p}^2-\alpha^2[m-2+3\operatorname{dn}^2(2\mathrm{i}z_0)]. \label{eq:akns1n1dcoeffs}
	\end{align}  
	The associated Riemann surface for this potential with $\tilde{p}=0 $ is given by
	\begin{align}
		&\omega^2=4(\lambda-\lambda_1)(\lambda-\lambda_2)(\lambda-\lambda_3)(\lambda-\lambda_4), \label{eq:akns1riemannsurface01}\\
		&\lambda_1= \tfrac{1}{2}(-s-c-d),\ \lambda_2=\tfrac{1}{2}(s+c-d),\nonumber \\
		&\lambda_3=\tfrac{1}{2}(s-c+d),\ \lambda_4=\tfrac{1}{2}(-s+c+d), \label{eq:akns1riemannsurface03} \\
		&s=-\mathrm{i}\alpha\sqrt{m}\operatorname{sn}(2\mathrm{i}z_0),\ c=\alpha \sqrt{m}\operatorname{cn}(2\mathrm{i}z_0),\ d=\alpha \operatorname{dn}(2\mathrm{i}z_0). \label{eq:akns1riemannsurface04}
	\end{align}
	The surface corresponding to $ \tilde{p}\ne0 $ is obtained by translation $ \lambda \rightarrow \lambda-\frac{\tilde{p}}{2} $. The surface is defined by $ \omega^2=\det V $, where $ V $ represents a matrix of the time-derivative part of the AKNS system \cite{Takahashi:2012aw}. 
	This surface is parametrized by 
	\begin{align}
		\lambda(z)&=\frac{\alpha\left[ \operatorname{dn}(\mathrm{i}(z+z_0))\operatorname{dn}(\mathrm{i}(z'+z_0))+\mathrm{i}m\operatorname{sn}(2\mathrm{i}z_0)\operatorname{cn}(2\mathrm{i}z_0) \right]}{-2\operatorname{dn}(2\mathrm{i}z_0)}, \label{eq:applambdazn1}\\
		\omega(z)&=\alpha \lambda'(z)=\frac{\alpha^2}{2}\left[ \operatorname{dn}^2(\mathrm{i}(z'+z_0))-\operatorname{dn}^2(\mathrm{i}(z+z_0)) \right], \label{eq:unifoomegan1}
	\end{align}
	where $ z':=K'-z $. For $ \tilde{p}\ne0 $, the Riemann surface is given by $\omega^2=4\prod_{i=1}^4(\lambda-\lambda_i+\frac{\tilde{p}}{2})$ and hence we should use $ \tilde{\lambda}(z)=\lambda(z)-\frac{\tilde{p}}{2} $. ($ \omega(z) $ does not change.)
	When $ z_0=\tilde{p}=0 $, we revisit the parametrization in Subsec.~\ref{subsec:mainakns}. \\
	\indent Now, let us write down the eigenstates of the BdG equation. If we parametrize  $ \epsilon $ in Eq.~(\ref{eq:zsbdgn1}) by
	\begin{align}
		\epsilon=-\tilde{\lambda}(z)=-\lambda(z)+\frac{\tilde{p}}{2}, \label{eq:epsilonbylambda}
	\end{align}
	then the two linearly independent solutions of the BdG equation for a given $ \epsilon $ are given by 
	\begin{align}
		&f_0(x,z):=\begin{pmatrix} u_0(x,z) \\ v_0(x,z) \end{pmatrix}=\mathrm{e}^{\mathrm{i}k(z)x}\mathrm{e}^{\mathrm{i}(\frac{1}{2}px-\frac{\pi\alpha x}{4K})\sigma_3}\nonumber \\
		&\quad\qquad\times\frac{\mathrm{i}\alpha \vartheta_2\vartheta_4}{\vartheta_3\vartheta_4(\frac{\alpha x}{2K})}\begin{pmatrix} \vartheta_1(\frac{\alpha x-\mathrm{i}(z+z_0)}{2K})/\vartheta_4(\frac{\mathrm{i}(z+z_0)}{2K}) \\ - \vartheta_1(\frac{\alpha x+\mathrm{i}(z'+z_0)}{2K})/\vartheta_4(\frac{\mathrm{i}(z'+z_0)}{2K}) \end{pmatrix}. \label{eq:fermionicsoln1}
	\end{align}
	and $ f_0(x,z') $, where we define the crystal momentum
	\begin{align}
		k(z):=-\frac{\mathrm{i}\alpha}{2}\left[Z(\mathrm{i}(z+z_0))-Z(\mathrm{i}(z'+z_0))\right]. \label{eq:appcrystammomenn1}
	\end{align} 
	If $ \psi_0(x-x_0)\mathrm{e}^{2\mathrm{i}\varphi_0} $ with $ x_0,\varphi_0\in\mathbb{R} $ is used, the solution is given by $ \mathrm{e}^{\mathrm{i}\varphi_0\sigma_3}f_0(x-x_0,z) $. 
	The Wronskian is calculated as 
	\begin{align}
		\det[f_0(x,z),f_0(x,z')]=-2\omega(z). \label{eq:wronski000}
	\end{align} 
\subsection{Periodicities and Symmetries}

	\begin{figure}[tb]
		\begin{center}
		\includegraphics{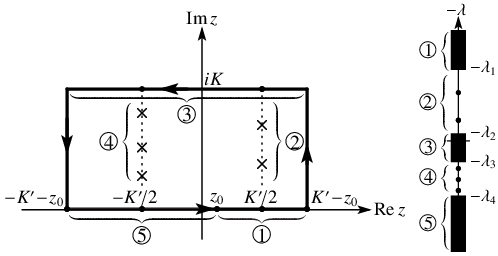}
		\caption{\label{fig:univar2} Relation between the uniformization variable $ z $ and the spectral parameter $ -\lambda $. The expressions of $ \lambda_i $'s are given by Eqs.~(\ref{eq:akns1riemannsurface03}) and (\ref{eq:akns1riemannsurface04}).  If we set $ z_0=0 $, it reduces to Fig.~\ref{fig:univar}, i.e., the case of real sn lattice. The rectangular contour is used for the completeness relation (\ref{eq:completeinmainsec}) and derivation of the GLM equation (Sec.~\ref{sec:ist}).}
		\end{center}
	\end{figure}
	\begin{figure}[tb]
		\begin{center}
		\includegraphics[width=.8\linewidth]{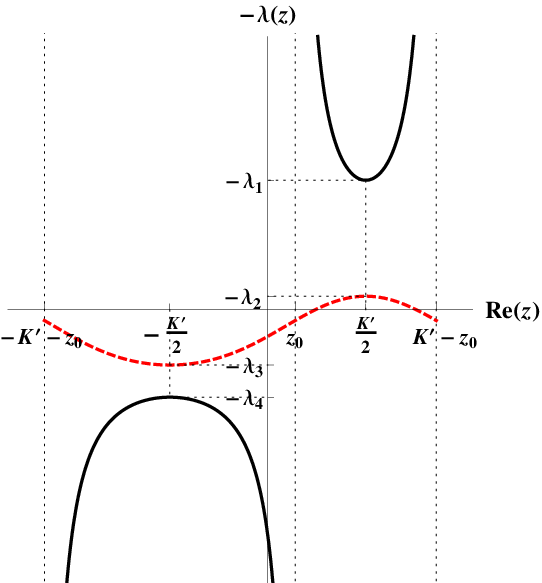} 
		\caption{\label{fig:lambdaz} $ -\lambda(z) $ on $ \operatorname{Im}z=0 $ (black solid line) and $ \operatorname{Im}z=K $ (red dashed line).}
		\end{center}
	\end{figure}
	\begin{figure}[tb]
		\begin{center}
		 \includegraphics[width=.8\linewidth]{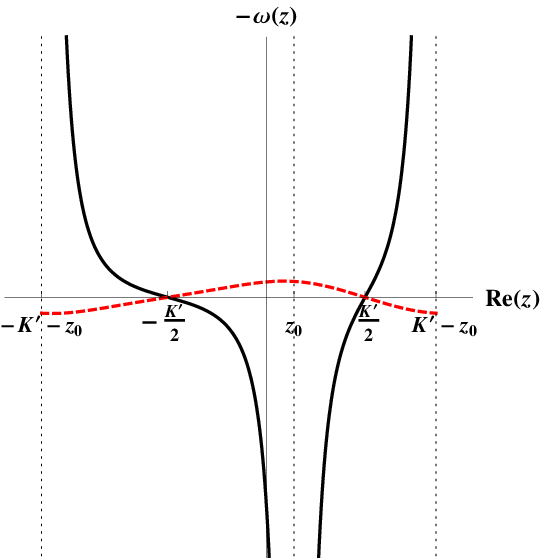} 
		\caption{\label{fig:omegaz} $ -\omega(z) $ on $ \operatorname{Im}z=0 $ (black solid line) and $ \operatorname{Im}z=K $ (red dashed line).}
		\end{center}
	\end{figure}
	\begin{figure}[tb]
		\begin{center}
		 \includegraphics[width=.8\linewidth]{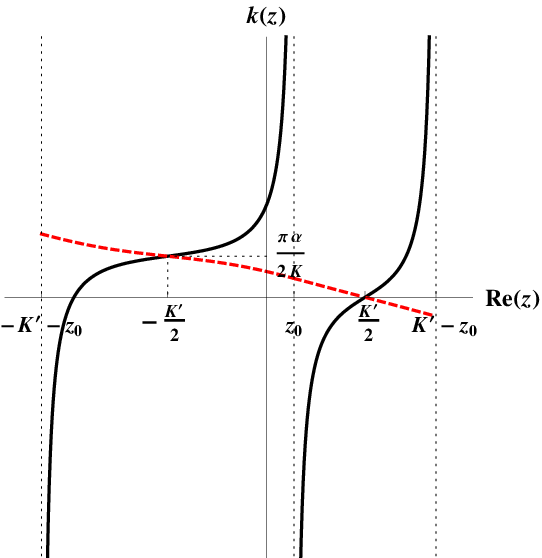}
		\caption{\label{fig:crystalkz}  $ k(z) $ on $ \operatorname{Im}z=0 $ (black solid line) and $ \operatorname{Im}z=K $ (red dashed line).}
		\end{center}
	\end{figure}

	\indent $ \lambda(z), \omega(z) $, and $ k(z) $ have the following (quasi-) periodicity, parity, and complex-conjugation relation: 
	\begin{align}
		\lambda(z)&=\lambda(z+2nK'+2\mathrm{i}lK)=\lambda(z')=\lambda(z^*)^*, \label{eq:unilambdasymmtn1} \\
		\omega(z)&=\omega(z+2nK'+2\mathrm{i}lK)=-\omega(z')=\omega(z^*)^*, \\
		k(z)&=k(z+2nK'+2\mathrm{i}lK)+\tfrac{\pi n\alpha}{K}=-k(z')=k(z^*)^*, \label{eq:unicryksymmtn1}
	\end{align}
	where $ l $ and $ n $ are integers.  $ \omega_3(z) $, which is defined in Eq.~(\ref{eq:mainomega5}) and used to solve the AKNS$_3$ equation, has the same symmetry as $ \omega(z) $. \\
	\indent In the algebro-geometric construction,  $ \lambda(z),\omega(z),k(z) $ are Abelian integrals appearing in the exponential part of the Baker-Akhiezer functions \cite{Krichever1977,BelokolosBobenkoEnolskiiItsMatveev,TanakaDate}. \\
	\indent $ \lambda(z) $ has real values on $ \operatorname{Im}z=nK $ and $ \operatorname{Re}z=\frac{1+2n}{2}K' $. 
	The scattering eigenstates exist on $ \operatorname{Im}z=nK $, and the gap corresponds to $ \operatorname{Re}z=\frac{1+2n}{2}K' $,  $ n\in\mathbb{Z} $. 
	The specific values are
	\begin{align}
		&\lambda\left(\tfrac{K'}{2}\right)=\lambda_1,\quad \lambda\left(\tfrac{K'}{2}+\mathrm{i}K\right)=\lambda_2,\nonumber \\
		&\lambda\left(-\tfrac{K'}{2}+\mathrm{i}K\right)=\lambda_3,\quad \lambda\left(-\tfrac{K'}{2}\right)=\lambda_4, \\
		&\omega\left(\pm\tfrac{K'}{2}\right)=\omega\left(\pm\tfrac{K'}{2}+\mathrm{i}K\right)=0.
	\end{align}
	If $ z_0 $ is restricted to  $ -\frac{K'}{2}<z_0<\frac{K'}{2} $, they satisfy $ \lambda_1<\lambda_2<\lambda_3<\lambda_4 $ and $ \lambda_1<0<\lambda_4 $. The spectrum is determined by $ \omega^2>0 $, and therefore $ \lambda<\lambda_1,\ \lambda_2<\lambda<\lambda_3, \lambda_4<\lambda $, which is equivalent to $ \epsilon<\frac{\tilde{p}}{2}-\lambda_1,\ \frac{\tilde{p}}{2}-\lambda_2<\epsilon<\frac{\tilde{p}}{2}-\lambda_3,\ \frac{\tilde{p}}{2}-\lambda_4<\epsilon $. When $ z_0=\tilde{p}=0 $, the real sn lattice is realized and $ \lambda_1=-\lambda_4 $ and $  \lambda_2=-\lambda_3 $ hold, and hence the spectrum is symmetric about the origin; see Figs.~\ref{fig:univar2}, \ref{fig:lambdaz}, \ref{fig:omegaz}, and \ref{fig:crystalkz}. \\ 
	\indent $ f_0(x,z) $ has the following double periodicity and complex conjugation relation: 
	\begin{align}
		&f_0(x,z)=(-1)^l f_0(x,z+2nK'+2\mathrm{i}lK), \label{eq:f0period} \\
		&f_0(x,z')=\sigma_1f_0(x,z^*)^*, \label{eq:f0ccrel}
	\end{align}
	where $ l, n \in \mathbb{Z} $. 
	For $ \operatorname{Im}z=lK $ with integer $ l $, which corresponds to scattering states, the relation
	\begin{align}
		f_0(x,z')=(-1)^l\sigma_1 f_0(x,z)^* \quad (\operatorname{Im}z=lK) \label{eq:appccrelf0}
	\end{align}
	holds.
	When $ \operatorname{Re}z=\pm\frac{K'}{2} $, which may become a discrete eigenvalue in the presence of solitons (see Fig.~\ref{fig:univar2}), the relation 
	\begin{align}
		f_0(x,z)^*=\sigma_1 f_0(x,z) \quad (\operatorname{Re}z=\pm\tfrac{K'}{2}) \label{eq:appccrelf2}
	\end{align}
	holds. 
	In order to cover all solutions of the BdG (or ZS) operator for all $ \epsilon $, we need to consider $ z $ in a rectangle with vertices $ (-K'-\mathrm{i}K,K'-\mathrm{i}K,K'+\mathrm{i}K,-K'+\mathrm{i}K) $, where $ \lambda(z) $ takes all complex values just twice. \\ 
	\indent $ f_0(x,z) $ satisfies the following completeness relation:
	\begin{align}
		&\int_{R}\frac{\mathrm{d}z}{4\pi\alpha}f_0(x,z)f_0(y,z')^T\sigma_1 =\delta(x-y)I_2, \label{eq:completeinmainsec}
	\end{align}
	where $ z':=K'-z $, and $ R $  represents the rectangular contour in Fig.~\ref{fig:univar2}. 
	The proof is given in Appendix~\ref{app:bdgforakns1}. \\
	\indent If $ \lambda(z) $ and $ k(z) $ are real, $ f_0(x,z) $ is a twisted Bloch function in the following sense.  $ \psi_0(x) $ is a twisted-periodic function satisfying
	\begin{align}
		\psi_0(x+L_0)=\psi_0(x)\mathrm{e}^{\mathrm{i}\theta},\quad L_0=\frac{2K}{\alpha},\ \theta=\frac{2Kp}{\alpha}-\pi. \label{eq:appqtwistn1}
	\end{align}
	Then, the corresponding eigenstate $ f_0(x,z) $ satisfies
	\begin{align}
		f_0(x+L_0,z)=\mathrm{e}^{\mathrm{i}k(z)L_0}\mathrm{e}^{(\mathrm{i}\theta/2)\sigma_3}f_0(x,z). \label{eq:appf0twistn1}
	\end{align}
	From this expression, one might think that the crystal momentum is defined up to $ \operatorname{mod}\frac{2\pi}{L_0} $. 
	In fact, it can be reduced to $ \operatorname{mod}\frac{\pi}{L_0} $. 
	The reason is as follows. For the twist angle $ \theta $ of $ \psi_0(x) $ in Eq.~(\ref{eq:appqtwistn1}), we can alternatively choose $ \theta+2\pi $. This makes no change in $ \psi_0(x) $, but the condition for $ f_0(x,z) $ is rewritten as
	\begin{align}
		f_0(x+L_0,z)=\mathrm{e}^{\mathrm{i}(k(z)-\frac{\pi}{L_0})L_0}\mathrm{e}^{(\mathrm{i}\theta/2+\mathrm{i}\pi)\sigma_3}f_0(x,z).
	\end{align}
	Thus, $ k(z) $ is shifted by $ \frac{\pi}{L_0} $ by this transformation. 
	On the other hand, the bosonic Bogoliubov quasiparticles, whose wavefunctions are given by the square of fermionic ones (Subsec.~\ref{subsec:bosonicbogoliubov}), have a crystal momentum $ 2k(z) $, which is defined only up to $ \operatorname{mod}\frac{2\pi}{L_0} $, the same as $ \psi_0(x) $. 
%
%
%
%
\section{IST with elliptic background}\label{sec:ist}
%
	\indent In this section, we formulate the IST in the presence of elliptic function background, and derive the soliton solutions. Although a more sophisticated way to derive these soliton solutions is reduction of general quasi-periodic Riemann theta solutions with $ g-1 $ periods of $ g $-fold quasi-periodic solution taken to be infinity \cite{Krichever1977,BelokolosBobenkoEnolskiiItsMatveev,TanakaDate}, an IST-based derivation can still provide a physical understanding from another view. 
%
\subsection{The tending-to-AKNS$_{\text{1}}$ boundary condition}
	We consider the scattering problem of the BdG or ZS operator
	\begin{align}
		\begin{pmatrix} -\mathrm{i}\partial_x & \psi(x) \\ \psi(x)^* & \mathrm{i}\partial_x  \end{pmatrix}\begin{pmatrix} u \\ v \end{pmatrix}=\epsilon \begin{pmatrix} u \\ v \end{pmatrix}, \label{eq:ismbdg01}
	\end{align}
	where $ \psi(x) $ asymptotically tends to the AKNS$_1$ potential for $ x\to\pm\infty $: 
	\begin{align}
		\psi(x) \rightarrow\begin{cases} \psi_0(x) & (x\rightarrow-\infty), \\ \psi_0(x-x_0)\mathrm{e}^{2\mathrm{i}\varphi_0} & (x\rightarrow+\infty). \end{cases} \label{eq:AKNS1boundary}
	\end{align}
	Here, $ \psi_0(x) $ is given by Eq.~(\ref{eq:akns1psi0n1}), and  $ x_0 $ and $ \varphi_0 $ represent the position and phase shifts of the background lattice induced by solitons and radiations. See Fig.~\ref{fig:istschematic}.
\subsection{Jost functions and scattering matrix}
	In the presence of $ \psi(x) $ with the above-mentioned asymptotic form, we define the left Jost function $ f_-(x,z) $ by the solution of Eq.~(\ref{eq:ismbdg01}) with $ \epsilon $ parametrized as Eq.~(\ref{eq:epsilonbylambda}) with the asymptotic form
	\begin{align}
		f_-(x,z)\to f_0(x,z) \quad (x\to-\infty). \label{eq:defleftjost}
	\end{align}
	The solution is uniquely defined by this asymptotic condition. Similarly, we define the right Jost function $ f_+(x,z) $ by 
	\begin{align}
		f_+(x,z)\to \mathrm{e}^{\mathrm{i}\varphi\sigma_3}f_0(x-x_0,z) \quad (x\to+\infty).
	\end{align}
	Because of the uniqueness of the solution under a given asymptotic form, the same relations as in Eqs.~(\ref{eq:f0period}) and (\ref{eq:f0ccrel}) hold:
	\begin{align}
		f_\pm(x,z)&=(-1)^lf_\pm(x,z+2nK'+2\mathrm{i}lK), \label{eq:fpmperiod}\\
		f_\pm(x,z')&=\sigma_1f_\pm(x,z^*)^*. \label{eq:fpminvo}
	\end{align}
	 We introduce the scattering matrix $ S(z) $ by the relation
	\begin{align}
		\begin{pmatrix} f_+(x,z) & f_+(x,z') \end{pmatrix}&=\begin{pmatrix} f_-(x,z) & f_-(x,z') \end{pmatrix}S(z), \label{eq:defscattmat} \\
		S(z)&=\begin{pmatrix} a(z) & b(z') \\ b(z) & a(z') \end{pmatrix},
	\end{align}
	which satisfies
	\begin{align}
		S(z)&=S(z+2nK'+2\mathrm{i}lK)=\sigma_1S(z')\sigma_1, \label{eq:scattperiod} \\
		S(z')&=S(z^*)^*, \label{eq:scattinvo} \\
		\det S(z)&=1, \label{eq:scattdet} \\
		S(z)^{-1}&= \sigma_2 S(z)^T\sigma_2. \label{eq:scattdet2}
	\end{align}
	Equations (\ref{eq:scattperiod}) and (\ref{eq:scattinvo}) are derived from Eqs. (\ref{eq:fpmperiod}) and (\ref{eq:fpminvo}). Equation (\ref{eq:scattdet}) is proved by the Wronskian. Equation (\ref{eq:scattdet2}) is  a general formula of $ 2\times 2 $ inverse matrix with determinant 1. In terms of $ a(z) $ and $ b(z) $, these relations are 
	\begin{align}
		&a(z)=a(z+2nK'+2\mathrm{i}lK),\ a(z')=a(z^*)^*, \\
		&b(z)=b(z+2nK'+2\mathrm{i}lK),\ b(z')=b(z^*)^*, \\
		&a(z)a(z')-b(z)b(z')=1. \label{eq:probacons}
	\end{align} 
	When $ \operatorname{Im}z=nK $, Eq.~(\ref{eq:probacons}) reduces to $ |a(z)|^2-|b(z)|^2=1 $. 
\subsection{Bound states}\label{subsec:bsnmrztn}
	The bound states appear at the zeros of $ a(z) $, since the coefficient of the exponentially divergent component in the Jost function vanishes. Since the BdG or ZS operator is self-adjoint, the discrete spectrum can appear for real $ \epsilon $, which corresponds to  $ z=\pm\frac{K'}{2}+\mathrm{i}\eta,\ 0<\eta<K $. Here we derive the normalization constant of bound states written by the scattering matrix. \\
	\indent Let $ z=z_j $ be a zero of $ a(z) $. Then, the bound state is given by $ f_+(x,z_j) $. Since $ a(z_j)=0 $, the left and right Jost function is related by
	\begin{align}
		f_+(x,z_j)=b(z_j)f_-(x,z_j'),
	\end{align}
	where $ z_j':=K'-z_j $. 
	We define the normalization constant
	\begin{align}
		c_j^{-2}:=\int_{-\infty}^\infty\mathrm{d}x f_+(x,z_j)^\dagger f_+(x,z_j).
	\end{align}
	Then $ c_jf_+(x,z_j) $ is normalized. Let us write the $ z $-derivative of a function $ f $ by dot $ \dot{f}=\partial f/\partial z $. Differentiating the BdG equation by $ z $ (and recalling $ \epsilon=-\lambda+\frac{\tilde{p}}{2} $), we find
	\begin{align}
		\partial_x\left[f_+(x,z_j)^\dagger\sigma_3\dot{f}_+(x,z_j)\right]=-\mathrm{i}\dot{\lambda}(z_j)f_+(x,z_j)^\dagger f_+(x,z_j). \label{eq:istboundnormal}
	\end{align}
	Integrating this and using Eqs.~(\ref{eq:wronski000}), (\ref{eq:appccrelf2}), and (\ref{eq:defleftjost}), we obtain $ -\mathrm{i}\dot{\lambda}(z_j)c_j^{-2}=2\omega(z_j)\dot{a}(z_j)b(z_j)^* $.
	Since $ \omega=\alpha\dot{\lambda} $, 
	\begin{align}
		c_j^{-2}=2\mathrm{i}\alpha\dot{a}(z_j)b(z_j)^*. \label{eq:ncbyscat01}
	\end{align}
\subsection{Integral representation of the Jost function}
	We introduce the integral representation for the left Jost function with a kernel $ \Gamma(x,y) $: 
	\begin{align}
		f_-(x,z)=f_0(x,z)+\int_{-\infty}^x\mathrm{d}y\Gamma(x,y)f_0(y,z). \label{eq:triang}
	\end{align}
	This expression is called the triangular representation in several references. 
	Following the same proof as Ref.~\cite{FaddeevTakhtajan} (see also Refs.~\cite{Takahashi:2012pk,Takahashi:2015hdt}), we obtain the equation for the kernel $ \Gamma $: 
	\begin{gather}
		\Gamma(x,x)-\sigma_3\Gamma(x,x)\sigma_3=U(x)-U_0(x), \\
		\frac{\partial \Gamma(x,y)}{\partial x}+\sigma_3\left( \frac{\partial \Gamma(x,y)}{\partial y}-\Gamma(x,y)U_0(x) \right)\sigma_3-U(x)\Gamma(x,y)=0,
	\end{gather}
	where $ U(x)=\left(\begin{smallmatrix}0&-\mathrm{i}\psi(x)\\ \mathrm{i}\psi(x)^*&0\end{smallmatrix}  \right) $ and $ U_0(x)=\left( \begin{smallmatrix}0&-\mathrm{i}\psi_0(x)\\ \mathrm{i}\psi_0(x)^*&0\end{smallmatrix} \right) $. 
	From this, $ \psi(x) $ is given by
	\begin{align}
		\psi(x) &= \psi_0(x)+2\mathrm{i}\Gamma_{12}(x,x), \label{eq:triang22} \\
		\psi(x)^* &= \psi_0(x)^*-2\mathrm{i}\Gamma_{21}(x,x).
	\end{align}
\subsection{The GLM equation}\label{subsec:glm}
	Let us derive the GLM equation. We start from the relation between right and left Jost functions,
	\begin{align}
		\frac{1}{a(z)}f_+(x,z)=f_-(x,z)+\frac{b(z)}{a(z)}f_-(x,z'),
	\end{align}
	which is the left column of Eq.~(\ref{eq:defscattmat}). 
	Substituting the integral representation (\ref{eq:triang}),
	\begin{align}
		&\frac{1}{a(z)}f_+(x,z)-f_0(x,z)=\int_{-\infty}^x\mathrm{d}y\Gamma(x,y)f_0(y,z)\nonumber \\
		&+\frac{b(z)}{a(z)}\left[ f_0(x,z')+\int_{-\infty}^x\mathrm{d}y\Gamma(x,y)f_0(y,z') \right]. \label{eq:deriveglm02}
	\end{align}
	We now evaluate $ \int_R\frac{\mathrm{d}z}{4\pi\alpha}[\text{Eq.~(\ref{eq:deriveglm02})}]f_0(w,z')^T\sigma_1 $ 
	for $ w<x $, where $ R $ is the rectangular contour in Fig.~\ref{fig:univar2}. 
	Let us introduce 
	\begin{align}
		\Omega_c(x,w):=\int_R\frac{\mathrm{d}z}{4\pi\alpha}\frac{b(z)}{a(z)}f_0(x,z')f_0(w,z')^T\sigma_1,
	\end{align}
	then 
	\begin{align}
		&\int_R\frac{\mathrm{d}z}{4\pi\alpha}[\text{R.H.S. of Eq.~(\ref{eq:deriveglm02})}]f_0(w,z')^T\sigma_1 \nonumber \\
		&=\Gamma(x,w)+\Omega_c(x,w)+\int_{-\infty}^x\mathrm{d}y\Gamma(x,y)\Omega_c(y,w).
	\end{align}
	Next, we evaluate the integration of the left hand side. Let us define
	\begin{align}
		\Omega_d(x,w)&:=-\frac{\mathrm{i}}{2\alpha}\sum_j\frac{b(z_j)}{\dot{a}(z_j)}f_0(x,z_j')f_0(w,z_j')^T\sigma_1 \nonumber \\
		&=\sum_jC_j^2f_0(x,z_j')f_0(w,z_j')^T\sigma_1,
	\end{align}
	where we write $ C_j:=|b(z_j)|c_j $ and Eq.~(\ref{eq:ncbyscat01}) is used. Using the residue theorem, we obtain
	\begin{align}
		&\int_R\frac{\mathrm{d}z}{4\pi\alpha}[\text{L.H.S. of Eq.~(\ref{eq:deriveglm02})}]f_0(w,z')^T\sigma_1 \nonumber \\
		&=-\Omega_d(x,w)-\int_{-\infty}^x\mathrm{d}y\Gamma(x,y)\Omega_d(y,w).
	\end{align}
	Summarizing, exchanging the dummy variables $ y $ and $ w $, the GLM equation for the kernel $ \Gamma $ is given by
	\begin{gather}
		\Gamma(x,y)+\Omega(x,y)+\int_{-\infty}^x\mathrm{d}w\Gamma(x,w)\Omega(w,y)=0 \quad (y<x), \label{eq:GLM001}\\
		\Omega(x,y):=\Omega_c(x,y)+\Omega_d(x,y).
	\end{gather}
	This equation solves the inverse problem, i.e., it determines the potential $ \psi(x) $ from the scattering data: the reflection coefficients $ r(z)=b(z)/a(z) $, the list of discrete eigenvalues $ z_1,\dots, z_n $, and the normalization constants of bound states $ C_1^2,\dots, C_n^2 $.  
\subsection{Integral appearing in reflectionless solutions}
	Here, we evaluate an integral  necessary to solve the GLM equation. Let us calculate
	\begin{align}
		M(x,z_j,z):=\int_{-\infty}^x\mathrm{d}xf_0(x,z_j')^T\sigma_1 f_0(x,z'),
	\end{align}
	where $ z_j $ is a zero of $ a(z) $ and hence written as $ z_j=\pm\frac{K'}{2}+\mathrm{i}\eta,\ 0<\eta<K $ and $ \lambda(z_j) $ is real. The other parameter $ z $ has no constraint except that the integrand must vanish at $ x\to-\infty $. We first note that if $ f_1,f_2 $ are eigenstates of the ZS operator with eigenvalues $ \epsilon_1,\epsilon_2 $, the relation $ f_1^\dagger f_2=\frac{(f_1^\dagger\sigma_3f_2)_x}{\mathrm{i}(\epsilon_2-\epsilon_1^*)} $ holds.
	Using this and Eq.~(\ref{eq:appccrelf2}), and recalling $ \epsilon=-\lambda+\frac{\tilde{p}}{2} $ [Eq.~(\ref{eq:epsilonbylambda})],
	\begin{align}
		M(x,z_j,z)=\frac{f_0(x,z_j')^T\sigma_1\sigma_3f_0(x,z')}{\mathrm{i}(\lambda(z_j)-\lambda(z))}. \label{eq:Mzjz002}
	\end{align}
	Now we derive an expression for $ M(x,z_j,z) $.  Using the three-term Weierstrass addition formula \cite{Kharchev201519}
	\begin{align}
		&\vartheta_1(a+c)\vartheta_1(a-c)\vartheta_4(b+d)\vartheta_4(b-d) \nonumber \\
		&-\vartheta_1(b+c)\vartheta_1(b-c)\vartheta_4(a+d)\vartheta_4(a-d) \nonumber \\
		=&\vartheta_1(a+b)\vartheta_1(a-b)\vartheta_4(c+d)\vartheta_4(c-d) \label{eq:weierstrass1144}
	\end{align}
	with $ a=\frac{\mathrm{i}(z_j+z'+2z_0)}{4K}, b=\frac{\mathrm{i}(z_j'+z+2z_0)}{4K}, c=\frac{2\alpha x+\mathrm{i}(z_j-z')}{4K}, d=\frac{\mathrm{i}(z_j-z')}{4K} $, the numerator of Eq.~(\ref{eq:Mzjz002}) is given by
	\begin{align}
		&f_0(x,z_j')^T\sigma_1\sigma_3f_0(x,z')= \mathrm{e}^{-\mathrm{i}[k(z_j)+k(z)]x}\times \nonumber \\
		&\frac{\alpha^2\vartheta_2^2\vartheta_4^2\vartheta_1(\frac{\mathrm{i}(2z_0+K')}{2K})\vartheta_1(\frac{\mathrm{i}(z-z_j)}{2K})\vartheta_4(\frac{\alpha x+\mathrm{i}(z_j-z')}{2K})}{\vartheta_3^2\vartheta_4(\frac{\alpha x}{2K})\vartheta_4(\frac{\mathrm{i}(z_j+z_0)}{2K})\vartheta_4(\frac{\mathrm{i}(z_j'+z_0)}{2K})\vartheta_4(\frac{\mathrm{i}(z+z_0)}{2K})\vartheta_4(\frac{\mathrm{i}(z'+z_0)}{2K})}.
	\end{align} 
	Similarly, the denominator of Eq.~(\ref{eq:Mzjz002}) is
	\begin{align}
		&\mathrm{i}(\lambda(z_j)-\lambda(z)) \nonumber \\
		&=-\frac{\alpha}{2}\frac{\vartheta_2\vartheta_4\vartheta_1(\frac{\mathrm{i}(2z_0+K')}{2K})\vartheta_1(\frac{\mathrm{i}(z_j-z')}{2K})\vartheta_1(\frac{\mathrm{i}(z-z_j)}{2K})}{\vartheta_3\vartheta_4(\frac{\mathrm{i}(z_j+z_0)}{2K})\vartheta_4(\frac{\mathrm{i}(z_j'+z_0)}{2K})\vartheta_4(\frac{\mathrm{i}(z+z_0)}{2K})\vartheta_4(\frac{\mathrm{i}(z'+z_0)}{2K})}.
	\end{align} 
	Therefore, we obtain
	\begin{align}
		M(x,z_j,z)&=-2\frac{\alpha\vartheta_2\vartheta_4\vartheta_4(\frac{\alpha x+\mathrm{i}(z_j-z')}{2K})}{\vartheta_3\vartheta_4(\frac{\alpha x}{2K})\vartheta_1(\frac{\mathrm{i}(z_j-z')}{2K})}\mathrm{e}^{-\mathrm{i}[k(z_j)+k(z)]x} \nonumber \\
		&=-2\frac{\alpha\vartheta_2\vartheta_4\vartheta_1(\frac{\alpha x+\mathrm{i}(z_j+z)}{2K})}{\vartheta_3\vartheta_4(\frac{\alpha x}{2K})\vartheta_4(\frac{\mathrm{i}(z_j+z)}{2K})}\mathrm{e}^{\mathrm{i}\frac{\pi\alpha}{2K}x}\mathrm{e}^{-\mathrm{i}[k(z_j)+k(z)]x}.
	\end{align}
\subsection{Reflectionless solution}\label{subsec:reflectionless}
	Now we solve the GLM equation for a reflectionless case $ \Omega_c=0 $ and $ \Omega=\Omega_d $. The solution can be obtained by imposing the following form for the kernel $ \Gamma $: 
	\begin{align}
		\Gamma(x,y)&=\sum_jC_j\begin{pmatrix} h_j(x) \\ h_j(x)^* \end{pmatrix}f_0(y,z_j')^T\sigma_1.
	\end{align}
	As we see below,  $ (u,v)=(h_j,h_j^*) $ is a normalized bound state. Substituting this to the GLM equation and performing the integration, we have
	 \begin{align}
	 	\begin{pmatrix} h_j(x) \\ h_j(x)^* \end{pmatrix}+ C_j\begin{pmatrix}u_0(x,z_j') \\ v_0(x,z_j') \end{pmatrix}+ \sum_{i}\begin{pmatrix} h_i(x) \\ h_i(x)^* \end{pmatrix}C_iC_jM(x,z_i,z_j)=0. \label{eq:fjdetermine2}
	 \end{align}
	We note that $ v_0(x,z_j')=u_0(x,z_j')^* $ and $ M(x,z_i,z_j)=M(x,z_i,z_j)^* $ from Eqs. (\ref{eq:f0period}), (\ref{eq:appccrelf2}), and (\ref{eq:Mzjz002}). Hence, the first and second component of Eq. (\ref{eq:fjdetermine2}) are equivalent. 
	Using the $ h_i $ satisfying Eq. (\ref{eq:fjdetermine2}), the potential and the Jost functions are given by (see Eqs. (\ref{eq:triang}) and (\ref{eq:triang22}))
	\begin{align}
		\psi(x)&=\psi_0(x)+2\mathrm{i}\sum_jh_j(x)C_ju_0(x,z_j'),  \label{eq:psirefless01} \\
		f_-(x,z')&=f_0(x,z')+\sum_i\begin{pmatrix} h_i(x) \\ h_i(x)^* \end{pmatrix}C_iM(x,z_i,z), \label{eq:uvrefless01}
	\end{align} 
	Multiplying $ C_j $ and substituting $ z=z_j $, we find $ (h_j,h_j^*)^T=-C_jf_-(x,z_j')=-c_j|b(z_j)|f_-(x,z_j') $, which is the normalized bound state (see Subsec.~\ref{subsec:bsnmrztn}). 
\subsection{Determinant expressions}\label{subsec:determinant}
	\indent Let us construct determinant expressions for the reflectionless solutions. 
	Let  $ \mathcal{E}(x) $ be a diagonal matrix with
	\begin{align}
		\mathcal{E}(x)&=\operatorname{diag}(e_1(x),\dots,e_n(x)), \\
		e_j(x)&=C_j\mathrm{e}^{-\mathrm{i}k(z_j)x}.
	\end{align}
%
%
	\indent Let $ \mathcal{M}(x) $ be an $ n\times n $ matrix with $ (i,j) $-components defined by
	\begin{align}
		\mathcal{M}_{ij}(x)&=-2\alpha\frac{ \vartheta_2\vartheta_4\vartheta_4(\frac{\alpha x+\mathrm{i}(z_i-z_j')}{2K})}{\vartheta_3\vartheta_4(\frac{\alpha x}{2K})\vartheta_1(\frac{\mathrm{i}(z_i-z_j')}{2K})}. 
	\end{align} 
	Then, the solution of Eq.~(\ref{eq:fjdetermine2}) is given by $ (h_1,\dots,h_n)=-(C_1u_1,\dots,C_nu_n)(I_n+\mathcal{EME})^{-1} $, with $ u_i=u(x,z_i') $.
	Using the Weierstrass addition formula (\ref{eq:weierstrass1144}) and the linear-algebraic formula $ a+y^\dagger A^{-1}x=\frac{a \det(A+a^{-1}xy^\dagger)}{\det A} $, where $ a $ is a scalar,  $ x,y $ are vectors, and $ A $ is a matrix, we rewrite Eqs. (\ref{eq:psirefless01}) and (\ref{eq:uvrefless01}). The resultant expressions are
	\begin{align}
		\psi(x)&=\psi_0(x)\frac{\det[I_n+\mathcal{EPAQE}]}{\det[I_n+\mathcal{EME}]}, \label{eq:psidet21} \\
		f_-(x,z')&=\frac{1}{\det[I_n+\mathcal{EME}]} \begin{pmatrix} u_0(x,z')\det[I_n+\mathcal{EP'UQ'E}] \\ v_0(x,z')\det[I_n+\mathcal{EP''VQ''E}] \end{pmatrix}, \label{eq:fermiondet21}
	\end{align}
	where we define $x$-independent diagonal matrices  $ \mathcal{P}$, $\mathcal{Q}$, $\mathcal{P}'$, $\mathcal{Q}'$, $\mathcal{P}''$, $\mathcal{Q}'' $ whose $ j $-th entries are given by 
	\begin{align}
		&\mathcal{P}_j=\frac{\vartheta_4(\frac{\mathrm{i}(z_0+z_j)}{2K})}{\vartheta_1(\frac{\mathrm{i}(z_0-z_j)}{2K})}, \quad \mathcal{Q}_j=\frac{\vartheta_1(\frac{\mathrm{i}(z_0-z_j')}{2K})}{\vartheta_4(\frac{\mathrm{i}(z_0+z_j')}{2K})}, \\
		&\mathcal{P}'_j=\frac{\vartheta_4(\frac{\mathrm{i}(z_0+z_j)}{2K})}{\vartheta_1(\frac{\mathrm{i}(z'-z_j)}{2K})}, \quad \mathcal{Q}'_j=\frac{\vartheta_1(\frac{\mathrm{i}(z'-z_j')}{2K})}{\vartheta_4(\frac{\mathrm{i}(z_0+z_j')}{2K})}, \\
		&\mathcal{P}''_j=\frac{\vartheta_1(\frac{\mathrm{i}(z_0-z_j)}{2K})}{\vartheta_4(\frac{\mathrm{i}(z+z_j)}{2K})}, \quad \mathcal{Q}''_j=\frac{\vartheta_4(\frac{\mathrm{i}(z+z_j')}{2K})}{\vartheta_1(\frac{\mathrm{i}(z_0-z_j')}{2K})}, 
	\end{align}
	and matrices $ \mathcal{A}(x),\mathcal{U}(x),\mathcal{V}(x) $ whose $ (i,j) $-components are
	\begin{align}
		&\mathcal{A}_{ij}(x)=-2\alpha\frac{\vartheta_2\vartheta_4\vartheta_1(\frac{\alpha x-\mathrm{i}(2z_0-z_i+z_j')}{2K})}{\vartheta_3\vartheta_1(\frac{\alpha x-2\mathrm{i}z_0}{2K})\vartheta_1(\frac{\mathrm{i}(z_i-z_j')}{2K})}, \\
		&\mathcal{U}_{ij}(x)=-2\alpha\frac{\vartheta_2\vartheta_4\vartheta_1(\frac{\alpha x-\mathrm{i}(z_0+z'-z_i+z_j')}{2K})}{\vartheta_3\vartheta_1(\frac{\alpha x-\mathrm{i}(z'+z_0)}{2K})\vartheta_1(\frac{\mathrm{i}(z_i-z_j')}{2K})}, \\
		&\mathcal{V}_{ij}(x)=-2\alpha\frac{\vartheta_2\vartheta_4\vartheta_1(\frac{\alpha x+\mathrm{i}(z_0+z+z_i-z_j')}{2K})}{\vartheta_3\vartheta_1(\frac{\alpha x-\mathrm{i}(z'+z_0)}{2K})\vartheta_1(\frac{\mathrm{i}(z_i-z_j')}{2K})}. 
	\end{align} 
	\indent We will prove in Sec.~\ref{sec:timeevo} that the time-dependent soliton solutions of the higher-order NLS (AKNS$_n$) equation (\ref{eq:timeevohighernls}) can be obtained by the simple replacement
	\begin{align}
		&e_j(x)=C_j\mathrm{e}^{-\mathrm{i}k(z_j)x} \nonumber \\
		\to\quad& e_j(t,x)=C_j \mathrm{e}^{-\mathrm{i}\omega_n(z)t-\mathrm{i}k(z_j)x}.
	\end{align}
	in $ \mathcal{E} $, where $ \omega_n $ is defined in Eq.~(\ref{eq:timeevoomeganz}). The velocity of the $ j $-th soliton is given by $ V_j=-\frac{\operatorname{Im}\omega_n(z_j)}{\operatorname{Im}k(z_j)} $. If we parametrize $ C_j $ as $ C_j=\frac{1}{\sqrt{\alpha}}\mathrm{e}^{-\operatorname{Im}k(z_j)x_j} $, then $ x_j $ represents the position of the $ j $-th soliton at $ t=0 $ up to an additive constant. \\
	\indent The reduction to the case where the background is the LO state, or the real sn lattice, is realized by setting $ z_0=\tilde{p}=0 $. In particular, the expressions of Subsec.~\ref{sec:nsolitonrealcase} are reproduced by writing $ \psi_0(x):=\psi_{\text{LO}}(x) $ and $ \mathcal{\tilde{A}}:=\mathcal{PAQ} $,  $  \mathcal{\tilde{U}}:=\mathcal{P'UQ'} $, and  $  \mathcal{\tilde{V}}:=\mathcal{P''VQ''} $.  The case of $ z_0\ne0, \tilde{p}\ne0 $ corresponds to the more general FFLO case, which is summarized in Subsec.~\ref{subsec:nsolitoncomplexcase}.

	
\subsection{Asymptotics}\label{subsec:asymptotics}
	Since $ \mathcal{E}(x)\to0 \ (\infty) $ in the limit $ x\to-\infty \ (+\infty) $, the asymptotic form of $ \psi(x) $ [Eq.~(\ref{eq:psidet21})] is
	\begin{align}
		\psi(x) \to \begin{cases} \psi_0(x) & (x\to-\infty), \\ \displaystyle \psi_0(x)\frac{(\prod_j\mathcal{P}_j\mathcal{Q}_j)\det\mathcal{A}}{\det\mathcal{M}} & (x\to+\infty). \end{cases}
	\end{align}
	Let us determine the asymptotic constants $ x_0 $ and $ \varphi_0 $ in Eq.~(\ref{eq:AKNS1boundary}).
	Using the determinant formula in Eq.~(\ref{eq:appthetaratiodet2}), we find 
	\begin{align}
		\lim_{x\rightarrow+\infty}\psi(x)= \prod_j \left(\mathcal{P}_j\mathcal{Q}_j \mathrm{e}^{\frac{p}{\alpha}(2z_j-K')} \right) \psi_0(x+\tfrac{\mathrm{i}\sum_j(2z_j-K')}{\alpha}).
	\end{align}
	We must not misidentify $ x_0=-\frac{\mathrm{i}\sum_j(2z_j-K')}{\alpha} $ from this expression, since $ \frac{\mathrm{i}\sum_j(2z_j-K')}{\alpha} $ is generally a complex number, unless all $ z_j $'s have a positive real part $ \frac{K'}{2} $. If there exists $ z_j $ with real part $ -\frac{K'}{2} $, we need a slight rewriting. \\
	\indent Let us write $ z_j=s_j\frac{K'}{2}+\mathrm{i}\eta_j $ with $ s_j=\pm 1 $ and $ 0<\eta_j<K $. Then,
	\begin{align}
		\mathcal{P}_j\mathcal{Q}_j=\begin{cases} \displaystyle\frac{\vartheta_1(\frac{\mathrm{i}(z_0-\frac{K'}{2})-\eta_j}{2K})^2}{\vartheta_1(\frac{\mathrm{i}(z_0-\frac{K'}{2})+\eta_j}{2K})^2}\mathrm{e}^{\frac{\pi}{2K}2\mathrm{i}\eta_j} & (s_j=+1), \\ \displaystyle \frac{\vartheta_1(\frac{\mathrm{i}(z_0+\frac{K'}{2})-\eta_j}{2K})^2}{\vartheta_1(\frac{\mathrm{i}(z_0+\frac{K'}{2})+\eta_j}{2K})^2}\mathrm{e}^{-\frac{2\pi z_0}{K}-\frac{\pi}{2K}2\mathrm{i}\eta_j} & (s_j=-1).  \end{cases} \label{eq:asymnsol0003}
	\end{align}
	Let us write $ \mathrm{i}\sum_j(2z_j-K')=-2\sum_j\eta_j-2\mathrm{i}K's^\# $, where  $ s^\#:=\sum_j\frac{1-s_j}{2} $ counts the number of $ z_j $'s having the real part $ -\frac{K'}{2} $. 
	Then, the main theta-functional part of $ \psi_0(x+\tfrac{\mathrm{i}\sum_j(2z_j-K')}{\alpha}) $ is rewritten as
	\begin{align}
		\frac{\vartheta_1(\frac{\alpha x-2\mathrm{i}z_0-2\sum_j\eta_j}{2K}-s^\#\tau)}{\vartheta_4(\frac{\alpha x-2\sum_j\eta_j}{2K}-s^\#\tau)}=\mathrm{e}^{\frac{2\pi z_0}{K}s^\#}\frac{\vartheta_1(\frac{\alpha x-2\mathrm{i}z_0-2\sum_j\eta_j}{2K})}{\vartheta_4(\frac{\alpha x-2\sum_j\eta_j}{2K})}. \label{eq:asymnsol0004}
	\end{align}
	The factors $ \mathrm{e}^{\frac{2\pi z_0}{K}s^\#} $ in Eq.~(\ref{eq:asymnsol0004}) and $ \mathrm{e}^{-\frac{2\pi z_0}{K}} $ in Eq. (\ref{eq:asymnsol0003}) are canceled out, and we obtain the asymptotic form
	\begin{align}
		\lim_{x\rightarrow+\infty}\psi(x)=\mathrm{e}^{2\mathrm{i}\varphi_0}\psi_0(x-x_0)
	\end{align}
	with
	\begin{align}
		x_0&=\frac{2\sum_j\eta_j}{\alpha}, \\
		\mathrm{e}^{2\mathrm{i}\varphi_0}&=\prod_j\frac{\mathrm{e}^{2\mathrm{i}\eta_j(\frac{p}{\alpha}+s_j\frac{\pi}{2K})}\vartheta_1(\frac{\mathrm{i}(z_0-s_j\frac{K'}{2})-\eta_j}{2K})^2}{\vartheta_1(\frac{\mathrm{i}(z_0-s_j\frac{K'}{2})+\eta_j}{2K})^2},
	\end{align}
	which represent the lattice translation and the phase shift induced by solitons. 

\section{ AKNS$_n$ covering AKNS$_{m<n}$}\label{sec:timeevo2}
	Here, we discuss a condition that the higher-order stationary AKNS equation has a solution for the lower-order one. We use the same notation as Ref.~\cite{Takahashi:2012aw}, and we write $ q=-\mathrm{i}\psi $ and $ r=\mathrm{i}\psi^* $. \\
	\indent Let us consider the stationary AKNS$_n$ equation 
	\begin{align}
		\sum_{j=1}^{n+2} c_jM_{12}^{(j)}=0, \label{eq:cjakns01}
	\end{align}
	where $ M_{12}^{(j)},\ j=1,2,\dots $ are the $ (1,2) $-component of the formal Laurent series solution $ M=\sum_{j=0}^\infty\frac{M^{(j)}}{(-2\lambda)^j} $  for the Lax equation $ M_x=[U,M] $ with $ M^{(0)}=\frac{\sigma_3}{2\mathrm{i}} $ and $  U=\left( \begin{smallmatrix} -\mathrm{i}\lambda & q \\ r & \mathrm{i}\lambda \end{smallmatrix} \right) $  \cite{FaddeevTakhtajan,Takahashi:2012aw}. Here, when we iteratively determine $ M^{(j)} $, the integration constants are fixed to keep the scaling property $ M(\alpha \lambda, \{ \alpha^{j+1} q^{(j)}(x), \alpha^{j+1} r^{(j)}(x) \})=M(\lambda, \{ q^{(j)}(x), r^{(j)}(x) \}) $. The first few $ M^{(j)} $'s are available in Ref.~\cite{Takahashi:2012aw}. \\
	\indent Equation~(\ref{eq:cjakns01}) has a solution of the lower-order AKNS$_{m<n}$ equation
	\begin{align}
		\sum_{j=1}^{m+2} d_jM_{12}^{(j)}=0, \label{eq:djakns01}
	\end{align}
	if the coefficients $ c_1,\dots, c_{n+2} $ and $ d_1,\dots,d_{m+2} $ satisfy the relation
	\begin{align}
		c_j=\sum_{k=0}^{n-m}d_{j-k}\alpha_{k+1} \quad (j=1,\dots,n+2), \label{eq:cjtodj01}
	\end{align}
	where $ \alpha_1,\dots,\alpha_{n-m+1} $ are arbitrary real constants and $ d_j $'s with extended indices are defined by
	\begin{align}
		d_j=\begin{cases} -2J_{1-j} & (1-n+m\le j \le 0) \\ d_j & (1\le j\le m+2) \\ 0 & (m+3\le j \le n+2),  \end{cases} \label{eq:extendeddjs}
	\end{align}
	where $ J_{1-j} $'s are integration constants in the stationary AKNS$_m$ equation determined by the following procedure: The infinite conservation laws in the AKNS system $ f_x=Uf $ and $ f_t=Vf $ can be obtained as \cite{WadatiSanukiKonno}
	\begin{align}
		0&=\partial_t(U_{11}+U_{12}\Gamma)+\partial_x(V_{11}+V_{12}\Gamma) \nonumber \\
		&=:\sum_{j=1}^\infty\frac{-\mathrm{i}}{(-2\lambda)^j}(\partial_tF_j+\partial_xJ_j), \label{eq:AKNSconserv}
	\end{align}
	where $ \Gamma=f_2/f_1 $ satisfies the Ricatti equation 
	\begin{align}
		\Gamma_x+U_{12}\Gamma^2+(U_{11}-U_{22})\Gamma-U_{21}=0.
	\end{align}
	Each order in Eq.~(\ref{eq:AKNSconserv}) gives the conservation law $ \partial_t F_j+\partial_x J_j=0 $ with the charge $ F_j $ and the current $ J_j $. When we consider the stationary solution ($\partial_t=0$), it reduces to $ \partial_x J_j=0 $, and hence $ J_j $ provides an integration constant. In the stationary AKNS$_m$ equation, only $ J_1,\dots, J_{m+1} $ are independent, since the equation is an $(m+1)$-th-order differential equation. The higher-order constants $ J_{m+2}, J_{m+3},\dots $ are iteratively determined by 
	\begin{align}
		\sum_{j=1}^{m+2}d_jJ_{j}&=0, \\
		\sum_{j=1}^{m+2}d_jJ_{j+l}&=\sum_{k=0}^{l-1}J_{1+k}J_{l-k} \quad (l\ge1).
	\end{align}
	We remark that Eq.~(\ref{eq:extendeddjs}) implies that the coefficients $ d_j $'s are regarded as ``negative-numbered'' integration constants. This guess can be justified by generating the first integrals using the Krichever's formal solution \cite{Krichever1977}. \\
	\indent The AKNS matrices $ U $ and $ V $ for Eqs.~(\ref{eq:cjakns01}) and (\ref{eq:djakns01}) are given by
	\begin{align}
		U=\begin{pmatrix} -\mathrm{i}\lambda & q \\ r & \mathrm{i}\lambda \end{pmatrix}, \quad V_c=\sum_{j=1}^{n+2}c_jV^{(j)},\quad V_d=\sum_{j=1}^{m+2}d_jV^{(j)} 
	\end{align}
	with $ V^{(j)}:=\sum_{k=0}^{j-1}(-2\lambda)^{j-1-k}M^{(k)} $. Using them, Eq.~(\ref{eq:cjakns01}) and  (\ref{eq:djakns01}) are given by $ \partial_x V_c=[U,V_c] $ and  $ \partial_x V_d=[U,V_d] $, respectively. If the coefficients satisfy the relation (\ref{eq:cjtodj01}), we can check the relation
	\begin{align}
		V_c=\left( \sum_{j=1}^{n-m+1}\alpha_j(-2\lambda)^{j-1} \right)V_d.
	\end{align}
	If we write the Riemann surfaces $ \omega_c^2=\det V_c $ and $ \omega_d^2=\det V_d $, they are related as
	\begin{align}
		\omega_c=\left( \sum_{j=1}^{n-m+1}\alpha_j(-2\lambda)^{j-1} \right)\omega_d. \label{eq:omegacdakns}
	\end{align}
	We have checked the validity of Eqs. (\ref{eq:cjtodj01})-(\ref{eq:omegacdakns}) for $ 1\le m<n\le 10 $ by Mathematica, though we do not give a general proof here. 
	
	For example, if we consider $ (n,m)=(3,1) $, i.e., the AKNS$_3$ and AKNS$_1$ equation, the above relation is
	\begin{align}
		\begin{pmatrix} c_1 \\ c_2 \\ c_3 \\  c_4 \\ c_5 \end{pmatrix}=\begin{pmatrix}d_1 & -2J_1 & -2J_2 \\ d_2 & d_1 & -2J_1 \\ d_3 & d_2 & d_1 \\ 0 & d_3 & d_2 \\ 0 & 0 & d_3\end{pmatrix}  \begin{pmatrix}\alpha_1 \\ \alpha _2 \\ \alpha_3 \end{pmatrix}
	\end{align}
	with
	\begin{align}
		J_1&=d_2 rq+\mathrm{i}d_3(r_xq-rq_x),  \label{eq:currentj1} \\
		J_2&=-d_1 rq+d_3(-r^2q^2+r_xq_x). \label{eq:currentj2}
	\end{align}
	The constants $ J_3,J_4,\dots $ are  successively determined by $ d_1J_1+d_2J_2+d_3J_3=0,\ d_1J_2+d_2J_3+d_3J_4=J_1^2, $ and so on. $ V_c $ and $ V_d $ are related as
	\begin{align}
		V_c=\left( \alpha_1-2\alpha_2\lambda+4\alpha_3\lambda^2 \right)V_d.
	\end{align}
	\indent The situation in Sec.~\ref{sec:mainresult} is reproduced by the reduction: $ d_3=1 $, $ c_2=c_4=0 $, and $ c_1=-\mu $. Then, we have $ \alpha_3=c_5,\ \alpha_2=-c_5d_2,\ \alpha_1=c_3+c_5(d_2^2-d_1) $. The chemical potential is given by
	\begin{align}
		\mu=-c_3d_1+c_5[d_1(d_1-d_2^2)-2d_2J_1+2J_2], \label{eq:chempotc3c5}
	\end{align}
	and the constraint between these coefficients is:
	\begin{align}
		d_2^3+\left( \tfrac{c_3}{c_5}-2d_1 \right)d_2-2J_1=0. \label{eq:constraintc3c5}
	\end{align}
	For the FF state $ q=\sqrt{\bar{\rho}}\mathrm{e}^{\mathrm{i}px},\ r=\sqrt{\bar{\rho}}\mathrm{e}^{-\mathrm{i}px} $, we have  $ J_1=d_2\bar{\rho}+2\bar{\rho}p, J_2=-d_1\bar{\rho}-\bar{\rho}^2+p^2\bar{\rho} $, and $ d_1=-d_2p-(2\bar{\rho}+p^2) $ then Eq.~(\ref{eq:chempotc3c5}) reproduces $ \mu_{\text{FF}} $ by using Eq.~(\ref{eq:constraintc3c5}). For the LO state $ q=r=\sqrt{m}\alpha \operatorname{sn}(\alpha x) $, we have $ d_2=J_1=0 $, $ d_1=-(m+1)\alpha^2, $ and $ J_2=m\alpha^4 $ (set $ \tilde{p}=z_0=0 $ in Eq.~(\ref{eq:akns1n1dcoeffs})), then Eq.~(\ref{eq:chempotc3c5}) reduces  to $ \mu_{\text{LO}} $. \\
	\indent For the FFLO state, using $ s,c,d $ of Eq.~(\ref{eq:akns1riemannsurface04}) and defining $ S_1=s^2+c^2+d^2 $, $ S_2=s^2c^2+c^2d^2+d^2s^2 $, and $ S_3=scd $, we get $ d_2=-2\tilde{p},\ d_1=\tilde{p}^2-S_1,\ J_1=2S_3, $ and $ J_2=S_2+2\tilde{p}S_3 $ . Equation~(\ref{eq:constraintc3c5}) reduces to $ \tilde{p}^3+\left( \tfrac{c_3}{2c_5}+S_1 \right)\tilde{p}+S_3=0 $, which determines $ \tilde{p} $. 
	The chemical potential (\ref{eq:chempotc3c5}) becomes $ \mu=c_3(S_1-7\tilde{p}^2)+c_5(S_1^2+2S_2-10S_1\tilde{p}^2-15\tilde{p}^4) $. 
	The uniformization variable is introduced as
	\begin{align}
		\omega_3(z)&=[\alpha_1-2\alpha_2\tilde{\lambda}(z)+4\alpha_3\tilde{\lambda}(z)^2]\omega(z) \nonumber \\
		&=[c_3+c_5(4\tilde{\lambda}(z)^2-4\tilde{p}\tilde{\lambda}(z)+3\tilde{p}^2+S_1)]\omega(z)
	\end{align}
	with $ \tilde{\lambda}(z)=\lambda(z)-\frac{\tilde{p}}{2} $. If $ z_0=0 $, the expressions reduce to the LO case. 

\section{Time evolution}\label{sec:timeevo}
	Finally, we solve the time-evolution problem of the higher-order NLS equation. While our main interest in Sec.~\ref{sec:mainresult} is the system $ H-\mu N =-\mu I_1+c_3I_3+c_5I_5 $, here we give a more general answer for the higher order NLS equations whose energy functional is given by $ \sum_{j=1}^{n+2} c_jI_j $ and the asymptotic form of $\psi$ is given by the tending-to-AKNS$_1$ boundary condition [Eq.~(\ref{eq:AKNS1boundary})]. \\
%
%
	\indent We now determine the time evolution of the AKNS$_n$ equation
	\begin{align}
		\mathrm{i}\partial_t \psi = \sum_{j=1}^{n+2} c_j(-\mathrm{i}M_{12}^{(j)}), \label{eq:timeevohighernls}
	\end{align}
	with the tending-to-AKNS$_1$ boundary condition (\ref{eq:AKNS1boundary}). If we set $ n=3$,  $  c_1=-\mu $, and $ c_2=c_4=0 $, Eq.~(\ref{eq:timeevohighernls}) reduces to Eq.~(\ref{eq:I3I5q}). We parametrize  $ d_1,d_2,d_3 $ in the same way as in Sec. \ref{sec:bdgwithakns1bg}. 
	 The coefficients $ c_1,\dots,c_{n+2} $ must satisfy the relation (\ref{eq:cjtodj01}), because the potential $ \psi $ asymptotically tends to the stationary AKNS$_1$ potential at spatial infinities $ x\to \pm\infty $. Following the result of the previous section, we introduce the uniformization variable
	\begin{align}
		\omega_n(z)=\omega(z)\left( \sum_{j=1}^{n}\alpha_j(-2\tilde{\lambda}(z))^{j-1} \right), \label{eq:timeevoomeganz}
	\end{align}
	where $ \tilde{\lambda}(z)=\lambda(z)-\frac{\tilde{p}}{2} $ with $ \tilde{p}=-\frac{d_2}{2d_3} $, and $ \lambda(z), \omega(z) $ are defined in Eqs. (\ref{eq:applambdazn1}) and (\ref{eq:unifoomegan1}). 
	Let us define the time-dependent right and left Jost functions by the asymptotic form
	\begin{align}
		f_+(t,x,z) &\to \mathrm{e}^{\mathrm{i}\varphi_0\sigma_3}f_0(x-x_0,z) \quad (x\to+\infty), \\
		f_-(t,x,z) &\to f_0(x,z) \quad (x\to-\infty).
	\end{align}
	We define the time-dependent scattering matrix by the relation
	\begin{align}
		f_+(t,x,z)=f_-(t,x,z)S(t,z).
	\end{align}
	We simply write $ f_\pm(0,x,z)=f_\pm(x,z) $ and $ S(0,z)=S(z) $. 
	Then, solving the time-derivative equation of the AKNS system $ \partial_t f=Vf $ at $ x=\pm\infty $, we find the time evolution of the scattering matrix
	\begin{align}
		S(t,z)=\mathrm{e}^{\mathrm{i}\omega_n(z)\sigma_3t}S(z)\mathrm{e}^{-\mathrm{i}\omega_n(z)\sigma_3t},
	\end{align}
	or equivalently,
	\begin{align}
		a(t,z)=a(z),\quad b(t,z)=\mathrm{e}^{-2\mathrm{i}\omega_n(z)t}b(z).
	\end{align}
	The time evolution of the normalization coefficient of the bound state $ C_j=|b(z_j)|c_j $ is
	\begin{align}
		C_j(t)=\mathrm{e}^{-\mathrm{i}\omega_n(z)t}C_j,
	\end{align}
	since $ C_j^2=|b(z_j)|^2c_j^2 $ has the same time dependence with $ b(z_j)/\dot{a}(z_j) $ due to Eq. (\ref{eq:ncbyscat01}). Solving the GLM equation (\ref{eq:GLM001}) for each time $ t $ with the use of the time evolution of the scattering data  $ a(t,z), b(t,z) $, and $ C_j(t) $, we can solve the initial-value problem of the AKNS$_n$ equation with the tending-to-AKNS$_1$ boundary condition, i.e., the problem with the soliton-lattice background. In particular, if we are interested in the reflectionless solution, we can obtain the time evolution by formally replacing $ C_j \to C_j(t) $ 
	in the equations of Subsec.~\ref{subsec:determinant}. 

%

\section{Summary and perspective}\label{sec:summary}
	We have introduced the integrable model of density-modulated quantum condensates as a linear combination of conserved quantities in the NLS hierarchy, and have provided an $ n $-soliton solution by formulating the IST with the elliptic-functional background.  
	The resulting exact soliton solutions exhibit various kinds of novel dynamics such as dark soliton billiards, stationary dislocations, gray solitons, and envelope solitons. Their behaviors are different from gap solitons and soliton trains.
	The tunneling phenomena of quasiparticle bound states have been also demonstrated. 
	Our result will be universal and useful to understand nonequilibrium and transport phenomena in non-uniform quantum matters. 
	These solitons will be realized using the phase imprinting \cite{PhysRevLett.83.5198,Nature4994262013,PhysRevLett.116.045304} or the barrier sweeping \cite{PhysRevLett.99.160405}, if a density-modulated state in ultracold atomic systems can be prepared. Recently, the density order in Dy atoms with the dipolar interaction is observed \cite{Nature5301942016,FerrierBarbutKadauSchmittWenzelPfau}. 

	\indent The author initiated this work because he was stimulated by the numerical simulation of soliton emission in the bose condensates with soft-core interaction in Ref.~\cite{KunimiConf}, and wanted to find an exactly tractable example of such solitons with spontaneously-modulated background. 
	The model was constructed based on the idea in Subsec.~\ref{subsec:ideainmainresult}. However, in order to achieve integrability, the model includes the terms whose physical meanings are not evident. Finding a more realistic model with solvability is left as a future problem. 
	In fact, as discussed in Subsec.~\ref{sec:nsolitonrealcase}, the soliton dynamics with soliton-lattice background will be realized even in the ordinary (not higher-order) NLS systems, if we can prepare the low-temperature state to suppress the instability. \\
	\indent The behavior of the soliton-lattice and multi-soliton solutions in the higher-order NLS system reminds us of fermionic condensates, rather than bosonic ones. This is quite natural, because it is known that the NLS hierarchy and the self-consistent BdG  solitons have a close relation \cite{Basar:2008ki,Correa:2009xa,Takahashi:2012aw,PhysRevD.92.034003}. \\
%
%
%
	\indent After submitting the first preprint in 2013, the author noticed several references which address similar issues and discuss related concepts \cite{KotlyarovIts1976,PhysRevD.90.125041,1402-4896-90-4-045205,PhysRevD.92.034003,DyachenkoZakharovZakharov,PhysRevD.92.105009,SmirnovMatveev}. \\
	\indent The next important future work is the construction of the self-consistent BdG solitons \cite{Takahashi:2012pk,PhysRevLett.111.121602} with elliptic backgrounds, employing the method of Ref.~\cite{Takahashi:2015nda}.
\section*{Acknowledgment}
	The author is grateful to M.~Kunimi, Y.~Kato, K.~Sakai, M.~Nitta, A.~S.~Ovchinnikov, J.~Kishine, Y.~Hidaka, K.~Kamikado, T.~Kanazawa, and T.~Noumi for valuable discussions. The author is also grateful to M.~T.~Batchelor, V.~V.~Bazhanov, and Z.~Tsuboi for their support to his survival in Canberra. \\
	\indent This paper was first submitted in May 2013, when the author was a visiting researcher of the Australian National University, supported by the JSPS Institutional Program for Young Researcher Overseas Visits. Later, the paper was thoroughly revised in the present affiliation.

\appendix
\makeatletter
\renewcommand{\theequation}{%
\thesection\arabic{equation}}
\@addtoreset{equation}{section}
\makeatother

\section{ Evaluation and minimization of energies for FF and LO states}\label{sec:minimizefflo}
	The energy density $ h(x) $ at a point $ x $ is defined by the integrand of Eq.~(\ref{eq:Hami}). The energy per particle is defined by  $ \mathcal{E}=\int_0^L\mathrm{d}xh(x)\Big/\int_0^L\mathrm{d}x|\psi|^2 $, where $ L $ is a period given by $ L=2\pi/p $ for the FF state and $ L=4K(m)/\alpha $ for the LO state, respectively. Let $ \mathcal{E}_{\text{FF}}(\bar{\rho},p) $ and  $ \mathcal{E}_{\text{LO}}(\bar{\rho},m) $ be the energies per particle for the FF and LO states. A straightforward calculation gives  
	\begin{align}
		\mathcal{E}_{\text{FF}}(\bar{\rho},p)&=c_3(p^2+\bar{\rho})+c_5(p^4+6p^2\bar{\rho}+2\bar{\rho}^2), \\
		\mathcal{E}_{\text{LO}}(\bar{\rho},m)&=c_3\frac{\bar{\rho}[m+(m+1)Q(m)]}{3Q(m)^2}\nonumber \\
		&+c_5\frac{\bar{\rho}^2[2m(m+1)+(m^2+4m+1)Q(m)]}{5Q(m)^3}. 
	\end{align}
	where $ Q(m):=1-\frac{E(m)}{K(m)} $. 
	The variational parameters $ p $ and $ m $ are to be chosen to minimize the above energies for fixed $ \bar{\rho} $. 
	Let $ p=p_g(\bar{\rho}) $ and $ m=m_g(\bar{\rho}) $ be such values. 
	They are determined as follows:
	\begin{align}
		p_g(\bar{\rho})&=\begin{cases} 0 & (\bar{\rho}>\frac{-c_3}{6c_5}) \\ \pm\sqrt{-(c_3+6c_5\bar{\rho})/(2c_5)} & (\bar{\rho}<\frac{-c_3}{6c_5}), \end{cases} \\ 
		m_g(\bar{\rho})&=\begin{cases} 1 & (\bar{\rho}>\frac{-5c_3}{18c_5}) \\ \text{inverse function of } \bar{\rho}_g(m) & (\bar{\rho}<\frac{-5c_3}{18c_5}), \end{cases} \\ 
		\bar{\rho}_g(m)&:=\frac{-5c_3[-2m+(1+m)Q(m)]Q(m)}{6c_5[-3m(1+m)+(1+4m+m^2)Q(m)]}. \label{eq:barrhominimizer} 
	\end{align}
	Here we have assumed $ c_3<0 $ and $ c_5>0 $. 
	Then, $ \mathcal{E}_{\text{FF}}(\bar{\rho}) $ and $ \mathcal{E}_{\text{LO}}(\bar{\rho}) $ appearing in Subsec.~\ref{subsec:maings} are defined as $ \mathcal{E}_{\text{FF}}(\bar{\rho})=\mathcal{E}_{\text{FF}}(\bar{\rho},p_g(\bar{\rho})) $ and $ \mathcal{E}_{\text{LO}}(\bar{\rho})=\mathcal{E}_{\text{LO}}(\bar{\rho},m_g(\bar{\rho})) $. The periods are given by $ 2\pi/p_g(\bar{\rho}) $ and $ 4K(m_g(\bar{\rho}))/\sqrt{\bar{\rho}/Q(m_g(\bar{\rho}))} $ for the FF and LO states, respectively. Figure~\ref{fig:energycompare} is made by these functions. \\ 
	\indent In Figs \ref{fig:bsbg}, \ref{fig:onesolvelo}, \ref{fig:onesoliton01}, and \ref{fig:bound}, we choose $ \alpha=\sqrt{\bar{\rho}_g(m)/Q(m)} $; i.e., the energy-minimizing LO states are always chosen in these figures.

\section{Convention of elliptic functions in this paper}\label{app:ellipconv}
	We use Mathematica's notations for the elliptic integrals and the Jacobi elliptic functions $ K(m)$, $E(m)$, $\Pi(n;\varphi|m)$, $\operatorname{am}(u|m)$, $\operatorname{sn}(u|m)$, $\operatorname{cn}(u|m)$, and $ \operatorname{dn}(u|m) $. We omit $ m $ when it is obvious. We write $ K=K(m),\ K'=K(1-m) $ and $ \tau=\mathrm{i}K'/K $. Exceptionally, the Jacobi zeta function $ Z(u|m) $ is defined in a different way from Mathematica (see below).  \\ 
	\indent For the theta functions, we use the following convention. Let us define
	\begin{align}
		\vartheta_{a,b}(u|\tau):=\sum_{n\in\mathbb{Z}}\mathrm{e}^{\mathrm{i}\pi\tau(n+a)^2}\mathrm{e}^{2\mathrm{i}\pi (n+a)(u+b)};
	\end{align}
	then 
	\begin{align}
		&\vartheta_3(u|\tau):= \vartheta_{0,0}(u|\tau),\quad \vartheta_4(u|\tau):=\vartheta_{0,\frac{1}{2}}(u|\tau), \\
		&\vartheta_2(u|\tau):= \vartheta_{\frac{1}{2},0}(u|\tau),\quad \vartheta_1(u|\tau):=-\vartheta_{\frac{1}{2},\frac{1}{2}}(u|\tau).
	\end{align}
	This convention is the same as that in Ref.~\cite{Kharchev201519}. 
	The relation with Mathematica's convention is $ [\vartheta_j(u,q)]_{\text{ used here}}=[\vartheta_j(\pi u,q)]_{\text{Mathematica}} $. 
	We also write $ \vartheta_j(u,q)=\vartheta_j(u|\tau) $ with the nome $ q=\mathrm{e}^{\mathrm{i}\pi \tau} $. They are written as $ \vartheta_j(u) $ when $ \tau $ or $ q $ is evident. The notation $ \vartheta_j=\vartheta_j(0) $ is also used. $ \vartheta_1(u) $ is odd and others are even. 
	The Jacobi elliptic functions in terms of thetas are $ \operatorname{sn}(2Ku)=\frac{\vartheta_3}{\vartheta_2}\frac{\vartheta_1(u)}{\vartheta_4(u)},\ \operatorname{cn}(2Ku)=\frac{\vartheta_4}{\vartheta_2}\frac{\vartheta_2(u)}{\vartheta_4(u)},\ \operatorname{dn}(2Ku)=\frac{\vartheta_4}{\vartheta_3}\frac{\vartheta_3(u)}{\vartheta_4(u)} $. The elliptic parameter is given by $ m=\vartheta_2^4/\vartheta_3^4 $. 

	We use the following definition for the Jacobi zeta function (the same convention as Toda's books, e.g., Ref. \cite{Todalattice}):
	\begin{align}
		Z(u|m)=\frac{1}{2K}\frac{\vartheta_4'(\frac{u}{2K})}{\vartheta_4(\frac{u}{2K})}=\frac{\mathrm{d}}{\mathrm{d}u}\log\vartheta_4(\tfrac{u}{2K}). \label{eq:Jacobizeta}
	\end{align}
	The parameter $ m $ is often omitted. It satisfies
	\begin{align}
		Z(-u)=-Z(u),\quad Z(u+2lK+2n\mathrm{i}K')=Z(u)-\frac{n\mathrm{i}\pi}{K}. \label{eq:zetaqperiod}
	\end{align}
	The following formulae are known:
	\begin{align}
		&\frac{\mathrm{d}}{\mathrm{d}u}Z(u|m)=\operatorname{dn}^2(u|m)-\frac{E(m)}{K(m)}, \label{eq:jacobizetaderv} \\
		&Z(u+v)-Z(u-v)-2Z(v)=-\frac{2m\operatorname{sn}^2u\operatorname{sn}v\operatorname{cn}v\operatorname{dn}v}{1-m\operatorname{sn}^2u\operatorname{sn}^2v}. \label{eq:zetaaddformla}
	\end{align}
	Substituting $ u=u+\mathrm{i}K' $ in Eq.~(\ref{eq:zetaaddformla}),
	\begin{align}
		\frac{\operatorname{sn}v\operatorname{cn}v\operatorname{dn}v}{\operatorname{sn}^2u-\operatorname{sn}^2v}=\frac{1}{2}\left( Z(u-v+\mathrm{i}K')-Z(u+v+\mathrm{i}K') \right)+Z(v).
	\end{align}
	Using the above formulae and $ \vartheta_4(z+\frac{\tau}{2})=\mathrm{i}\mathrm{e}^{-\mathrm{i}\pi(z+\tau/4)}\vartheta_1(z) $, we obtain the integral formula
	\begin{align}
		\int\mathrm{d}u\frac{\operatorname{sn}v\operatorname{cn}v\operatorname{dn}v}{\operatorname{sn}^2u-\operatorname{sn}^2v}
		&=\frac{1}{2}\log\frac{\vartheta_1(\frac{u-v}{2K})}{\vartheta_1(\frac{u+v}{2K})}+uZ(v)+\text{const}. \label{eq:zetaintegral}
	\end{align}
	The const only depends on $ v $. \\

\section{Fermionic eigenstates for AKNS$_1$ background}\label{app:bdgforakns1}
	In this appendix, we provide a detailed derivation for the expressions in Sec.~\ref{sec:bdgwithakns1bg}, i.e., the fermionic BdG (ZS) eigenstates expressed by theta functions when the general AKNS$_1$ potentials exist. 
	Here, we refer to higher-order NLS equations as ``AKNS$_g$ equations'', in accordance with Refs. \cite{Correa:2009xa,Takahashi:2012aw}.  $ g=1 $ corresponds to the normal NLS equation and $ g=3 $ is considered in Sec.~\ref{sec:mainresult}. \\
	\indent For convenience of comparison with Ref.~\cite{Takahashi:2012aw}, we write $ \psi=\mathrm{i}q,\ r=q^* $, and $ \epsilon=-\lambda $. Then, the BdG equation reduces to the spatial-derivative part of the AKNS form
	\begin{align}
		\partial_x \begin{pmatrix}u \\ v \end{pmatrix}=U\begin{pmatrix}u \\ v \end{pmatrix},\ U=\begin{pmatrix}-\mathrm{i}\lambda & q \\ r & \mathrm{i}\lambda  \end{pmatrix}. \label{eq:appakns1sppart}
	\end{align}
	\indent The stationary $\text{AKNS}_1$ equation is given by
	\begin{align}
		d_1q+d_2(-\mathrm{i}q_x)+d_3(-q_{xx}+2|q|^2q)=0, \label{eq:akns1app0}
	\end{align}
	where $ d_i $'s are real. We can eliminate the $ d_2 $-term by gauge transformation $ q \rightarrow  q \mathrm{e}^{\mathrm{i}\tilde{p}x} $ with $ \tilde{p}=-\frac{c_2}{2c_3} $, and the resulting equation is 
	\begin{align}
		-\mu q-q_{xx}+2|q|^2q=0. \label{eq:akns1app}
	\end{align}
	with $ \mu=\tilde{p}^2-\frac{d_1}{d_3} $. 
	 If $ (q,u,v,\lambda) $ is a solution of Eq.~(\ref{eq:appakns1sppart}),  
	$ (q\mathrm{e}^{\mathrm{i}\tilde{p}x},u\mathrm{e}^{\mathrm{i}\tilde{p}x/2},v\mathrm{e}^{-\mathrm{i}\tilde{p}x/2},\lambda-\frac{\tilde{p}}{2}) $ 
	is also a solution. Thus, the solutions for $ d_2\ne 0 $ are easily constructed from those for $ d_2=0 $. So, henceforth we only consider $ q(x) $ described by Eq.~(\ref{eq:akns1app}) without loss of generality.

\subsection{Solution of the AKNS$_{\text{1}}$  equation}
	By U(1)-gauge and translational symmetries, we obtain two integration constants for Eq. (\ref{eq:akns1app}):
	\begin{align}
		j=\frac{q^*q_x-qq^*_x}{2\mathrm{i}},\quad j_m=|q_x|^2+\mu|q|^2-|q|^4, \label{eq:akns1appcurrnts}
	\end{align}
	which are Eqs. (\ref{eq:currentj1}) and (\ref{eq:currentj2}) with $ (d_1,d_2,d_3)=(-\mu,0,1) $, and represent the currents of the number and momentum densities. Writing $ q=\sqrt{\rho}\mathrm{e}^{\mathrm{i}S} $, 
	\begin{align}
		j=\rho S_x,\quad \frac{\rho_x^2}{4}=-j^2+j_m\rho-\mu\rho^2+\rho^3.
	\end{align}
	Thus the phase is given by $ S=j\int\frac{\mathrm{d}x}{\rho} $. If the second expression is factorized as
	\begin{align}
		&\frac{\rho_x^2}{4}=(\rho-\rho_1)(\rho-\rho_2)(\rho-\rho_3),  \label{eq:akns1riemannsrfcresolvent} \\
		&\mu=\rho_1+\rho_2+\rho_3,\ j_m=\rho_1\rho_2+\rho_2\rho_3+\rho_3\rho_1,\ j^2=\rho_1\rho_2\rho_3, \label{eq:akns1riemannsrfcresolvent02}
	\end{align}
	then the solution is
	\begin{align}
		\frac{\rho(x)-\rho_1}{\rho_2-\rho_1}=\operatorname{sn}^2\left(\sqrt{\rho_3-\rho_1}(x-x_0) \bigg|\frac{\rho_2-\rho_1}{\rho_3-\rho_1} \right).
	\end{align}
	If we choose $ \rho_i $'s such that $ 0\le\rho_1\le\rho_2\le\rho_3 $ and $ x_0 $ is real,  $ \rho(x) $ is bounded and periodic, and takes the minimum (maximum) value $ \rho_1 \ (\rho_2) $. Henceforth we set $ x_0=0 $. 
	Let us write $ \alpha=\sqrt{\rho_3-\rho_1},\ m=\frac{\rho_2-\rho_1}{\rho_3-\rho_1} $, which satisfy $ \alpha\ge0,\ 0\le m\le1 $. Furthermore, let $ z_0 $ be a real number satisfying $ -\frac{K'}{2}<z_0<\frac{K'}{2} $, and we introduce the parametrization:
	\begin{align}
		&\rho_1=-m\alpha^2\operatorname{sn}^2(2\mathrm{i}z_0|m),\quad 
		\rho_2=m\alpha^2\operatorname{cn}^2(2\mathrm{i}z_0|m),\quad \nonumber \\
		&\rho_3=\alpha^2\operatorname{dn}^2(2\mathrm{i}z_0|m). \label{eq:appz0paramet}
	\end{align}
	Since the mass current is given by  $ j^2=\rho_1\rho_2\rho_3 $, we obtain
	\begin{align}
		j=-\mathrm{i}m\alpha^3\operatorname{sn}(2\mathrm{i}z_0)\operatorname{cn}(2\mathrm{i}z_0)\operatorname{dn}(2\mathrm{i}z_0).
	\end{align}
	The relation $ \operatorname{sgn}z_0=\operatorname{sgn}j $ holds by this choice of sign. $ \rho(x) $ is rewritten as
	\begin{align}
		\rho(x) &= m\alpha^2[\operatorname{sn}^2(\alpha x|m)-\operatorname{sn}^2(2\mathrm{i}z_0|m)] \nonumber \\
		&=\alpha^2[\operatorname{dn}^2(2\mathrm{i}z_0|m)-\operatorname{dn}^2(\alpha x|m)]. \label{eq:akns1den}
	\end{align}
	\indent The phase is integrated by the formula (\ref{eq:zetaintegral}):
	\begin{align}
		\mathrm{i}S&=\mathrm{i}\int^x\frac{j\mathrm{d}x}{\rho}=\alpha\int^x\mathrm{d}x\frac{\operatorname{sn}(2\mathrm{i}z_0)\operatorname{cn}(2\mathrm{i}z_0)\operatorname{dn}(2\mathrm{i}z_0)}{\operatorname{sn}^2(\alpha x)-\operatorname{sn}^2(2\mathrm{i}z_0)} \nonumber \\
		&=\frac{1}{2}\log\frac{\vartheta_1(\frac{\alpha x-2\mathrm{i}z_0}{2K})}{\vartheta_1(\frac{\alpha x+2\mathrm{i}z_0}{2K})}+\alpha x Z(2\mathrm{i}z_0)+2\mathrm{i}\varphi_0,
	\end{align}
	where $ Z(2\mathrm{i}z_0) $ is the Jacobi zeta function (see Appendix~\ref{app:ellipconv}) and $ 2\varphi_0 $ is a real constant. This integration can be also performed by the elliptic integral of the third kind (see Eq.~(\ref{eq:mainfflophase})).
	Thus,
	\begin{align}
		\mathrm{e}^{\pm\mathrm{i}S}=\mathrm{e}^{\pm(2\mathrm{i}\varphi_0+\alpha x Z(2\mathrm{i}z_0))} \sqrt{\frac{\vartheta_1(\frac{\alpha x\mp2\mathrm{i}z_0}{2K})}{\vartheta_1(\frac{\alpha x\pm2\mathrm{i}z_0}{2K})}}. \label{eq:akns1phase}
	\end{align}
	Note that $ \vartheta_j(z,q)^*=\vartheta_j(z^*,q) \ (j=1,2,3,4) $ holds if the nome $ q=\mathrm{e}^{-\pi K'/K} $ is real. \\
	\indent Rewriting the density (\ref{eq:akns1den}) in terms of theta functions, and
	using the addition formula $ \vartheta_1(v+w)\vartheta_1(v-w)\vartheta_4^2=\vartheta_1(v)^2\vartheta_4(w)^2-\vartheta_4(v)^2\vartheta_1(w)^2  \ \leftrightarrow $ 
	\begin{align}
		\frac{\vartheta_1(v+w)\vartheta_1(v-w)\vartheta_4^2}{\vartheta_4(v)^2\vartheta_4(w)^2}=\frac{\vartheta_1(v)^2}{\vartheta_4(v)^2}-\frac{\vartheta_1(w)^2}{\vartheta_4(w)^2}, \label{eq:thetaaddition001}
	\end{align}
	we obtain 
	\begin{align}
		\sqrt{\rho}=\frac{\alpha\vartheta_2\vartheta_4}{\vartheta_3\vartheta_4(\frac{\alpha x}{2K})\vartheta_4(\frac{2\mathrm{i}z_0}{2K})}\sqrt{\vartheta_1(\tfrac{\alpha x+2\mathrm{i}z_0}{2K})\vartheta_1(\tfrac{\alpha x-2\mathrm{i}z_0}{2K})}. \label{eq:akns1dens}
	\end{align}
	From Eqs. (\ref{eq:akns1phase}) and (\ref{eq:akns1dens}),
	\begin{align}
		p&=-\mathrm{i}\alpha Z(2\mathrm{i}z_0),\label{eq:sGPsol00}\\
		q&=\sqrt{\rho}\mathrm{e}^{\mathrm{i}S}=\mathrm{e}^{\mathrm{i}(2\varphi_0+px)}\alpha\frac{\vartheta_2\vartheta_4\vartheta_1(\frac{\alpha x-2\mathrm{i}z_0}{2K})}{\vartheta_3\vartheta_4(\frac{2\mathrm{i}z_0}{2K})\vartheta_4(\frac{\alpha x}{2K})}, \label{eq:sGPsol01} \\
		q^*&=\sqrt{\rho}\mathrm{e}^{-\mathrm{i}S}=\mathrm{e}^{-\mathrm{i}(2\varphi_0+px)}\alpha\frac{\vartheta_2\vartheta_4\vartheta_1(\frac{\alpha x+2\mathrm{i}z_0}{2K})}{\vartheta_3\vartheta_4(\frac{2\mathrm{i}z_0}{2K})\vartheta_4(\frac{\alpha x}{2K})}. \label{eq:sGPsol02}
	\end{align}
	It provides the general solution of Eq. (\ref{eq:akns1app}). The case $ c_2\ne 0 $ [Eq. (\ref{eq:akns1app0})] can be included by the modification 
	\begin{align}
		p&=-\mathrm{i}\alpha Z(2\mathrm{i}z_0)+\tilde{p}. \label{eq:sGPsol00p}
	\end{align}
	with $ \tilde{p}=-\frac{d_2}{2d_3} $.  Recalling the relation $ \psi=\mathrm{i}q $ and setting $ \varphi_0=0 $, we obtain Eq.~(\ref{eq:akns1psi0n1}).
\subsection{Eigenstates of the BdG or ZS operator}
\subsubsection{Parametrization of $ \lambda $ by uniformization variable $ z $}
	Generally,  the stationary $ \text{AKNS}_g $ equation can be solved by the  $ g $-variable Riemann theta functions, and it has an associated genus-$ g $ Riemann surface \cite{BelokolosBobenkoEnolskiiItsMatveev}. The Riemann surface $ (\omega,\lambda)\in \mathbb{C}^2 $ is given by $ \omega^2=\det V $, where $ V $  is the matrix appearing in the time-derivative equation in the AKNS formalism. The spectrum of the ZS operator, or the BdG operator in condensed-matter context, can be determined by the condition $ \omega^2>0 $ \cite{Takahashi:2012aw}. Although a given Riemann theta solution with genus $ g $ can also become a solution for higher-order AKNS$_{g'}$ equation s.t. $ g'>g $ (see Sec.~\ref{sec:timeevo2}), the corresponding Riemann surface should be constructed  using the AKNS form for the smallest $ g $, as noted in Ref.~\cite{Takahashi:2012aw}. \\ 
	\indent The matrix $ U,V $ giving the AKNS$_1$ equation with $ d_1=-\mu,\ d_2=0, d_3=1 $ is (now consider $ r=q^* $) 
	\begin{align}
		U&=\begin{pmatrix} -\mathrm{i}\lambda & q \\ r & \mathrm{i}\lambda \end{pmatrix}, \\
		V&=-\mu V^{(1)}+V^{(3)}=\begin{pmatrix} -2\mathrm{i}\lambda^2+\frac{\mathrm{i}\mu}{2}-\mathrm{i}qr & 2 \lambda q+\mathrm{i}q_x \\ 2\lambda r-\mathrm{i}r_x& 2\mathrm{i}\lambda^2-\frac{\mathrm{i}\mu}{2}+\mathrm{i}qr \end{pmatrix}. \label{eq:appaknsV01}
	\end{align}
	The associated Riemann surface is
	\begin{align}
		\omega^2=\det V = 4\lambda^4-2\mu\lambda^2+4j\lambda+\frac{\mu^2}{4}-j_m, \label{eq:detvAkns1}
	\end{align}
	where $ j $ and $ j_m $ are defined in Eq. (\ref{eq:akns1appcurrnts}). 
	Using Eqs.~(\ref{eq:akns1riemannsrfcresolvent02}) and (\ref{eq:appz0paramet}), the RHS of Eq.~(\ref{eq:detvAkns1}) is factorized as Eqs. (\ref{eq:akns1riemannsurface01})-(\ref{eq:akns1riemannsurface04}).
	We note that the quartic polynomial in the RHS of Eq. (\ref{eq:akns1riemannsurface01}) has the resolvent cubic polynomial given by the RHS in Eq. (\ref{eq:akns1riemannsrfcresolvent}). \\ 
	\indent A uniformization variable is introduced as follows. Let $ \lambda(z) $ be a solution of the differential equation
	\begin{align}
		\alpha^2\lambda'(z)^2=4\prod_{i=1,2,3,4}(\lambda(z)-\lambda_i). \label{eq:unifintroduce}
	\end{align}
	Then, we can parametrize the Riemann surface (\ref{eq:detvAkns1}) or (\ref{eq:akns1riemannsurface01}) by $ (\omega,\lambda)=(\alpha\lambda'(z),\lambda(z)) $. Equation~(\ref{eq:applambdazn1}) provides the solution of Eq.~(\ref{eq:unifintroduce}). The symmetries of $ \lambda(z) $ and $ \omega(z) $ in $ z $ plane are summarized in Sec~\ref{sec:bdgwithakns1bg}. 

\subsubsection{Eigenstates of the BdG or ZS operator for AKNS$_1$ potentials}
	Now let us provide the expression of BdG eigenstates in the presence of general AKNS$_1$ potentials. Though the formal symbolic expression of eigenstates using the AKNS matrices $ U $ and $ V $ is given in Ref.~\cite{Takahashi:2012aw}, rewriting it by theta functions is essential to formulate the IST. \\
	\indent By the addition formula, the square of $ \lambda $ [Eq.~(\ref{eq:applambdazn1})] is
	\begin{align}
		\lambda(z)^2&=\frac{\alpha^2}{4}\left[ \operatorname{dn}^2(\mathrm{i}(z+z_0))+\operatorname{dn}^2(\mathrm{i}(z'+z_0)) \right. \nonumber \\
		&\left. \qquad\qquad+\operatorname{dn}^2(2\mathrm{i}z_0)+m-2 \right]. \label{eq:lambdasqaure01}
	\end{align}
	Using this and Eqs.~(\ref{eq:unifoomegan1}) and (\ref{eq:appz0paramet}) and $ \mu=\rho_1+\rho_2+\rho_3=\alpha^2[m-2+3\operatorname{dn}^2(2\mathrm{i}z_0)] $,
	\begin{align}
		2\lambda^2-\frac{\mu}{2}+\rho_3\pm\omega=\begin{cases} \alpha^2\operatorname{dn}^2(\mathrm{i}(z'+z_0)) \\ \alpha^2\operatorname{dn}^2(\mathrm{i}(z+z_0)). \end{cases}\label{eq:lambdasqaure02}
	\end{align}
	Thus, 
	\begin{align}
		\mathrm{i}V_{11}\pm\omega&=2\lambda^2-\frac{\mu}{2}+\rho(x)\pm\omega \nonumber \\
		&=\begin{cases} m\alpha^2\left[ \operatorname{sn}^2(\alpha x)-\operatorname{sn}^2(\mathrm{i}(z'+z_0))\right]  \\ m\alpha^2\left[ \operatorname{sn}^2(\alpha x)-\operatorname{sn}^2(\mathrm{i}(z+z_0))\right],  \end{cases} \label{eq:appiv11pmomega01}
	\end{align}
	where $ V_{11} $ denotes the top-left component of Eq.~(\ref{eq:appaknsV01}). 
	Using the addition formula (\ref{eq:thetaaddition001}), it is rewritten as
	\begin{align}
		\mathrm{i}V_{11}+\omega&=\alpha^2\frac{\vartheta_2^2\vartheta_4^2\vartheta_1(\frac{\alpha x+\mathrm{i}(z'+z_0)}{2K})\vartheta_1(\frac{\alpha x-\mathrm{i}(z'+z_0)}{2K})}{\vartheta_3^2\vartheta_4(\frac{\alpha x}{2K})^2\vartheta_4(\frac{\mathrm{i}(z'+z_0)}{2K})^2}, \label{eq:appiv11pmomega02} \\ 
		\mathrm{i}V_{11}-\omega&=\alpha^2\frac{\vartheta_2^2\vartheta_4^2\vartheta_1(\frac{\alpha x+\mathrm{i}(z+z_0)}{2K})\vartheta_1(\frac{\alpha x-\mathrm{i}(z+z_0)}{2K})}{\vartheta_3^2\vartheta_4(\frac{\alpha x}{2K})^2\vartheta_4(\frac{\mathrm{i}(z+z_0)}{2K})^2}. \label{eq:appiv11pmomega03}
	\end{align}
	We can determine the expressions of $ V_{12} $ and $ V_{21} $ using theta functions from the following facts: (i) $ \omega^2=\det V \ \leftrightarrow \ V_{12}V_{21}=(\mathrm{i}V_{11}+\omega)(\mathrm{i}V_{11}-\omega) $, (ii) $ V_{12}=V_{21}^* $ for real $ \lambda $,  and (iii) $ V_{12}=2\lambda q+\mathrm{i}q_x $ is invariant under the exchange $ z \leftrightarrow z' $ and have the same twisted periodicity with $ q(x) $ [Eq. (\ref{eq:appqtwistn1})]. The resultant is
	\begin{align}
		&V_{12} = \nonumber \\
		&-\mathrm{i}\mathrm{e}^{2\mathrm{i}\varphi_0}m\alpha^2\frac{\vartheta_3^2\vartheta_4^2\vartheta_1(\frac{\alpha x-\mathrm{i}(z+z_0)}{2K})\vartheta_1(\frac{\alpha x-\mathrm{i}(z'+z_0)}{2K})}{\vartheta_2^2\vartheta_4(\frac{\alpha x}{2K})^2\vartheta_4(\frac{\mathrm{i}(z+z_0)}{2K})\vartheta_4(\frac{\mathrm{i}(z'+z_0)}{2K})}\mathrm{e}^{\mathrm{i}px-\frac{\mathrm{i}\pi\alpha x}{2K}},\label{eq:appiv11pmomega04} \\
		&V_{21} =\nonumber \\
		&\mathrm{i}\mathrm{e}^{-2\mathrm{i}\varphi_0}m\alpha^2\frac{\vartheta_3^2\vartheta_4^2\vartheta_1(\frac{\alpha x+\mathrm{i}(z+z_0)}{2K})\vartheta_1(\frac{\alpha x+\mathrm{i}(z'+z_0)}{2K})}{\vartheta_2^2\vartheta_4(\frac{\alpha x}{2K})^2\vartheta_4(\frac{\mathrm{i}(z+z_0)}{2K})\vartheta_4(\frac{\mathrm{i}(z'+z_0)}{2K})}\mathrm{e}^{-\mathrm{i}px+\frac{\mathrm{i}\pi\alpha x}{2K}}. \label{eq:appiv11pmomega05}
	\end{align}
	By partial fraction decomposition,
	\begin{align}
		&\omega\left( \frac{U_{12}}{V_{12}}+\frac{U_{21}}{V_{21}} \right) \nonumber \\
		&=\frac{-j+\mu\lambda-4\lambda^3+2\lambda\omega}{\mathrm{i}V_{11}-\omega}-\frac{-j+\mu\lambda-4\lambda^3-2\lambda\omega}{\mathrm{i}V_{11}+\omega}.  \label{eq:apppardecomps}
	\end{align}
	The numerators of the above are in fact expressed as
	\begin{align}
		&-j+\mu\lambda-4\lambda^3\pm 2\lambda\omega= \nonumber \\
		&\begin{cases} -\mathrm{i}m\alpha^3\operatorname{sn}(\mathrm{i}(z+z_0))\operatorname{cn}(\mathrm{i}(z+z_0))\operatorname{dn}(\mathrm{i}(z+z_0)), \\ -\mathrm{i}m\alpha^3\operatorname{sn}(\mathrm{i}(z'+z_0))\operatorname{cn}(\mathrm{i}(z'+z_0))\operatorname{dn}(\mathrm{i}(z'+z_0)), \end{cases} \label{eq:appnumersimplify}
	\end{align}
	because $ \omega=\alpha\lambda' $ and Eq. (\ref{eq:detvAkns1}) implies
	\begin{align}
		&-j+\mu\lambda-4\lambda^3\pm 2\lambda\omega=\frac{\alpha}{2}( \pm2\lambda^2-\omega)',
	\end{align}
	which can be calculated by using Eqs.~(\ref{eq:lambdasqaure01}) and (\ref{eq:unifoomegan1}). 
	From Eqs.~(\ref{eq:appiv11pmomega01}), (\ref{eq:apppardecomps}), (\ref{eq:appnumersimplify}), and the formula (\ref{eq:zetaintegral}),
	\begin{align}
		\mathrm{i}\omega\!\int^x\!\mathrm{d}x\left( \frac{U_{12}}{V_{12}}+\frac{U_{21}}{V_{21}} \right)&=\frac{1}{2}\log\frac{\vartheta_1(\frac{\alpha x-\mathrm{i}(z+z_0)}{2K})\vartheta_1(\frac{\alpha x+\mathrm{i}(z'+z_0)}{2K})}{\vartheta_1(\frac{\alpha x+\mathrm{i}(z+z_0)}{2K})\vartheta_1(\frac{\alpha x-\mathrm{i}(z'+z_0)}{2K})}\nonumber \\
		&\quad\,+\alpha x\left[ Z(\mathrm{i}(z+z_0))-Z(\mathrm{i}(z'+z_0)) \right]. \label{eq:appiv11pmomega06}
	\end{align}
	Using the formula of Ref.~\cite{Takahashi:2012aw} and Eqs. (\ref{eq:appiv11pmomega02})-(\ref{eq:appiv11pmomega05}), and (\ref{eq:appiv11pmomega06}), the square of fermionic eigenstates is given by
	\begin{align}
		&u^2=V_{12}\sqrt{\frac{\mathrm{i}V_{11}-\omega}{\mathrm{i}V_{11}+\omega}}\exp\left[ \mathrm{i}\omega\int^x\mathrm{d}x\left( \frac{U_{12}}{V_{12}}+\frac{U_{21}}{V_{21}} \right) \right] \nonumber \\
		&=-\mathrm{i}\mathrm{e}^{2\mathrm{i}\varphi_0}\alpha^2\frac{\vartheta_2^2\vartheta_4^2\vartheta_1(\frac{\alpha x-\mathrm{i}(z+z_0)}{2K})^2}{\vartheta_3^2\vartheta_4(\frac{\alpha x}{2K})^2\vartheta_4(\frac{\mathrm{i}(z+z_0)}{2K})^2}\mathrm{e}^{\mathrm{i}(2k(z)+p)x-\frac{\mathrm{i}\pi\alpha}{2K}x}, \\
		&v^2=-V_{21}\sqrt{\frac{\mathrm{i}V_{11}+\omega}{\mathrm{i}V_{11}-\omega}}\exp\left[ \mathrm{i}\omega\int^x\mathrm{d}x\left( \frac{U_{12}}{V_{12}}+\frac{U_{21}}{V_{21}} \right) \right] \nonumber \\
		&=-\mathrm{i}\mathrm{e}^{-2\mathrm{i}\varphi_0}\alpha^2\frac{\vartheta_2^2\vartheta_4^2\vartheta_1(\frac{\alpha x+\mathrm{i}(z'+z_0)}{2K})^2}{\vartheta_3^2\vartheta_4(\frac{\alpha x}{2K})^2\vartheta_4(\frac{\mathrm{i}(z'+z_0)}{2K})^2}\mathrm{e}^{\mathrm{i}(2k(z)-p)x+\frac{\mathrm{i}\pi\alpha }{2K}x}, 
	\end{align}
	where we have defined the crystal momentum  $ k(z) $ by Eq.~(\ref{eq:appcrystammomenn1}). 
	Taking the square roots of the above expressions, and setting a factor to satisfy the BdG equation, we obtain
	\begin{align}
		&f_0(x,z):=\begin{pmatrix} u(x,z) \\ v(x,z) \end{pmatrix}=\mathrm{e}^{\mathrm{i}k(z)x+\mathrm{i}(\varphi_0+\frac{1}{2}px-\frac{\pi\alpha x}{4K})\sigma_3}\nonumber \\
		&\quad\qquad\times\frac{\mathrm{i}\alpha \vartheta_2\vartheta_4}{\vartheta_3\vartheta_4(\frac{\alpha x}{2K})}\begin{pmatrix} \vartheta_1(\frac{\alpha x-\mathrm{i}(z+z_0)}{2K})/\vartheta_4(\frac{\mathrm{i}(z+z_0)}{2K}) \\ - \vartheta_1(\frac{\alpha x+\mathrm{i}(z'+z_0)}{2K})/\vartheta_4(\frac{\mathrm{i}(z'+z_0)}{2K}) \end{pmatrix}. \label{eq:fermionicsol}
	\end{align}
	If we set $ \varphi_0=0 $ and rewrite $ (u,v)\to (u_0,v_0) $, it gives Eq.~(\ref{eq:fermionicsoln1}). \\
	\indent For a given $ \lambda=\lambda(z) $, the two linearly independent solutions of the BdG equation are $ f_0(x,z) $ and $ f_0(x,z') $ unless $ \lambda=\lambda_1,\lambda_2,\lambda_3,\lambda_4 $. For degenerate points $ \lambda=\lambda_1,\lambda_2,\lambda_3,\lambda_4 \ \leftrightarrow \ z=\pm\frac{K'}{2},\ \pm\frac{K'}{2}+\mathrm{i}K$, two linearly independent solutions are given by $ f_0(x,z) $ and $ \frac{\mathrm{d} f_0(x,z)}{\mathrm{d} z} $. \\
		\indent The solution for $ d_2\ne 0 $ can be obtained by modifying the spectral parameter parametrization  $ (\lambda,\omega)=(\lambda(z)-\frac{\tilde{p}}{2},\alpha \lambda'(z)) $ and using $ p $ of Eq.~(\ref{eq:sGPsol00p}) in Eq.~(\ref{eq:fermionicsol}). 
	 The periodicity and symmetry of $ f_0(x,z) $ are summarized in Sec.~\ref{sec:bdgwithakns1bg}.

\subsection{Completeness relation}
	\indent Here, we derive the completeness relation of the BdG eigenstates (\ref{eq:completeinmainsec}), which is necessary when we derive the GLM equation. To avoid mathematical difficulty of the infinite system, we first consider a finite-length system, and take the limit to infinity. \\
	\indent Let us consider the finite-length system in  $ [-\frac{L}{2},\frac{L}{2}],\ L>0 $ with periodic boundary condition.  From Eq.~(\ref{eq:appqtwistn1}), in order for the density and phase of $ \psi_0(x) $ to be continuous, 
	\begin{align}
		L=N_0L_0,\quad \theta=\frac{2\pi M_0}{N_0},\quad  N_0 \in \mathbb{N},\ M_0 \in\mathbb{Z}. 
	\end{align}
	The parameters $ m, \alpha, $ and $ z_0 $ must be chosen to satisfy these discretization conditions. From Eq.~(\ref{eq:appf0twistn1}), the discretization condition for fermionic eigenstates is given by 
	\begin{align}
		k(z)=\frac{\pi(2N-M_0)}{L},\quad N\in\mathbb{Z}. \label{eq:compreldisck}
	\end{align}
	This condition implies that, if the eigenstates are labeled by crystal momentum, they are equally spaced. Therefore, if we use this labeling, we need no weight function when we replace a summation by an integral in the infinite-length limit. \\ 
	\indent Let us refer to the three bands {\normalsize\textcircled{\small 1}}, {\normalsize\textcircled{\small 3}}, and {\normalsize\textcircled{\small 5}} in Fig.~\ref{fig:univar2} as C, M, and V, respectively. (The names originate from conduction, mid-gap, and valence bands, respectively). In each band, $ k(z) $ is monotonic (Fig.~\ref{fig:crystalkz}). In the C and V bands,  $ k(z) $ goes from $ -\infty $ to $ +\infty $ monotonically. In the M band,  $ k(z) $ monotonically decreases.
	Since $ k(z) $ in each band is monotonic, we can use it as a label of eigenstates. Let $ f_0(x,k,\mathrm{b}) $ be an eigenstate labeled by the crystal momentum $ k $ and the band index $ \mathrm{b}=\mathrm{C},\mathrm{M} $, and $ \mathrm{V} $. Then, following the conventional wisdom of self-adjoint operators, the completeness relation is given by
	\begin{align}
		\sum_{\mathrm{b}=\mathrm{C,M,V}}\sum_k\frac{f_0(x,k,\mathrm{b})f_0(y,k,\mathrm{b})^\dagger}{\mathcal{N}(k,\mathrm{b})}=\delta(x-y)I_2, \label{eq:appcompdisc}
	\end{align}
	where $ \mathcal{N}(k,\mathrm{b})=\int_{-L/2}^{L/2}\mathrm{d}x f_0^\dagger f_0 $ is a normalization. We want to take an infinite-length limit of this expression. First, let us rewrite $ \mathcal{N}(k,\mathrm{b}) $. 
	From Eqs.~(\ref{eq:appcrystammomenn1}) and (\ref{eq:jacobizetaderv}), we can prove
	\begin{align}
		\frac{\mathrm{d}k}{\mathrm{d}z}=\frac{\overline{\mathrm{i}V_{11}}}{\alpha}=\frac{\alpha}{2}\left( \operatorname{dn}^2(\mathrm{i}(z+z_0))+\operatorname{dn}^2(\mathrm{i}(z'+z_0)) \right)-\frac{\alpha E}{K}, \label{eq:appdkdz}
	\end{align}
	where $ \overline{\mathrm{i}V_{11}}:=\frac{2\alpha}{K}\int_{-K/\alpha}^{K/\alpha}\mathrm{d}x (\mathrm{i}V_{11}) $ is an averaged value of $ \mathrm{i}V_{11} $, which can be calculated using Eq.~(\ref{eq:appiv11pmomega01}).
	Furthermore, following the discussion of Sec. 2.5 of Ref.~\cite{Takahashi:2012aw} 
	and using Eq.~(\ref{eq:appiv11pmomega01}), we can check
	\begin{align}
		|u|^2+|v|^2=\begin{cases} 2\mathrm{i}V_{11} & (z\in\mathbb{R}), \\ -2\mathrm{i}V_{11} & (z\in \mathbb{R}+\mathrm{i}K). \end{cases}
	\end{align}
	Integrating this over $ [-L/2,L/2] $ and using (\ref{eq:appdkdz}), 
	\begin{align}
		\mathcal{N}(k,\mathrm{b})=\begin{cases} 2\alpha L\frac{\mathrm{d}k}{\mathrm{d}z} & (\mathrm{b}=\mathrm{C,V}), \\ - 2\alpha L\frac{\mathrm{d}k}{\mathrm{d}z} & (\mathrm{b}=\mathrm{M}). \end{cases} \label{eq:complnc}
	\end{align}
	Taking the limit $ L\to\infty $ and changing the integration variable from $ k $ to $ z $, the summation is replaced by the integral
	\begin{align}
		\sum_{\mathrm{b}=\mathrm{C,M,V}}\sum_k \ \to \  L\left( \int_{-K'-z_0}^{K'-z_0}-\int_{-K'-z_0+\mathrm{i}K}^{K'-z_0+\mathrm{i}K} \right)\frac{\mathrm{d}z}{2\pi}\frac{\mathrm{d}k}{\mathrm{d}z}, \label{eq:compsumtoint}
	\end{align}
	where the minus sign for the M band comes from the fact that $ k(z) $ is a decreasing function in this region. Using Eqs.~(\ref{eq:complnc}) and (\ref{eq:compsumtoint}), the infinite-length limit of Eq.~(\ref{eq:appcompdisc}) is 
	\begin{align}
		\left( \int_{-K'-z_0}^{K'-z_0}+\int_{-K'-z_0+\mathrm{i}K}^{K'-z_0+\mathrm{i}K} \right)\frac{\mathrm{d}z}{4\pi\alpha}f_0(x,z)f_0(y,z)^\dagger=\delta(x-y)I_2.
	\end{align}
	Using Eq.~(\ref{eq:appccrelf0}), and adding vertical contours $ \int_{K'-z_0}^{K'-z_0+\mathrm{i}K} $ and $ \int_{-K'-z_0+\mathrm{i}K}^{-K'-z_0} $, which cancel because of the periodicity of $ f_0(x,z) $, we obtain Eq.~(\ref{eq:completeinmainsec}).

\section{Theta ratio determinant}
	Let $ x,\xi_1,\dots,\xi_n,\ \eta_1,\dots,\eta_n $ be complex numbers. We want to prove:
	\begin{align}
		&\det\left( \frac{\vartheta_r(x+\xi_i+\eta_j)}{\vartheta_1(\xi_i+\eta_j)} \right)_{1\le i,j\le n}= \nonumber \\
		&\frac{\vartheta_r(x)^{n-1}\vartheta_r(x+\sum_{i=1}^n(\xi_i+\eta_i))\prod_{i<j}\vartheta_1(\xi_i-\xi_j)\vartheta_1(\eta_i-\eta_j)}{\prod_{i,j=1}^n\vartheta_1(\xi_i+\eta_j)}, \label{eq:appthetaratiodet}
	\end{align}
	where $ r $ can be any of $ 1,2,3, $ and $ 4 $. A corollary of Eq.~(\ref{eq:appthetaratiodet}) is
	\begin{align}
		\frac{\det\left( \frac{\vartheta_r(x+\xi_i+\xi_j)}{\vartheta_r(x)\vartheta_1(\xi_i+\xi_j)} \right)_{1\le i,j\le n}}{\det\left( \frac{\vartheta_{r'}(y+\xi_i+\xi_j)}{\vartheta_{r'}(y)\vartheta_1(\xi_i+\xi_j)} \right)_{1\le i,j\le n}}=\frac{\vartheta_{r'}(y)\vartheta_{r}(x+2\xi_1+\dots+2\xi_n)}{\vartheta_r(x)\vartheta_{r'}(y+2\xi_1+\dots+2\xi_n)}, \label{eq:appthetaratiodet2}
	\end{align}
	where $ y $ is complex and $ r' $ is any of $ 1,2,3, $ and $ 4 $. This is used in the asymptotics of soliton solutions (Subsec.~\ref{subsec:asymptotics}). 

	\textit{Proof of Eq.~(\ref{eq:appthetaratiodet}):} We prove by induction.  $ n=1 $ is trivial. $ n=2 $ is proved by using the Weierstrass formula \cite{Kharchev201519}. We now assume the theorem up to matrices of size $ n-1 $. For brevity, let $ D_n(x;\begin{smallmatrix}\xi_1,\dots,\xi_n \\ \eta_1,\dots,\eta_n \end{smallmatrix}) $ denote the left-hand side of Eq.~(\ref{eq:appthetaratiodet}). 
	The Desnanot-Jacobi formula tells us that
	\begin{align}
		D_n(x;\begin{smallmatrix}\xi_1,\dots,\xi_n \\ \eta_1,\dots,\eta_n \end{smallmatrix})=&\frac{D_{n-1}(x;\begin{smallmatrix}\xi_1,\dots,\xi_{n-1} \\ \eta_1,\dots,\eta_{n-1} \end{smallmatrix})D_{n-1}(x;\begin{smallmatrix}\xi_1,\dots,\xi_{n-2},\xi_n \\ \eta_1,\dots,\eta_{n-2},\eta_n \end{smallmatrix})}{D_{n-2}(x;\begin{smallmatrix}\xi_1,\dots,\xi_{n-2} \\ \eta_1,\dots,\eta_{n-2} \end{smallmatrix})} \nonumber \\
		&-\frac{D_{n-1}(x;\begin{smallmatrix}\xi_1,\dots,\xi_{n-1} \\ \eta_1,\dots,\eta_{n-2},\eta_n \end{smallmatrix})D_{n-1}(x;\begin{smallmatrix}\xi_1,\dots,\xi_{n-2},\xi_n \\ \eta_1,\dots,\eta_{n-1} \end{smallmatrix})}{D_{n-2}(x;\begin{smallmatrix}\xi_1,\dots,\xi_{n-2} \\ \eta_1,\dots,\eta_{n-2} \end{smallmatrix})}.
	\end{align}
	Calculating the right-hand side with the help of the Weierstrass formula, we complete the proof. 


\begin{thebibliography}{94}%
\makeatletter
\providecommand \@ifxundefined [1]{%
 \@ifx{#1\undefined}
}%
\providecommand \@ifnum [1]{%
 \ifnum #1\expandafter \@firstoftwo
 \else \expandafter \@secondoftwo
 \fi
}%
\providecommand \@ifx [1]{%
 \ifx #1\expandafter \@firstoftwo
 \else \expandafter \@secondoftwo
 \fi
}%
\providecommand \natexlab [1]{#1}%
\providecommand \enquote  [1]{``#1''}%
\providecommand \bibnamefont  [1]{#1}%
\providecommand \bibfnamefont [1]{#1}%
\providecommand \citenamefont [1]{#1}%
\providecommand \href@noop [0]{\@secondoftwo}%
\providecommand \href [0]{\begingroup \@sanitize@url \@href}%
\providecommand \@href[1]{\@@startlink{#1}\@@href}%
\providecommand \@@href[1]{\endgroup#1\@@endlink}%
\providecommand \@sanitize@url [0]{\catcode `\\12\catcode `\$12\catcode
  `\&12\catcode `\#12\catcode `\^12\catcode `\_12\catcode `\%12\relax}%
\providecommand \@@startlink[1]{}%
\providecommand \@@endlink[0]{}%
\providecommand \url  [0]{\begingroup\@sanitize@url \@url }%
\providecommand \@url [1]{\endgroup\@href {#1}{\urlprefix }}%
\providecommand \urlprefix  [0]{URL }%
\providecommand \Eprint [0]{\href }%
\providecommand \doibase [0]{http://dx.doi.org/}%
\providecommand \selectlanguage [0]{\@gobble}%
\providecommand \bibinfo  [0]{\@secondoftwo}%
\providecommand \bibfield  [0]{\@secondoftwo}%
\providecommand \translation [1]{[#1]}%
\providecommand \BibitemOpen [0]{}%
\providecommand \bibitemStop [0]{}%
\providecommand \bibitemNoStop [0]{.\EOS\space}%
\providecommand \EOS [0]{\spacefactor3000\relax}%
\providecommand \BibitemShut  [1]{\csname bibitem#1\endcsname}%
\let\auto@bib@innerbib\@empty
\bibitem [{\citenamefont {Andreev}\ and\ \citenamefont
  {Lifshitz}(1969)}]{AndreevLifshitz1969}%
  \BibitemOpen
  \bibfield  {author} {\bibinfo {author} {\bibfnamefont {A.~F.}\ \bibnamefont
  {Andreev}}\ and\ \bibinfo {author} {\bibfnamefont {I.~M.}\ \bibnamefont
  {Lifshitz}},\ }\href@noop {} {\bibfield  {journal} {\bibinfo  {journal} {Sov.
  Phys. JETP}\ }\textbf {\bibinfo {volume} {29}},\ \bibinfo {pages} {1107}
  (\bibinfo {year} {1969})}\BibitemShut {NoStop}%
\bibitem [{\citenamefont {Chester}(1970)}]{Chester1970}%
  \BibitemOpen
  \bibfield  {author} {\bibinfo {author} {\bibfnamefont {G.~V.}\ \bibnamefont
  {Chester}},\ }\href@noop {} {\bibfield  {journal} {\bibinfo  {journal} {Phys.
  Rev. A}\ }\textbf {\bibinfo {volume} {2}},\ \bibinfo {pages} {256} (\bibinfo
  {year} {1970})}\BibitemShut {NoStop}%
\bibitem [{\citenamefont {Leggett}(1970)}]{Leggett1970}%
  \BibitemOpen
  \bibfield  {author} {\bibinfo {author} {\bibfnamefont {A.~J.}\ \bibnamefont
  {Leggett}},\ }\href@noop {} {\bibfield  {journal} {\bibinfo  {journal} {Phys.
  Rev. Lett.}\ }\textbf {\bibinfo {volume} {25}},\ \bibinfo {pages} {1543}
  (\bibinfo {year} {1970})}\BibitemShut {NoStop}%
\bibitem [{\citenamefont {Kim}\ and\ \citenamefont
  {Chan}(2004{\natexlab{a}})}]{EKimMHWChanN}%
  \BibitemOpen
  \bibfield  {author} {\bibinfo {author} {\bibfnamefont {E.}~\bibnamefont
  {Kim}}\ and\ \bibinfo {author} {\bibfnamefont {M.~H.~W.}\ \bibnamefont
  {Chan}},\ }\href@noop {} {\bibfield  {journal} {\bibinfo  {journal}
  {Nature(London)}\ }\textbf {\bibinfo {volume} {427}},\ \bibinfo {pages} {225}
  (\bibinfo {year} {2004}{\natexlab{a}})}\BibitemShut {NoStop}%
\bibitem [{\citenamefont {Kim}\ and\ \citenamefont
  {Chan}(2004{\natexlab{b}})}]{EKimMHWChanS}%
  \BibitemOpen
  \bibfield  {author} {\bibinfo {author} {\bibfnamefont {E.}~\bibnamefont
  {Kim}}\ and\ \bibinfo {author} {\bibfnamefont {M.~H.~W.}\ \bibnamefont
  {Chan}},\ }\href@noop {} {\bibfield  {journal} {\bibinfo  {journal}
  {Science}\ }\textbf {\bibinfo {volume} {305}},\ \bibinfo {pages} {1941}
  (\bibinfo {year} {2004}{\natexlab{b}})}\BibitemShut {NoStop}%
\bibitem [{\citenamefont {Kim}\ and\ \citenamefont
  {Chan}(2012)}]{DYKimMHWChan}%
  \BibitemOpen
  \bibfield  {author} {\bibinfo {author} {\bibfnamefont {D.~Y.}\ \bibnamefont
  {Kim}}\ and\ \bibinfo {author} {\bibfnamefont {M.~H.~W.}\ \bibnamefont
  {Chan}},\ }\href@noop {} {\bibfield  {journal} {\bibinfo  {journal} {Phys.
  Rev. Lett.}\ }\textbf {\bibinfo {volume} {109}},\ \bibinfo {pages} {155301}
  (\bibinfo {year} {2012})}\BibitemShut {NoStop}%
\bibitem [{\citenamefont {Henkel}\ \emph {et~al.}(2010)\citenamefont {Henkel},
  \citenamefont {Nath},\ and\ \citenamefont {Pohl}}]{HenkelNathPohl}%
  \BibitemOpen
  \bibfield  {author} {\bibinfo {author} {\bibfnamefont {N.}~\bibnamefont
  {Henkel}}, \bibinfo {author} {\bibfnamefont {R.}~\bibnamefont {Nath}}, \ and\
  \bibinfo {author} {\bibfnamefont {T.}~\bibnamefont {Pohl}},\ }\href@noop {}
  {\bibfield  {journal} {\bibinfo  {journal} {Phys. Rev. Lett.}\ }\textbf
  {\bibinfo {volume} {104}},\ \bibinfo {pages} {195302} (\bibinfo {year}
  {2010})}\BibitemShut {NoStop}%
\bibitem [{\citenamefont {Cinti}\ \emph {et~al.}(2010)\citenamefont {Cinti},
  \citenamefont {Jain}, \citenamefont {Boninsegni}, \citenamefont {Micheli},
  \citenamefont {Zoller},\ and\ \citenamefont
  {Pupillo}}]{CintiJainBoninsegniMicheliZollerPupillo}%
  \BibitemOpen
  \bibfield  {author} {\bibinfo {author} {\bibfnamefont {F.}~\bibnamefont
  {Cinti}}, \bibinfo {author} {\bibfnamefont {P.}~\bibnamefont {Jain}},
  \bibinfo {author} {\bibfnamefont {M.}~\bibnamefont {Boninsegni}}, \bibinfo
  {author} {\bibfnamefont {A.}~\bibnamefont {Micheli}}, \bibinfo {author}
  {\bibfnamefont {P.}~\bibnamefont {Zoller}}, \ and\ \bibinfo {author}
  {\bibfnamefont {G.}~\bibnamefont {Pupillo}},\ }\href@noop {} {\bibfield
  {journal} {\bibinfo  {journal} {Phys. Rev. Lett.}\ }\textbf {\bibinfo
  {volume} {105}},\ \bibinfo {pages} {135301} (\bibinfo {year}
  {2010})}\BibitemShut {NoStop}%
\bibitem [{\citenamefont {Fulde}\ and\ \citenamefont
  {Ferrell}(1964)}]{FuldeFerrell}%
  \BibitemOpen
  \bibfield  {author} {\bibinfo {author} {\bibfnamefont {P.}~\bibnamefont
  {Fulde}}\ and\ \bibinfo {author} {\bibfnamefont {R.~A.}\ \bibnamefont
  {Ferrell}},\ }\href@noop {} {\bibfield  {journal} {\bibinfo  {journal} {Phys.
  Rev.}\ }\textbf {\bibinfo {volume} {135}},\ \bibinfo {pages} {A550} (\bibinfo
  {year} {1964})}\BibitemShut {NoStop}%
\bibitem [{\citenamefont {Larkin}\ and\ \citenamefont
  {Ovchinnikov}(1965)}]{LarkinOvchinnikov}%
  \BibitemOpen
  \bibfield  {author} {\bibinfo {author} {\bibfnamefont {A.~I.}\ \bibnamefont
  {Larkin}}\ and\ \bibinfo {author} {\bibfnamefont {Y.~N.}\ \bibnamefont
  {Ovchinnikov}},\ }\href@noop {} {\bibfield  {journal} {\bibinfo  {journal}
  {Sov. Phys. JETP}\ }\textbf {\bibinfo {volume} {20}},\ \bibinfo {pages} {762}
  (\bibinfo {year} {1965})}\BibitemShut {NoStop}%
\bibitem [{\citenamefont {Machida}\ and\ \citenamefont
  {Nakanishi}(1984)}]{Machida:1984zz}%
  \BibitemOpen
  \bibfield  {author} {\bibinfo {author} {\bibfnamefont {K.}~\bibnamefont
  {Machida}}\ and\ \bibinfo {author} {\bibfnamefont {H.}~\bibnamefont
  {Nakanishi}},\ }\href@noop {} {\bibfield  {journal} {\bibinfo  {journal}
  {Phys. Rev. B}\ }\textbf {\bibinfo {volume} {30}},\ \bibinfo {pages} {122}
  (\bibinfo {year} {1984})}\BibitemShut {NoStop}%
\bibitem [{\citenamefont {Mizushima}\ \emph {et~al.}(2005)\citenamefont
  {Mizushima}, \citenamefont {Machida},\ and\ \citenamefont
  {Ichioka}}]{MizushimaMachidaIchioka}%
  \BibitemOpen
  \bibfield  {author} {\bibinfo {author} {\bibfnamefont {T.}~\bibnamefont
  {Mizushima}}, \bibinfo {author} {\bibfnamefont {K.}~\bibnamefont {Machida}},
  \ and\ \bibinfo {author} {\bibfnamefont {M.}~\bibnamefont {Ichioka}},\
  }\href@noop {} {\bibfield  {journal} {\bibinfo  {journal} {Phys.\ Rev.\
  Lett.}\ }\textbf {\bibinfo {volume} {94}},\ \bibinfo {pages} {060404}
  (\bibinfo {year} {2005})}\BibitemShut {NoStop}%
\bibitem [{\citenamefont {Radovan}\ \emph {et~al.}(2003)\citenamefont
  {Radovan}, \citenamefont {Fortune}, \citenamefont {Murphy}, \citenamefont
  {Hannahs}, \citenamefont {Palm}, \citenamefont {Tozer},\ and\ \citenamefont
  {Hall}}]{Radovan2003}%
  \BibitemOpen
  \bibfield  {author} {\bibinfo {author} {\bibfnamefont {H.~A.}\ \bibnamefont
  {Radovan}}, \bibinfo {author} {\bibfnamefont {N.~A.}\ \bibnamefont
  {Fortune}}, \bibinfo {author} {\bibfnamefont {T.~P.}\ \bibnamefont {Murphy}},
  \bibinfo {author} {\bibfnamefont {S.~T.}\ \bibnamefont {Hannahs}}, \bibinfo
  {author} {\bibfnamefont {E.~C.}\ \bibnamefont {Palm}}, \bibinfo {author}
  {\bibfnamefont {S.~W.}\ \bibnamefont {Tozer}}, \ and\ \bibinfo {author}
  {\bibfnamefont {D.}~\bibnamefont {Hall}},\ }\href@noop {} {\bibfield
  {journal} {\bibinfo  {journal} {Nature (London)}\ }\textbf {\bibinfo {volume}
  {425}},\ \bibinfo {pages} {51} (\bibinfo {year} {2003})}\BibitemShut
  {NoStop}%
\bibitem [{\citenamefont {Kakuyanagi}\ \emph {et~al.}(2005)\citenamefont
  {Kakuyanagi}, \citenamefont {Saitoh}, \citenamefont {Kumagai}, \citenamefont
  {Takashima}, \citenamefont {Nohara}, \citenamefont {Takagi},\ and\
  \citenamefont {Matsuda}}]{KakuyanagiPRL2005}%
  \BibitemOpen
  \bibfield  {author} {\bibinfo {author} {\bibfnamefont {K.}~\bibnamefont
  {Kakuyanagi}}, \bibinfo {author} {\bibfnamefont {M.}~\bibnamefont {Saitoh}},
  \bibinfo {author} {\bibfnamefont {K.}~\bibnamefont {Kumagai}}, \bibinfo
  {author} {\bibfnamefont {S.}~\bibnamefont {Takashima}}, \bibinfo {author}
  {\bibfnamefont {M.}~\bibnamefont {Nohara}}, \bibinfo {author} {\bibfnamefont
  {H.}~\bibnamefont {Takagi}}, \ and\ \bibinfo {author} {\bibfnamefont
  {Y.}~\bibnamefont {Matsuda}},\ }\href@noop {} {\bibfield  {journal} {\bibinfo
   {journal} {Phys. Rev. Lett.}\ }\textbf {\bibinfo {volume} {94}},\ \bibinfo
  {pages} {047602} (\bibinfo {year} {2005})}\BibitemShut {NoStop}%
\bibitem [{\citenamefont {Zwierlein}\ \emph {et~al.}(2006)\citenamefont
  {Zwierlein}, \citenamefont {Schirotzek}, \citenamefont {Schunck},\ and\
  \citenamefont {Ketterle}}]{ZwierleinScience2006}%
  \BibitemOpen
  \bibfield  {author} {\bibinfo {author} {\bibfnamefont {M.~W.}\ \bibnamefont
  {Zwierlein}}, \bibinfo {author} {\bibfnamefont {A.}~\bibnamefont
  {Schirotzek}}, \bibinfo {author} {\bibfnamefont {C.~H.}\ \bibnamefont
  {Schunck}}, \ and\ \bibinfo {author} {\bibfnamefont {W.}~\bibnamefont
  {Ketterle}},\ }\href@noop {} {\bibfield  {journal} {\bibinfo  {journal}
  {Science}\ }\textbf {\bibinfo {volume} {311}},\ \bibinfo {pages} {492}
  (\bibinfo {year} {2006})}\BibitemShut {NoStop}%
\bibitem [{\citenamefont {Partridge}\ \emph {et~al.}(2006)\citenamefont
  {Partridge}, \citenamefont {Li}, \citenamefont {Kamar}, \citenamefont
  {Liao},\ and\ \citenamefont {Hulet}}]{PartridgeScience2006}%
  \BibitemOpen
  \bibfield  {author} {\bibinfo {author} {\bibfnamefont {G.~B.}\ \bibnamefont
  {Partridge}}, \bibinfo {author} {\bibfnamefont {W.}~\bibnamefont {Li}},
  \bibinfo {author} {\bibfnamefont {R.~I.}\ \bibnamefont {Kamar}}, \bibinfo
  {author} {\bibfnamefont {Y.}~\bibnamefont {Liao}}, \ and\ \bibinfo {author}
  {\bibfnamefont {R.~G.}\ \bibnamefont {Hulet}},\ }\href@noop {} {\bibfield
  {journal} {\bibinfo  {journal} {Science}\ }\textbf {\bibinfo {volume}
  {311}},\ \bibinfo {pages} {503} (\bibinfo {year} {2006})}\BibitemShut
  {NoStop}%
\bibitem [{\citenamefont {Liao}\ \emph {et~al.}(2010)\citenamefont {Liao},
  \citenamefont {Rittner}, \citenamefont {Paprotta}, \citenamefont {Li},
  \citenamefont {Partridge}, \citenamefont {Hulet}, \citenamefont {Baur},\ and\
  \citenamefont {Mueller}}]{LiaoNature2010}%
  \BibitemOpen
  \bibfield  {author} {\bibinfo {author} {\bibfnamefont {Y.}~\bibnamefont
  {Liao}}, \bibinfo {author} {\bibfnamefont {A.~S.~C.}\ \bibnamefont
  {Rittner}}, \bibinfo {author} {\bibfnamefont {T.}~\bibnamefont {Paprotta}},
  \bibinfo {author} {\bibfnamefont {W.}~\bibnamefont {Li}}, \bibinfo {author}
  {\bibfnamefont {G.~B.}\ \bibnamefont {Partridge}}, \bibinfo {author}
  {\bibfnamefont {R.~G.}\ \bibnamefont {Hulet}}, \bibinfo {author}
  {\bibfnamefont {S.~K.}\ \bibnamefont {Baur}}, \ and\ \bibinfo {author}
  {\bibfnamefont {E.~J.}\ \bibnamefont {Mueller}},\ }\href@noop {} {\bibfield
  {journal} {\bibinfo  {journal} {Nature (London)}\ }\textbf {\bibinfo {volume}
  {467}},\ \bibinfo {pages} {567} (\bibinfo {year} {2010})}\BibitemShut
  {NoStop}%
\bibitem [{\citenamefont {Bedaque}\ \emph {et~al.}(2003)\citenamefont
  {Bedaque}, \citenamefont {Caldas},\ and\ \citenamefont
  {Rupak}}]{PhysRevLett.91.247002}%
  \BibitemOpen
  \bibfield  {author} {\bibinfo {author} {\bibfnamefont {P.~F.}\ \bibnamefont
  {Bedaque}}, \bibinfo {author} {\bibfnamefont {H.}~\bibnamefont {Caldas}}, \
  and\ \bibinfo {author} {\bibfnamefont {G.}~\bibnamefont {Rupak}},\ }\href
  {\doibase 10.1103/PhysRevLett.91.247002} {\bibfield  {journal} {\bibinfo
  {journal} {Phys. Rev. Lett.}\ }\textbf {\bibinfo {volume} {91}},\ \bibinfo
  {pages} {247002} (\bibinfo {year} {2003})}\BibitemShut {NoStop}%
\bibitem [{\citenamefont {Caldas}(2004)}]{PhysRevA.69.063602}%
  \BibitemOpen
  \bibfield  {author} {\bibinfo {author} {\bibfnamefont {H.}~\bibnamefont
  {Caldas}},\ }\href {\doibase 10.1103/PhysRevA.69.063602} {\bibfield
  {journal} {\bibinfo  {journal} {Phys. Rev. A}\ }\textbf {\bibinfo {volume}
  {69}},\ \bibinfo {pages} {063602} (\bibinfo {year} {2004})}\BibitemShut
  {NoStop}%
\bibitem [{\citenamefont {De~Silva}\ and\ \citenamefont
  {Mueller}(2006)}]{PhysRevLett.97.070402}%
  \BibitemOpen
  \bibfield  {author} {\bibinfo {author} {\bibfnamefont {T.~N.}\ \bibnamefont
  {De~Silva}}\ and\ \bibinfo {author} {\bibfnamefont {E.~J.}\ \bibnamefont
  {Mueller}},\ }\href {\doibase 10.1103/PhysRevLett.97.070402} {\bibfield
  {journal} {\bibinfo  {journal} {Phys. Rev. Lett.}\ }\textbf {\bibinfo
  {volume} {97}},\ \bibinfo {pages} {070402} (\bibinfo {year}
  {2006})}\BibitemShut {NoStop}%
\bibitem [{\citenamefont {Caldas}(2007)}]{1742-5468-2007-11-P11012}%
  \BibitemOpen
  \bibfield  {author} {\bibinfo {author} {\bibfnamefont {H.}~\bibnamefont
  {Caldas}},\ }\href {http://stacks.iop.org/1742-5468/2007/i=11/a=P11012}
  {\bibfield  {journal} {\bibinfo  {journal} {J. Stat. Mech.}\ }\textbf
  {\bibinfo {volume} {2007}},\ \bibinfo {pages} {P11012} (\bibinfo {year}
  {2007})}\BibitemShut {NoStop}%
\bibitem [{\citenamefont {{Y.-il Shin}}\ \emph {et~al.}(2008)\citenamefont
  {{Y.-il Shin}}, \citenamefont {Schunck}, \citenamefont {Schirotzek},\ and\
  \citenamefont {Ketterle}}]{ShinNature2008}%
  \BibitemOpen
  \bibfield  {author} {\bibinfo {author} {\bibnamefont {{Y.-il Shin}}},
  \bibinfo {author} {\bibfnamefont {C.~H.}\ \bibnamefont {Schunck}}, \bibinfo
  {author} {\bibfnamefont {A.}~\bibnamefont {Schirotzek}}, \ and\ \bibinfo
  {author} {\bibfnamefont {W.}~\bibnamefont {Ketterle}},\ }\href@noop {}
  {\bibfield  {journal} {\bibinfo  {journal} {Nature (London)}\ }\textbf
  {\bibinfo {volume} {451}},\ \bibinfo {pages} {689} (\bibinfo {year}
  {2008})}\BibitemShut {NoStop}%
\bibitem [{\citenamefont {Nambu}\ and\ \citenamefont
  {Jona-Lasinio}(1961)}]{Nambu:1961tp}%
  \BibitemOpen
  \bibfield  {author} {\bibinfo {author} {\bibfnamefont {Y.}~\bibnamefont
  {Nambu}}\ and\ \bibinfo {author} {\bibfnamefont {G.}~\bibnamefont
  {Jona-Lasinio}},\ }\href@noop {} {\bibfield  {journal} {\bibinfo  {journal}
  {Phys. Rev.}\ }\textbf {\bibinfo {volume} {122}},\ \bibinfo {pages} {345}
  (\bibinfo {year} {1961})}\BibitemShut {NoStop}%
\bibitem [{\citenamefont {Gross}\ and\ \citenamefont
  {Neveu}(1974)}]{Gross:1974jv}%
  \BibitemOpen
  \bibfield  {author} {\bibinfo {author} {\bibfnamefont {D.~J.}\ \bibnamefont
  {Gross}}\ and\ \bibinfo {author} {\bibfnamefont {A.}~\bibnamefont {Neveu}},\
  }\href@noop {} {\bibfield  {journal} {\bibinfo  {journal} {Phys. Rev. D}\
  }\textbf {\bibinfo {volume} {10}},\ \bibinfo {pages} {3235} (\bibinfo {year}
  {1974})}\BibitemShut {NoStop}%
\bibitem [{\citenamefont {Thies}\ and\ \citenamefont
  {Urlichs}(2003)}]{Thies:2003kk}%
  \BibitemOpen
  \bibfield  {author} {\bibinfo {author} {\bibfnamefont {M.}~\bibnamefont
  {Thies}}\ and\ \bibinfo {author} {\bibfnamefont {K.}~\bibnamefont
  {Urlichs}},\ }\href {\doibase 10.1103/PhysRevD.67.125015} {\bibfield
  {journal} {\bibinfo  {journal} {Phys. Rev. D}\ }\textbf {\bibinfo {volume}
  {67}},\ \bibinfo {pages} {125015} (\bibinfo {year} {2003})}\BibitemShut
  {NoStop}%
\bibitem [{\citenamefont {Schnetz}\ \emph {et~al.}(2004)\citenamefont
  {Schnetz}, \citenamefont {Thies},\ and\ \citenamefont
  {Urlichs}}]{Schnetz:2004vr}%
  \BibitemOpen
  \bibfield  {author} {\bibinfo {author} {\bibfnamefont {O.}~\bibnamefont
  {Schnetz}}, \bibinfo {author} {\bibfnamefont {M.}~\bibnamefont {Thies}}, \
  and\ \bibinfo {author} {\bibfnamefont {K.}~\bibnamefont {Urlichs}},\ }\href
  {\doibase 10.1016/j.aop.2004.06.009} {\bibfield  {journal} {\bibinfo
  {journal} {Ann. Phys.}\ }\textbf {\bibinfo {volume} {314}},\ \bibinfo {pages}
  {425} (\bibinfo {year} {2004})}\BibitemShut {NoStop}%
\bibitem [{\citenamefont {Ba{\c{s}}ar}\ and\ \citenamefont
  {Dunne}(2008{\natexlab{a}})}]{Basar:2008im}%
  \BibitemOpen
  \bibfield  {author} {\bibinfo {author} {\bibfnamefont {G.}~\bibnamefont
  {Ba{\c{s}}ar}}\ and\ \bibinfo {author} {\bibfnamefont {G.~V.}\ \bibnamefont
  {Dunne}},\ }\href@noop {} {\bibfield  {journal} {\bibinfo  {journal} {Phys.
  Rev. Lett.}\ }\textbf {\bibinfo {volume} {100}},\ \bibinfo {pages} {200404}
  (\bibinfo {year} {2008}{\natexlab{a}})}\BibitemShut {NoStop}%
\bibitem [{\citenamefont {Ba{\c{s}}ar}\ and\ \citenamefont
  {Dunne}(2008{\natexlab{b}})}]{Basar:2008ki}%
  \BibitemOpen
  \bibfield  {author} {\bibinfo {author} {\bibfnamefont {G.}~\bibnamefont
  {Ba{\c{s}}ar}}\ and\ \bibinfo {author} {\bibfnamefont {G.~V.}\ \bibnamefont
  {Dunne}},\ }\href@noop {} {\bibfield  {journal} {\bibinfo  {journal} {Phys.
  Rev. D}\ }\textbf {\bibinfo {volume} {78}},\ \bibinfo {pages} {065022}
  (\bibinfo {year} {2008}{\natexlab{b}})}\BibitemShut {NoStop}%
\bibitem [{\citenamefont {Ba{\c{s}}ar}\ \emph {et~al.}(2009)\citenamefont
  {Ba{\c{s}}ar}, \citenamefont {Dunne},\ and\ \citenamefont
  {Thies}}]{Basar:2009fg}%
  \BibitemOpen
  \bibfield  {author} {\bibinfo {author} {\bibfnamefont {G.}~\bibnamefont
  {Ba{\c{s}}ar}}, \bibinfo {author} {\bibfnamefont {G.~V.}\ \bibnamefont
  {Dunne}}, \ and\ \bibinfo {author} {\bibfnamefont {M.}~\bibnamefont
  {Thies}},\ }\href@noop {} {\bibfield  {journal} {\bibinfo  {journal} {Phys.
  Rev. D}\ }\textbf {\bibinfo {volume} {79}},\ \bibinfo {pages} {105012}
  (\bibinfo {year} {2009})}\BibitemShut {NoStop}%
\bibitem [{\citenamefont {Correa}\ \emph {et~al.}(2009)\citenamefont {Correa},
  \citenamefont {Dunne},\ and\ \citenamefont {Plyushchay}}]{Correa:2009xa}%
  \BibitemOpen
  \bibfield  {author} {\bibinfo {author} {\bibfnamefont {F.}~\bibnamefont
  {Correa}}, \bibinfo {author} {\bibfnamefont {G.~V.}\ \bibnamefont {Dunne}}, \
  and\ \bibinfo {author} {\bibfnamefont {M.~S.}\ \bibnamefont {Plyushchay}},\
  }\href@noop {} {\bibfield  {journal} {\bibinfo  {journal} {Ann. Phys.}\
  }\textbf {\bibinfo {volume} {324}},\ \bibinfo {pages} {2522} (\bibinfo {year}
  {2009})}\BibitemShut {NoStop}%
\bibitem [{\citenamefont {Hatsuda}\ and\ \citenamefont
  {Kunihiro}(1994)}]{Hatsuda:1994pi}%
  \BibitemOpen
  \bibfield  {author} {\bibinfo {author} {\bibfnamefont {T.}~\bibnamefont
  {Hatsuda}}\ and\ \bibinfo {author} {\bibfnamefont {T.}~\bibnamefont
  {Kunihiro}},\ }\href@noop {} {\bibfield  {journal} {\bibinfo  {journal}
  {Phys. Rept.}\ }\textbf {\bibinfo {volume} {247}},\ \bibinfo {pages} {221}
  (\bibinfo {year} {1994})}\BibitemShut {NoStop}%
\bibitem [{\citenamefont {Adhikari}\ and\ \citenamefont
  {Salasnich}(2008)}]{PhysRevA.78.043616}%
  \BibitemOpen
  \bibfield  {author} {\bibinfo {author} {\bibfnamefont {S.~K.}\ \bibnamefont
  {Adhikari}}\ and\ \bibinfo {author} {\bibfnamefont {L.}~\bibnamefont
  {Salasnich}},\ }\href {\doibase 10.1103/PhysRevA.78.043616} {\bibfield
  {journal} {\bibinfo  {journal} {Phys. Rev. A}\ }\textbf {\bibinfo {volume}
  {78}},\ \bibinfo {pages} {043616} (\bibinfo {year} {2008})}\BibitemShut
  {NoStop}%
\bibitem [{\citenamefont {Pieri}\ and\ \citenamefont
  {Strinati}(2003)}]{PieriStrinati}%
  \BibitemOpen
  \bibfield  {author} {\bibinfo {author} {\bibfnamefont {P.}~\bibnamefont
  {Pieri}}\ and\ \bibinfo {author} {\bibfnamefont {G.~C.}\ \bibnamefont
  {Strinati}},\ }\href@noop {} {\bibfield  {journal} {\bibinfo  {journal}
  {Phys.~Rev.~Lett.}\ }\textbf {\bibinfo {volume} {91}},\ \bibinfo {pages}
  {030401} (\bibinfo {year} {2003})}\BibitemShut {NoStop}%
\bibitem [{\citenamefont {Hakim}(1997)}]{Hakim1997}%
  \BibitemOpen
  \bibfield  {author} {\bibinfo {author} {\bibfnamefont {V.}~\bibnamefont
  {Hakim}},\ }\href@noop {} {\bibfield  {journal} {\bibinfo  {journal} {Phys.\
  Rev.\ E}\ }\textbf {\bibinfo {volume} {55}},\ \bibinfo {pages} {2835}
  (\bibinfo {year} {1997})}\BibitemShut {NoStop}%
\bibitem [{\citenamefont {Engels}\ and\ \citenamefont
  {Atherton}(2007)}]{PhysRevLett.99.160405}%
  \BibitemOpen
  \bibfield  {author} {\bibinfo {author} {\bibfnamefont {P.}~\bibnamefont
  {Engels}}\ and\ \bibinfo {author} {\bibfnamefont {C.}~\bibnamefont
  {Atherton}},\ }\href {\doibase 10.1103/PhysRevLett.99.160405} {\bibfield
  {journal} {\bibinfo  {journal} {Phys. Rev. Lett.}\ }\textbf {\bibinfo
  {volume} {99}},\ \bibinfo {pages} {160405} (\bibinfo {year}
  {2007})}\BibitemShut {NoStop}%
\bibitem [{\citenamefont {Togawa}\ \emph {et~al.}(2012)\citenamefont {Togawa},
  \citenamefont {Koyama}, \citenamefont {Takayanagi}, \citenamefont {Mori},
  \citenamefont {Kousaka}, \citenamefont {Akimitsu}, \citenamefont {Nishihara},
  \citenamefont {Inoue}, \citenamefont {Ovchinnikov},\ and\ \citenamefont
  {Kishine}}]{PhysRevLett.108.107202}%
  \BibitemOpen
  \bibfield  {author} {\bibinfo {author} {\bibfnamefont {Y.}~\bibnamefont
  {Togawa}}, \bibinfo {author} {\bibfnamefont {T.}~\bibnamefont {Koyama}},
  \bibinfo {author} {\bibfnamefont {K.}~\bibnamefont {Takayanagi}}, \bibinfo
  {author} {\bibfnamefont {S.}~\bibnamefont {Mori}}, \bibinfo {author}
  {\bibfnamefont {Y.}~\bibnamefont {Kousaka}}, \bibinfo {author} {\bibfnamefont
  {J.}~\bibnamefont {Akimitsu}}, \bibinfo {author} {\bibfnamefont
  {S.}~\bibnamefont {Nishihara}}, \bibinfo {author} {\bibfnamefont
  {K.}~\bibnamefont {Inoue}}, \bibinfo {author} {\bibfnamefont {A.~S.}\
  \bibnamefont {Ovchinnikov}}, \ and\ \bibinfo {author} {\bibfnamefont
  {J.}~\bibnamefont {Kishine}},\ }\href {\doibase
  10.1103/PhysRevLett.108.107202} {\bibfield  {journal} {\bibinfo  {journal}
  {Phys. Rev. Lett.}\ }\textbf {\bibinfo {volume} {108}},\ \bibinfo {pages}
  {107202} (\bibinfo {year} {2012})}\BibitemShut {NoStop}%
\bibitem [{\citenamefont {Borisov}\ \emph {et~al.}(2009)\citenamefont
  {Borisov}, \citenamefont {Kishine}, \citenamefont {Bostrem},\ and\
  \citenamefont {Ovchinnikov}}]{PhysRevB.79.134436}%
  \BibitemOpen
  \bibfield  {author} {\bibinfo {author} {\bibfnamefont {A.~B.}\ \bibnamefont
  {Borisov}}, \bibinfo {author} {\bibfnamefont {J.}~\bibnamefont {Kishine}},
  \bibinfo {author} {\bibfnamefont {I.~G.}\ \bibnamefont {Bostrem}}, \ and\
  \bibinfo {author} {\bibfnamefont {A.~S.}\ \bibnamefont {Ovchinnikov}},\
  }\href {\doibase 10.1103/PhysRevB.79.134436} {\bibfield  {journal} {\bibinfo
  {journal} {Phys. Rev. B}\ }\textbf {\bibinfo {volume} {79}},\ \bibinfo
  {pages} {134436} (\bibinfo {year} {2009})}\BibitemShut {NoStop}%
\bibitem [{\citenamefont {Novoa}\ \emph {et~al.}(2008)\citenamefont {Novoa},
  \citenamefont {Malomed}, \citenamefont {Michinel},\ and\ \citenamefont
  {P\'erez-Garc\'{\i}a}}]{PhysRevLett.101.144101}%
  \BibitemOpen
  \bibfield  {author} {\bibinfo {author} {\bibfnamefont {D.}~\bibnamefont
  {Novoa}}, \bibinfo {author} {\bibfnamefont {B.~A.}\ \bibnamefont {Malomed}},
  \bibinfo {author} {\bibfnamefont {H.}~\bibnamefont {Michinel}}, \ and\
  \bibinfo {author} {\bibfnamefont {V.~M.}\ \bibnamefont
  {P\'erez-Garc\'{\i}a}},\ }\href {\doibase 10.1103/PhysRevLett.101.144101}
  {\bibfield  {journal} {\bibinfo  {journal} {Phys. Rev. Lett.}\ }\textbf
  {\bibinfo {volume} {101}},\ \bibinfo {pages} {144101} (\bibinfo {year}
  {2008})}\BibitemShut {NoStop}%
\bibitem [{\citenamefont {Muruganandam}\ and\ \citenamefont
  {Adhikari}(2011)}]{MuruganandamAdhikari}%
  \BibitemOpen
  \bibfield  {author} {\bibinfo {author} {\bibfnamefont {P.}~\bibnamefont
  {Muruganandam}}\ and\ \bibinfo {author} {\bibfnamefont {S.~K.}\ \bibnamefont
  {Adhikari}},\ }\href@noop {} {\bibfield  {journal} {\bibinfo  {journal} {J.
  Phys. B}\ }\textbf {\bibinfo {volume} {44}},\ \bibinfo {pages} {121001}
  (\bibinfo {year} {2011})}\BibitemShut {NoStop}%
\bibitem [{\citenamefont {Zakharov}\ and\ \citenamefont
  {Shabat}(1973)}]{ZakharovShabat2}%
  \BibitemOpen
  \bibfield  {author} {\bibinfo {author} {\bibfnamefont {V.~E.}\ \bibnamefont
  {Zakharov}}\ and\ \bibinfo {author} {\bibfnamefont {A.~B.}\ \bibnamefont
  {Shabat}},\ }\href@noop {} {\bibfield  {journal} {\bibinfo  {journal} {Sov.
  Phys. JETP}\ }\textbf {\bibinfo {volume} {37}},\ \bibinfo {pages} {823}
  (\bibinfo {year} {1973})}\BibitemShut {NoStop}%
\bibitem [{\citenamefont {Faddeev}\ and\ \citenamefont
  {Takhtajan}(1987)}]{FaddeevTakhtajan}%
  \BibitemOpen
  \bibfield  {author} {\bibinfo {author} {\bibfnamefont {L.~D.}\ \bibnamefont
  {Faddeev}}\ and\ \bibinfo {author} {\bibfnamefont {L.~A.}\ \bibnamefont
  {Takhtajan}},\ }\href@noop {} {\emph {\bibinfo {title} {Hamiltonian Methods
  in the Theory of Solitons}}}\ (\bibinfo  {publisher} {Springer},\ \bibinfo
  {address} {Berlin},\ \bibinfo {year} {1987})\BibitemShut {NoStop}%
\bibitem [{\citenamefont {Pomeau}\ and\ \citenamefont
  {Rica}(1994)}]{PomeauRica}%
  \BibitemOpen
  \bibfield  {author} {\bibinfo {author} {\bibfnamefont {Y.}~\bibnamefont
  {Pomeau}}\ and\ \bibinfo {author} {\bibfnamefont {S.}~\bibnamefont {Rica}},\
  }\href@noop {} {\bibfield  {journal} {\bibinfo  {journal} {Phys. Rev. Lett.}\
  }\textbf {\bibinfo {volume} {72}},\ \bibinfo {pages} {2426} (\bibinfo {year}
  {1994})}\BibitemShut {NoStop}%
\bibitem [{\citenamefont {Josserand}\ \emph {et~al.}(2007)\citenamefont
  {Josserand}, \citenamefont {Pomeau},\ and\ \citenamefont
  {Rica}}]{JosserandPomeauRica}%
  \BibitemOpen
  \bibfield  {author} {\bibinfo {author} {\bibfnamefont {C.}~\bibnamefont
  {Josserand}}, \bibinfo {author} {\bibfnamefont {Y.}~\bibnamefont {Pomeau}}, \
  and\ \bibinfo {author} {\bibfnamefont {S.}~\bibnamefont {Rica}},\ }\href@noop
  {} {\bibfield  {journal} {\bibinfo  {journal} {Phys. Rev. Lett.}\ }\textbf
  {\bibinfo {volume} {98}},\ \bibinfo {pages} {195301} (\bibinfo {year}
  {2007})}\BibitemShut {NoStop}%
\bibitem [{\citenamefont {Sep{\'u}lveda}\ \emph {et~al.}(2008)\citenamefont
  {Sep{\'u}lveda}, \citenamefont {Josserand},\ and\ \citenamefont
  {Rica}}]{SepulvedaJosserandRica}%
  \BibitemOpen
  \bibfield  {author} {\bibinfo {author} {\bibfnamefont {N.}~\bibnamefont
  {Sep{\'u}lveda}}, \bibinfo {author} {\bibfnamefont {C.}~\bibnamefont
  {Josserand}}, \ and\ \bibinfo {author} {\bibfnamefont {S.}~\bibnamefont
  {Rica}},\ }\href@noop {} {\bibfield  {journal} {\bibinfo  {journal} {Phys.
  Rev. B}\ }\textbf {\bibinfo {volume} {77}},\ \bibinfo {pages} {054513}
  (\bibinfo {year} {2008})}\BibitemShut {NoStop}%
\bibitem [{\citenamefont {Kunimi}\ \emph {et~al.}(2011)\citenamefont {Kunimi},
  \citenamefont {Nagai},\ and\ \citenamefont {Kato}}]{KunimiNagaiKato}%
  \BibitemOpen
  \bibfield  {author} {\bibinfo {author} {\bibfnamefont {M.}~\bibnamefont
  {Kunimi}}, \bibinfo {author} {\bibfnamefont {Y.}~\bibnamefont {Nagai}}, \
  and\ \bibinfo {author} {\bibfnamefont {Y.}~\bibnamefont {Kato}},\ }\href@noop
  {} {\bibfield  {journal} {\bibinfo  {journal} {Phys. Rev. B}\ }\textbf
  {\bibinfo {volume} {84}},\ \bibinfo {pages} {094521} (\bibinfo {year}
  {2011})}\BibitemShut {NoStop}%
\bibitem [{\citenamefont {Kunimi}\ \emph {et~al.}(2012)\citenamefont {Kunimi},
  \citenamefont {Kobayashi},\ and\ \citenamefont {Kato}}]{KunimiConf}%
  \BibitemOpen
  \bibfield  {author} {\bibinfo {author} {\bibfnamefont {M.}~\bibnamefont
  {Kunimi}}, \bibinfo {author} {\bibfnamefont {M.}~\bibnamefont {Kobayashi}}, \
  and\ \bibinfo {author} {\bibfnamefont {Y.}~\bibnamefont {Kato}},\ }\href@noop
  {} {\bibfield  {journal} {\bibinfo  {journal} {J. Phys.: Conf. Ser.}\
  }\textbf {\bibinfo {volume} {400}},\ \bibinfo {pages} {012037} (\bibinfo
  {year} {2012})}\BibitemShut {NoStop}%
\bibitem [{\citenamefont {Kunimi}\ and\ \citenamefont
  {Kato}(2012)}]{PhysRevB.86.060510}%
  \BibitemOpen
  \bibfield  {author} {\bibinfo {author} {\bibfnamefont {M.}~\bibnamefont
  {Kunimi}}\ and\ \bibinfo {author} {\bibfnamefont {Y.}~\bibnamefont {Kato}},\
  }\href {\doibase 10.1103/PhysRevB.86.060510} {\bibfield  {journal} {\bibinfo
  {journal} {Phys. Rev. B}\ }\textbf {\bibinfo {volume} {86}},\ \bibinfo
  {pages} {060510} (\bibinfo {year} {2012})}\BibitemShut {NoStop}%
\bibitem [{\citenamefont {Swift}\ and\ \citenamefont
  {Hohenberg}(1977)}]{SwiftHohenberg}%
  \BibitemOpen
  \bibfield  {author} {\bibinfo {author} {\bibfnamefont {J.}~\bibnamefont
  {Swift}}\ and\ \bibinfo {author} {\bibfnamefont {P.~C.}\ \bibnamefont
  {Hohenberg}},\ }\href@noop {} {\bibfield  {journal} {\bibinfo  {journal}
  {Phys. Rev. A}\ }\textbf {\bibinfo {volume} {15}},\ \bibinfo {pages} {319}
  (\bibinfo {year} {1977})}\BibitemShut {NoStop}%
\bibitem [{\citenamefont {Richter}\ and\ \citenamefont
  {Barashenkov}(2005)}]{RichterBarashenkov}%
  \BibitemOpen
  \bibfield  {author} {\bibinfo {author} {\bibfnamefont {R.}~\bibnamefont
  {Richter}}\ and\ \bibinfo {author} {\bibfnamefont {I.~V.}\ \bibnamefont
  {Barashenkov}},\ }\href@noop {} {\bibfield  {journal} {\bibinfo  {journal}
  {Phys. Rev. Lett.}\ }\textbf {\bibinfo {volume} {94}},\ \bibinfo {pages}
  {184503} (\bibinfo {year} {2005})}\BibitemShut {NoStop}%
\bibitem [{\citenamefont {Buzdin}\ and\ \citenamefont
  {Kachkachi}(1997)}]{BuzdinKachkachi}%
  \BibitemOpen
  \bibfield  {author} {\bibinfo {author} {\bibfnamefont {A.~I.}\ \bibnamefont
  {Buzdin}}\ and\ \bibinfo {author} {\bibfnamefont {H.}~\bibnamefont
  {Kachkachi}},\ }\href@noop {} {\bibfield  {journal} {\bibinfo  {journal}
  {Phys. Lett. A}\ }\textbf {\bibinfo {volume} {225}},\ \bibinfo {pages} {341}
  (\bibinfo {year} {1997})}\BibitemShut {NoStop}%
\bibitem [{\citenamefont {Krichever}(1977)}]{Krichever1977}%
  \BibitemOpen
  \bibfield  {author} {\bibinfo {author} {\bibfnamefont {I.~M.}\ \bibnamefont
  {Krichever}},\ }\href@noop {} {\bibfield  {journal} {\bibinfo  {journal}
  {Funktsional. Anal. i Prilozhen}\ }\textbf {\bibinfo {volume} {11:1}},\
  \bibinfo {pages} {15} (\bibinfo {year} {1977})},\ \bibinfo {note}
  {[Functional Anal. Appl. {\bf 11:1} 12-26 (1977)]}\BibitemShut {NoStop}%
\bibitem [{\citenamefont {Belokolos}\ \emph {et~al.}(1994)\citenamefont
  {Belokolos}, \citenamefont {Bobenko}, \citenamefont {Enol'skii},
  \citenamefont {Its},\ and\ \citenamefont
  {Matveev}}]{BelokolosBobenkoEnolskiiItsMatveev}%
  \BibitemOpen
  \bibfield  {author} {\bibinfo {author} {\bibfnamefont {E.~D.}\ \bibnamefont
  {Belokolos}}, \bibinfo {author} {\bibfnamefont {A.~I.}\ \bibnamefont
  {Bobenko}}, \bibinfo {author} {\bibfnamefont {V.~Z.}\ \bibnamefont
  {Enol'skii}}, \bibinfo {author} {\bibfnamefont {A.~R.}\ \bibnamefont {Its}},
  \ and\ \bibinfo {author} {\bibfnamefont {V.~B.}\ \bibnamefont {Matveev}},\
  }\href@noop {} {\emph {\bibinfo {title} {Algebro-Geometric Approach to
  Nonlinear Integrable Equations}}}\ (\bibinfo  {publisher} {Springer},\
  \bibinfo {address} {Berlin},\ \bibinfo {year} {1994})\BibitemShut {NoStop}%
\bibitem [{\citenamefont {Tanaka}\ and\ \citenamefont
  {Date}(1979)}]{TanakaDate}%
  \BibitemOpen
  \bibfield  {author} {\bibinfo {author} {\bibfnamefont {S.}~\bibnamefont
  {Tanaka}}\ and\ \bibinfo {author} {\bibfnamefont {E.}~\bibnamefont {Date}},\
  }\href@noop {} {\emph {\bibinfo {title} {KdV houteisiki (The KdV
  equation)}}}\ (\bibinfo  {publisher} {Kinokuniya Shoten},\ \bibinfo {address}
  {Tokyo},\ \bibinfo {year} {1979})\ \bibinfo {note} {[written in
  Japanese]}\BibitemShut {NoStop}%
\bibitem [{\citenamefont {Gesztesy}\ and\ \citenamefont
  {Holden}(2003)}]{GesztesyHolden}%
  \BibitemOpen
  \bibfield  {author} {\bibinfo {author} {\bibfnamefont {F.}~\bibnamefont
  {Gesztesy}}\ and\ \bibinfo {author} {\bibfnamefont {H.}~\bibnamefont
  {Holden}},\ }\href@noop {} {\emph {\bibinfo {title} {Soliton Equations and
  Their Algebro-Geometric Solutions}}}\ (\bibinfo  {publisher} {Cambridge},\
  \bibinfo {address} {Cambridge},\ \bibinfo {year} {2003})\BibitemShut
  {NoStop}%
\bibitem [{\citenamefont {Ablowitz}\ \emph {et~al.}(1974)\citenamefont
  {Ablowitz}, \citenamefont {Kaup}, \citenamefont {Newell},\ and\ \citenamefont
  {Segur}}]{AKNS1974}%
  \BibitemOpen
  \bibfield  {author} {\bibinfo {author} {\bibfnamefont {M.~J.}\ \bibnamefont
  {Ablowitz}}, \bibinfo {author} {\bibfnamefont {D.~J.}\ \bibnamefont {Kaup}},
  \bibinfo {author} {\bibfnamefont {A.~C.}\ \bibnamefont {Newell}}, \ and\
  \bibinfo {author} {\bibfnamefont {H.}~\bibnamefont {Segur}},\ }\href@noop {}
  {\bibfield  {journal} {\bibinfo  {journal} {Stud. Appl. Math.}\ }\textbf
  {\bibinfo {volume} {53}},\ \bibinfo {pages} {249} (\bibinfo {year}
  {1974})}\BibitemShut {NoStop}%
\bibitem [{\citenamefont {Takahashi}\ \emph {et~al.}(2012)\citenamefont
  {Takahashi}, \citenamefont {Tsuchiya}, \citenamefont {Yoshii},\ and\
  \citenamefont {Nitta}}]{Takahashi:2012aw}%
  \BibitemOpen
  \bibfield  {author} {\bibinfo {author} {\bibfnamefont {D.~A.}\ \bibnamefont
  {Takahashi}}, \bibinfo {author} {\bibfnamefont {S.}~\bibnamefont {Tsuchiya}},
  \bibinfo {author} {\bibfnamefont {R.}~\bibnamefont {Yoshii}}, \ and\ \bibinfo
  {author} {\bibfnamefont {M.}~\bibnamefont {Nitta}},\ }\href@noop {}
  {\bibfield  {journal} {\bibinfo  {journal} {Phys. Lett. B}\ }\textbf
  {\bibinfo {volume} {718}},\ \bibinfo {pages} {632} (\bibinfo {year}
  {2012})}\BibitemShut {NoStop}%
\bibitem [{\citenamefont {Brazovskii}\ \emph {et~al.}(1980)\citenamefont
  {Brazovskii}, \citenamefont {S.A.Gordyunin},\ and\ \citenamefont
  {Kirova}}]{BrazovskiiGordyuninKirova}%
  \BibitemOpen
  \bibfield  {author} {\bibinfo {author} {\bibfnamefont {S.~A.}\ \bibnamefont
  {Brazovskii}}, \bibinfo {author} {\bibnamefont {S.A.Gordyunin}}, \ and\
  \bibinfo {author} {\bibfnamefont {N.~N.}\ \bibnamefont {Kirova}},\
  }\href@noop {} {\bibfield  {journal} {\bibinfo  {journal} {JETP~Lett.}\
  }\textbf {\bibinfo {volume} {31}},\ \bibinfo {pages} {456} (\bibinfo {year}
  {1980})}\BibitemShut {NoStop}%
\bibitem [{\citenamefont {Horovitz}(1981)}]{Horovitz:1981}%
  \BibitemOpen
  \bibfield  {author} {\bibinfo {author} {\bibfnamefont {B.}~\bibnamefont
  {Horovitz}},\ }\href@noop {} {\bibfield  {journal} {\bibinfo  {journal}
  {Phys. Rev. Lett.}\ }\textbf {\bibinfo {volume} {46}},\ \bibinfo {pages}
  {742} (\bibinfo {year} {1981})}\BibitemShut {NoStop}%
\bibitem [{\citenamefont {Hara}\ and\ \citenamefont {Nagai}(1986)}]{HaraNagai}%
  \BibitemOpen
  \bibfield  {author} {\bibinfo {author} {\bibfnamefont {J.}~\bibnamefont
  {Hara}}\ and\ \bibinfo {author} {\bibfnamefont {K.}~\bibnamefont {Nagai}},\
  }\href@noop {} {\bibfield  {journal} {\bibinfo  {journal} {Prog. Theor.
  Phys.}\ }\textbf {\bibinfo {volume} {76}},\ \bibinfo {pages} {1237} (\bibinfo
  {year} {1986})}\BibitemShut {NoStop}%
\bibitem [{\citenamefont {Dalfovo}\ \emph {et~al.}(1999)\citenamefont
  {Dalfovo}, \citenamefont {Giorgini}, \citenamefont {Pitaevskii},\ and\
  \citenamefont {Stringari}}]{DalfovoGiorginiPitaevskiiStringari}%
  \BibitemOpen
  \bibfield  {author} {\bibinfo {author} {\bibfnamefont {F.}~\bibnamefont
  {Dalfovo}}, \bibinfo {author} {\bibfnamefont {S.}~\bibnamefont {Giorgini}},
  \bibinfo {author} {\bibfnamefont {L.~P.}\ \bibnamefont {Pitaevskii}}, \ and\
  \bibinfo {author} {\bibfnamefont {S.}~\bibnamefont {Stringari}},\ }\href@noop
  {} {\bibfield  {journal} {\bibinfo  {journal} {Rev. Mod. Phys.}\ }\textbf
  {\bibinfo {volume} {71}},\ \bibinfo {pages} {463} (\bibinfo {year}
  {1999})}\BibitemShut {NoStop}%
\bibitem [{\citenamefont {Kaup}(1976)}]{Kaup1976}%
  \BibitemOpen
  \bibfield  {author} {\bibinfo {author} {\bibfnamefont {D.~J.}\ \bibnamefont
  {Kaup}},\ }\href@noop {} {\bibfield  {journal} {\bibinfo  {journal} {J. Math.
  Anal. Appl.}\ }\textbf {\bibinfo {volume} {54}},\ \bibinfo {pages} {849}
  (\bibinfo {year} {1976})}\BibitemShut {NoStop}%
\bibitem [{\citenamefont {Chen}\ \emph {et~al.}(1998)\citenamefont {Chen},
  \citenamefont {Chen},\ and\ \citenamefont {Huang}}]{ChenChenHuang}%
  \BibitemOpen
  \bibfield  {author} {\bibinfo {author} {\bibfnamefont {X.-J.}\ \bibnamefont
  {Chen}}, \bibinfo {author} {\bibfnamefont {Z.-D.}\ \bibnamefont {Chen}}, \
  and\ \bibinfo {author} {\bibfnamefont {N.-N.}\ \bibnamefont {Huang}},\
  }\href@noop {} {\bibfield  {journal} {\bibinfo  {journal} {J.\ Phys.\ A}\
  }\textbf {\bibinfo {volume} {31}},\ \bibinfo {pages} {6929} (\bibinfo {year}
  {1998})}\BibitemShut {NoStop}%
\bibitem [{\citenamefont {Takahashi}\ and\ \citenamefont
  {Nitta}(2015)}]{Takahashi2015101}%
  \BibitemOpen
  \bibfield  {author} {\bibinfo {author} {\bibfnamefont {D.~A.}\ \bibnamefont
  {Takahashi}}\ and\ \bibinfo {author} {\bibfnamefont {M.}~\bibnamefont
  {Nitta}},\ }\href {\doibase http://dx.doi.org/10.1016/j.aop.2014.12.009}
  {\bibfield  {journal} {\bibinfo  {journal} {Ann. Phys.}\ }\textbf {\bibinfo
  {volume} {354}},\ \bibinfo {pages} {101 } (\bibinfo {year}
  {2015})}\BibitemShut {NoStop}%
\bibitem [{\citenamefont {Nitta}\ and\ \citenamefont
  {Takahashi}(2015)}]{PhysRevD.91.025018}%
  \BibitemOpen
  \bibfield  {author} {\bibinfo {author} {\bibfnamefont {M.}~\bibnamefont
  {Nitta}}\ and\ \bibinfo {author} {\bibfnamefont {D.~A.}\ \bibnamefont
  {Takahashi}},\ }\href {\doibase 10.1103/PhysRevD.91.025018} {\bibfield
  {journal} {\bibinfo  {journal} {Phys. Rev. D}\ }\textbf {\bibinfo {volume}
  {91}},\ \bibinfo {pages} {025018} (\bibinfo {year} {2015})}\BibitemShut
  {NoStop}%
\bibitem [{\citenamefont {Takahashi}\ \emph {et~al.}(2015)\citenamefont
  {Takahashi}, \citenamefont {Kobayashi},\ and\ \citenamefont
  {Nitta}}]{PhysRevB.91.184501}%
  \BibitemOpen
  \bibfield  {author} {\bibinfo {author} {\bibfnamefont {D.~A.}\ \bibnamefont
  {Takahashi}}, \bibinfo {author} {\bibfnamefont {M.}~\bibnamefont
  {Kobayashi}}, \ and\ \bibinfo {author} {\bibfnamefont {M.}~\bibnamefont
  {Nitta}},\ }\href {\doibase 10.1103/PhysRevB.91.184501} {\bibfield  {journal}
  {\bibinfo  {journal} {Phys. Rev. B}\ }\textbf {\bibinfo {volume} {91}},\
  \bibinfo {pages} {184501} (\bibinfo {year} {2015})}\BibitemShut {NoStop}%
\bibitem [{\citenamefont {Watanabe}\ and\ \citenamefont
  {Brauner}(2011)}]{Watanabe:2011ec}%
  \BibitemOpen
  \bibfield  {author} {\bibinfo {author} {\bibfnamefont {H.}~\bibnamefont
  {Watanabe}}\ and\ \bibinfo {author} {\bibfnamefont {T.}~\bibnamefont
  {Brauner}},\ }\href {\doibase 10.1103/PhysRevD.84.125013} {\bibfield
  {journal} {\bibinfo  {journal} {Phys. Rev. D}\ }\textbf {\bibinfo {volume}
  {84}},\ \bibinfo {pages} {125013} (\bibinfo {year} {2011})}\BibitemShut
  {NoStop}%
\bibitem [{\citenamefont {Watanabe}\ and\ \citenamefont
  {Murayama}(2012)}]{Watanabe:2012hr}%
  \BibitemOpen
  \bibfield  {author} {\bibinfo {author} {\bibfnamefont {H.}~\bibnamefont
  {Watanabe}}\ and\ \bibinfo {author} {\bibfnamefont {H.}~\bibnamefont
  {Murayama}},\ }\href {\doibase 10.1103/PhysRevLett.108.251602} {\bibfield
  {journal} {\bibinfo  {journal} {Phys. Rev. Lett.}\ }\textbf {\bibinfo
  {volume} {108}},\ \bibinfo {pages} {251602} (\bibinfo {year}
  {2012})}\BibitemShut {NoStop}%
\bibitem [{\citenamefont {Hidaka}(2013)}]{Hidaka:2012ym}%
  \BibitemOpen
  \bibfield  {author} {\bibinfo {author} {\bibfnamefont {Y.}~\bibnamefont
  {Hidaka}},\ }\href {\doibase 10.1103/PhysRevLett.110.091601} {\bibfield
  {journal} {\bibinfo  {journal} {Phys. Rev. Lett.}\ }\textbf {\bibinfo
  {volume} {110}},\ \bibinfo {pages} {091601} (\bibinfo {year}
  {2013})}\BibitemShut {NoStop}%
\bibitem [{\citenamefont {Watanabe}\ and\ \citenamefont
  {Murayama}(2014)}]{PhysRevLett.112.191804}%
  \BibitemOpen
  \bibfield  {author} {\bibinfo {author} {\bibfnamefont {H.}~\bibnamefont
  {Watanabe}}\ and\ \bibinfo {author} {\bibfnamefont {H.}~\bibnamefont
  {Murayama}},\ }\href {\doibase 10.1103/PhysRevLett.112.191804} {\bibfield
  {journal} {\bibinfo  {journal} {Phys. Rev. Lett.}\ }\textbf {\bibinfo
  {volume} {112}},\ \bibinfo {pages} {191804} (\bibinfo {year}
  {2014})}\BibitemShut {NoStop}%
\bibitem [{\citenamefont {Kobayashi}\ and\ \citenamefont
  {Nitta}(2014)}]{PhysRevLett.113.120403}%
  \BibitemOpen
  \bibfield  {author} {\bibinfo {author} {\bibfnamefont {M.}~\bibnamefont
  {Kobayashi}}\ and\ \bibinfo {author} {\bibfnamefont {M.}~\bibnamefont
  {Nitta}},\ }\href {\doibase 10.1103/PhysRevLett.113.120403} {\bibfield
  {journal} {\bibinfo  {journal} {Phys. Rev. Lett.}\ }\textbf {\bibinfo
  {volume} {113}},\ \bibinfo {pages} {120403} (\bibinfo {year}
  {2014})}\BibitemShut {NoStop}%
\bibitem [{See()}]{SeeAnimation}%
  \BibitemOpen
  \href@noop {} {}\bibinfo {note} {See Supplemental Material at [URL will be
  inserted by publisher] for gif animation files of the soliton
  solutions.}\BibitemShut {Stop}%
\bibitem [{\citenamefont {Kuznetsov}\ and\ \citenamefont
  {Mikha{\u{i}}lov}(1975)}]{KuznetsovMikhailov}%
  \BibitemOpen
  \bibfield  {author} {\bibinfo {author} {\bibfnamefont {E.~A.}\ \bibnamefont
  {Kuznetsov}}\ and\ \bibinfo {author} {\bibfnamefont {A.~V.}\ \bibnamefont
  {Mikha{\u{i}}lov}},\ }\href@noop {} {\bibfield  {journal} {\bibinfo
  {journal} {Sov. Phys. JETP}\ }\textbf {\bibinfo {volume} {40}},\ \bibinfo
  {pages} {855} (\bibinfo {year} {1975})}\BibitemShut {NoStop}%
\bibitem [{\citenamefont {Burger}\ \emph {et~al.}(1999)\citenamefont {Burger},
  \citenamefont {Bongs}, \citenamefont {Dettmer}, \citenamefont {Ertmer},
  \citenamefont {Sengstock}, \citenamefont {Sanpera}, \citenamefont
  {Shlyapnikov},\ and\ \citenamefont {Lewenstein}}]{PhysRevLett.83.5198}%
  \BibitemOpen
  \bibfield  {author} {\bibinfo {author} {\bibfnamefont {S.}~\bibnamefont
  {Burger}}, \bibinfo {author} {\bibfnamefont {K.}~\bibnamefont {Bongs}},
  \bibinfo {author} {\bibfnamefont {S.}~\bibnamefont {Dettmer}}, \bibinfo
  {author} {\bibfnamefont {W.}~\bibnamefont {Ertmer}}, \bibinfo {author}
  {\bibfnamefont {K.}~\bibnamefont {Sengstock}}, \bibinfo {author}
  {\bibfnamefont {A.}~\bibnamefont {Sanpera}}, \bibinfo {author} {\bibfnamefont
  {G.~V.}\ \bibnamefont {Shlyapnikov}}, \ and\ \bibinfo {author} {\bibfnamefont
  {M.}~\bibnamefont {Lewenstein}},\ }\href {\doibase
  10.1103/PhysRevLett.83.5198} {\bibfield  {journal} {\bibinfo  {journal}
  {Phys. Rev. Lett.}\ }\textbf {\bibinfo {volume} {83}},\ \bibinfo {pages}
  {5198} (\bibinfo {year} {1999})}\BibitemShut {NoStop}%
\bibitem [{\citenamefont {Muryshev}\ \emph {et~al.}(1999)\citenamefont
  {Muryshev}, \citenamefont {van Linden van~den Heuvell},\ and\ \citenamefont
  {Shlyapnikov}}]{PhysRevA.60.R2665}%
  \BibitemOpen
  \bibfield  {author} {\bibinfo {author} {\bibfnamefont {A.~E.}\ \bibnamefont
  {Muryshev}}, \bibinfo {author} {\bibfnamefont {H.~B.}\ \bibnamefont {van
  Linden van~den Heuvell}}, \ and\ \bibinfo {author} {\bibfnamefont {G.~V.}\
  \bibnamefont {Shlyapnikov}},\ }\href {\doibase 10.1103/PhysRevA.60.R2665}
  {\bibfield  {journal} {\bibinfo  {journal} {Phys. Rev. A}\ }\textbf {\bibinfo
  {volume} {60}},\ \bibinfo {pages} {R2665} (\bibinfo {year}
  {1999})}\BibitemShut {NoStop}%
\bibitem [{\citenamefont {Fedichev}\ \emph {et~al.}(1999)\citenamefont
  {Fedichev}, \citenamefont {Muryshev},\ and\ \citenamefont
  {Shlyapnikov}}]{PhysRevA.60.3220}%
  \BibitemOpen
  \bibfield  {author} {\bibinfo {author} {\bibfnamefont {P.~O.}\ \bibnamefont
  {Fedichev}}, \bibinfo {author} {\bibfnamefont {A.~E.}\ \bibnamefont
  {Muryshev}}, \ and\ \bibinfo {author} {\bibfnamefont {G.~V.}\ \bibnamefont
  {Shlyapnikov}},\ }\href {\doibase 10.1103/PhysRevA.60.3220} {\bibfield
  {journal} {\bibinfo  {journal} {Phys. Rev. A}\ }\textbf {\bibinfo {volume}
  {60}},\ \bibinfo {pages} {3220} (\bibinfo {year} {1999})}\BibitemShut
  {NoStop}%
\bibitem [{\citenamefont {Muryshev}\ \emph {et~al.}(2002)\citenamefont
  {Muryshev}, \citenamefont {Shlyapnikov}, \citenamefont {Ertmer},
  \citenamefont {Sengstock},\ and\ \citenamefont
  {Lewenstein}}]{PhysRevLett.89.110401}%
  \BibitemOpen
  \bibfield  {author} {\bibinfo {author} {\bibfnamefont {A.}~\bibnamefont
  {Muryshev}}, \bibinfo {author} {\bibfnamefont {G.~V.}\ \bibnamefont
  {Shlyapnikov}}, \bibinfo {author} {\bibfnamefont {W.}~\bibnamefont {Ertmer}},
  \bibinfo {author} {\bibfnamefont {K.}~\bibnamefont {Sengstock}}, \ and\
  \bibinfo {author} {\bibfnamefont {M.}~\bibnamefont {Lewenstein}},\ }\href
  {\doibase 10.1103/PhysRevLett.89.110401} {\bibfield  {journal} {\bibinfo
  {journal} {Phys. Rev. Lett.}\ }\textbf {\bibinfo {volume} {89}},\ \bibinfo
  {pages} {110401} (\bibinfo {year} {2002})}\BibitemShut {NoStop}%
\bibitem [{\citenamefont {Takahashi}\ and\ \citenamefont
  {Nitta}(2013)}]{Takahashi:2012pk}%
  \BibitemOpen
  \bibfield  {author} {\bibinfo {author} {\bibfnamefont {D.~A.}\ \bibnamefont
  {Takahashi}}\ and\ \bibinfo {author} {\bibfnamefont {M.}~\bibnamefont
  {Nitta}},\ }\href@noop {} {\bibfield  {journal} {\bibinfo  {journal} {Phys.
  Rev. Lett.}\ }\textbf {\bibinfo {volume} {110}},\ \bibinfo {pages} {131601}
  (\bibinfo {year} {2013})}\BibitemShut {NoStop}%
\bibitem [{\citenamefont {Takahashi}(2016{\natexlab{a}})}]{Takahashi:2015hdt}%
  \BibitemOpen
  \bibfield  {author} {\bibinfo {author} {\bibfnamefont {D.~A.}\ \bibnamefont
  {Takahashi}},\ }\href@noop {} {\bibfield  {journal} {\bibinfo  {journal}
  {Prog. Theor. Exp. Phys.}\ }\textbf {\bibinfo {volume} {2016}},\ \bibinfo
  {pages} {043I01} (\bibinfo {year} {2016}{\natexlab{a}})}\BibitemShut
  {NoStop}%
\bibitem [{\citenamefont {Kharchev}\ and\ \citenamefont
  {Zabrodin}(2015)}]{Kharchev201519}%
  \BibitemOpen
  \bibfield  {author} {\bibinfo {author} {\bibfnamefont {S.}~\bibnamefont
  {Kharchev}}\ and\ \bibinfo {author} {\bibfnamefont {A.}~\bibnamefont
  {Zabrodin}},\ }\href {\doibase
  http://dx.doi.org/10.1016/j.geomphys.2015.03.010} {\bibfield  {journal}
  {\bibinfo  {journal} {Journal of Geometry and Physics}\ }\textbf {\bibinfo
  {volume} {94}},\ \bibinfo {pages} {19 } (\bibinfo {year} {2015})},\ \bibinfo
  {note} {arXiv:1502.04603}\BibitemShut {NoStop}%
\bibitem [{\citenamefont {Wadati}\ \emph {et~al.}(1975)\citenamefont {Wadati},
  \citenamefont {Sanuki},\ and\ \citenamefont {Konno}}]{WadatiSanukiKonno}%
  \BibitemOpen
  \bibfield  {author} {\bibinfo {author} {\bibfnamefont {M.}~\bibnamefont
  {Wadati}}, \bibinfo {author} {\bibfnamefont {H.}~\bibnamefont {Sanuki}}, \
  and\ \bibinfo {author} {\bibfnamefont {K.}~\bibnamefont {Konno}},\
  }\href@noop {} {\bibfield  {journal} {\bibinfo  {journal} {Prog. Theor.
  Phys.}\ }\textbf {\bibinfo {volume} {53}},\ \bibinfo {pages} {419} (\bibinfo
  {year} {1975})}\BibitemShut {NoStop}%
\bibitem [{\citenamefont {Yefsah}\ \emph {et~al.}(2013)\citenamefont {Yefsah},
  \citenamefont {Sommer}, \citenamefont {Ku}, \citenamefont {Cheuk},
  \citenamefont {Ji}, \citenamefont {Bakr},\ and\ \citenamefont
  {Zwierlein}}]{Nature4994262013}%
  \BibitemOpen
  \bibfield  {author} {\bibinfo {author} {\bibfnamefont {T.}~\bibnamefont
  {Yefsah}}, \bibinfo {author} {\bibfnamefont {A.~T.}\ \bibnamefont {Sommer}},
  \bibinfo {author} {\bibfnamefont {M.~J.~H.}\ \bibnamefont {Ku}}, \bibinfo
  {author} {\bibfnamefont {L.~W.}\ \bibnamefont {Cheuk}}, \bibinfo {author}
  {\bibfnamefont {W.}~\bibnamefont {Ji}}, \bibinfo {author} {\bibfnamefont
  {W.~S.}\ \bibnamefont {Bakr}}, \ and\ \bibinfo {author} {\bibfnamefont
  {M.~W.}\ \bibnamefont {Zwierlein}},\ }\href@noop {} {\bibfield  {journal}
  {\bibinfo  {journal} {Nature}\ }\textbf {\bibinfo {volume} {499}},\ \bibinfo
  {pages} {426} (\bibinfo {year} {2013})}\BibitemShut {NoStop}%
\bibitem [{\citenamefont {Ku}\ \emph {et~al.}(2016)\citenamefont {Ku},
  \citenamefont {Mukherjee}, \citenamefont {Yefsah},\ and\ \citenamefont
  {Zwierlein}}]{PhysRevLett.116.045304}%
  \BibitemOpen
  \bibfield  {author} {\bibinfo {author} {\bibfnamefont {M.~J.~H.}\
  \bibnamefont {Ku}}, \bibinfo {author} {\bibfnamefont {B.}~\bibnamefont
  {Mukherjee}}, \bibinfo {author} {\bibfnamefont {T.}~\bibnamefont {Yefsah}}, \
  and\ \bibinfo {author} {\bibfnamefont {M.~W.}\ \bibnamefont {Zwierlein}},\
  }\href {\doibase 10.1103/PhysRevLett.116.045304} {\bibfield  {journal}
  {\bibinfo  {journal} {Phys. Rev. Lett.}\ }\textbf {\bibinfo {volume} {116}},\
  \bibinfo {pages} {045304} (\bibinfo {year} {2016})}\BibitemShut {NoStop}%
\bibitem [{\citenamefont {Kadau}\ \emph {et~al.}(2016)\citenamefont {Kadau},
  \citenamefont {Schmitt}, \citenamefont {Wenzel}, \citenamefont {Wink},
  \citenamefont {Maier}, \citenamefont {Ferrier-Barbut},\ and\ \citenamefont
  {Pfau}}]{Nature5301942016}%
  \BibitemOpen
  \bibfield  {author} {\bibinfo {author} {\bibfnamefont {H.}~\bibnamefont
  {Kadau}}, \bibinfo {author} {\bibfnamefont {M.}~\bibnamefont {Schmitt}},
  \bibinfo {author} {\bibfnamefont {M.}~\bibnamefont {Wenzel}}, \bibinfo
  {author} {\bibfnamefont {C.}~\bibnamefont {Wink}}, \bibinfo {author}
  {\bibfnamefont {T.}~\bibnamefont {Maier}}, \bibinfo {author} {\bibfnamefont
  {I.}~\bibnamefont {Ferrier-Barbut}}, \ and\ \bibinfo {author} {\bibfnamefont
  {T.}~\bibnamefont {Pfau}},\ }\href@noop {} {\bibfield  {journal} {\bibinfo
  {journal} {Nature}\ }\textbf {\bibinfo {volume} {530}},\ \bibinfo {pages}
  {194} (\bibinfo {year} {2016})}\BibitemShut {NoStop}%
\bibitem [{Fer()}]{FerrierBarbutKadauSchmittWenzelPfau}%
  \BibitemOpen
  \href@noop {} {}\bibinfo {note} {I. Ferrier-Barbut, H. Kadau, M. Schmitt, M.
  Wenzel, and T. Pfau, arXiv:1601.03318.}\BibitemShut {Stop}%
\bibitem [{\citenamefont {Hidaka}\ \emph {et~al.}(2015)\citenamefont {Hidaka},
  \citenamefont {Kamikado}, \citenamefont {Kanazawa},\ and\ \citenamefont
  {Noumi}}]{PhysRevD.92.034003}%
  \BibitemOpen
  \bibfield  {author} {\bibinfo {author} {\bibfnamefont {Y.}~\bibnamefont
  {Hidaka}}, \bibinfo {author} {\bibfnamefont {K.}~\bibnamefont {Kamikado}},
  \bibinfo {author} {\bibfnamefont {T.}~\bibnamefont {Kanazawa}}, \ and\
  \bibinfo {author} {\bibfnamefont {T.}~\bibnamefont {Noumi}},\ }\href
  {\doibase 10.1103/PhysRevD.92.034003} {\bibfield  {journal} {\bibinfo
  {journal} {Phys. Rev. D}\ }\textbf {\bibinfo {volume} {92}},\ \bibinfo
  {pages} {034003} (\bibinfo {year} {2015})}\BibitemShut {NoStop}%
\bibitem [{Kot()}]{KotlyarovIts1976}%
  \BibitemOpen
  \href@noop {} {}\bibinfo {note} {V, Kotlyarov and A. Its,
  arXiv:1401.4445.}\BibitemShut {Stop}%
\bibitem [{\citenamefont {Arancibia}\ \emph {et~al.}(2014)\citenamefont
  {Arancibia}, \citenamefont {Correa}, \citenamefont {Jakubsk\'y},
  \citenamefont {Mateos~Guilarte},\ and\ \citenamefont
  {Plyushchay}}]{PhysRevD.90.125041}%
  \BibitemOpen
  \bibfield  {author} {\bibinfo {author} {\bibfnamefont {A.}~\bibnamefont
  {Arancibia}}, \bibinfo {author} {\bibfnamefont {F.}~\bibnamefont {Correa}},
  \bibinfo {author} {\bibfnamefont {V.}~\bibnamefont {Jakubsk\'y}}, \bibinfo
  {author} {\bibfnamefont {J.}~\bibnamefont {Mateos~Guilarte}}, \ and\ \bibinfo
  {author} {\bibfnamefont {M.~S.}\ \bibnamefont {Plyushchay}},\ }\href
  {\doibase 10.1103/PhysRevD.90.125041} {\bibfield  {journal} {\bibinfo
  {journal} {Phys. Rev. D}\ }\textbf {\bibinfo {volume} {90}},\ \bibinfo
  {pages} {125041} (\bibinfo {year} {2014})}\BibitemShut {NoStop}%
\bibitem [{\citenamefont {Liu}\ \emph {et~al.}(2015)\citenamefont {Liu},
  \citenamefont {Tian}, \citenamefont {Sun},\ and\ \citenamefont
  {Wang}}]{1402-4896-90-4-045205}%
  \BibitemOpen
  \bibfield  {author} {\bibinfo {author} {\bibfnamefont {D.-Y.}\ \bibnamefont
  {Liu}}, \bibinfo {author} {\bibfnamefont {B.}~\bibnamefont {Tian}}, \bibinfo
  {author} {\bibfnamefont {W.-R.}\ \bibnamefont {Sun}}, \ and\ \bibinfo
  {author} {\bibfnamefont {Y.-P.}\ \bibnamefont {Wang}},\ }\href
  {http://stacks.iop.org/1402-4896/90/i=4/a=045205} {\bibfield  {journal}
  {\bibinfo  {journal} {Physica Scripta}\ }\textbf {\bibinfo {volume} {90}},\
  \bibinfo {pages} {045205} (\bibinfo {year} {2015})}\BibitemShut {NoStop}%
\bibitem [{Dya()}]{DyachenkoZakharovZakharov}%
  \BibitemOpen
  \href@noop {} {}\bibinfo {note} {S. A. Dyachenko, D. Zakharov, and V.
  Zakharov, arXiv:1505.05806.}\BibitemShut {Stop}%
\bibitem [{\citenamefont {Arancibia}\ and\ \citenamefont
  {Plyushchay}(2015)}]{PhysRevD.92.105009}%
  \BibitemOpen
  \bibfield  {author} {\bibinfo {author} {\bibfnamefont {A.}~\bibnamefont
  {Arancibia}}\ and\ \bibinfo {author} {\bibfnamefont {M.~S.}\ \bibnamefont
  {Plyushchay}},\ }\href {\doibase 10.1103/PhysRevD.92.105009} {\bibfield
  {journal} {\bibinfo  {journal} {Phys. Rev. D}\ }\textbf {\bibinfo {volume}
  {92}},\ \bibinfo {pages} {105009} (\bibinfo {year} {2015})}\BibitemShut
  {NoStop}%
\bibitem [{Smi()}]{SmirnovMatveev}%
  \BibitemOpen
  \href@noop {} {}\bibinfo {note} {A. O. Smirnov and V. B. Matveev,
  arXiv:1509.01134.}\BibitemShut {Stop}%
\bibitem [{\citenamefont {Dunne}\ and\ \citenamefont
  {Thies}(2013)}]{PhysRevLett.111.121602}%
  \BibitemOpen
  \bibfield  {author} {\bibinfo {author} {\bibfnamefont {G.~V.}\ \bibnamefont
  {Dunne}}\ and\ \bibinfo {author} {\bibfnamefont {M.}~\bibnamefont {Thies}},\
  }\href {\doibase 10.1103/PhysRevLett.111.121602} {\bibfield  {journal}
  {\bibinfo  {journal} {Phys. Rev. Lett.}\ }\textbf {\bibinfo {volume} {111}},\
  \bibinfo {pages} {121602} (\bibinfo {year} {2013})}\BibitemShut {NoStop}%
\bibitem [{\citenamefont {Takahashi}(2016{\natexlab{b}})}]{Takahashi:2015nda}%
  \BibitemOpen
  \bibfield  {author} {\bibinfo {author} {\bibfnamefont {D.~A.}\ \bibnamefont
  {Takahashi}},\ }\href@noop {} {\bibfield  {journal} {\bibinfo  {journal}
  {Phys. Rev. B}\ }\textbf {\bibinfo {volume} {93}},\ \bibinfo {pages} {024512}
  (\bibinfo {year} {2016}{\natexlab{b}})}\BibitemShut {NoStop}%
\bibitem [{\citenamefont {Toda}(1989)}]{Todalattice}%
  \BibitemOpen
  \bibfield  {author} {\bibinfo {author} {\bibfnamefont {M.}~\bibnamefont
  {Toda}},\ }\href@noop {} {\emph {\bibinfo {title} {Theory of Nonlinear
  Lattices}}},\ \bibinfo {edition} {2nd}\ ed.\ (\bibinfo  {publisher}
  {Springer},\ \bibinfo {address} {Berlin},\ \bibinfo {year}
  {1989})\BibitemShut {NoStop}%
\end{thebibliography}
%
\end{document}